\documentclass[twocolumn]{aastex62}
\graphicspath{{./}{figures/}}

\usepackage{color,subfigure,lineno} 

\graphicspath{{HCH_figs/} {quasar_non0_parallax/} {quasar_non0_pm/}}
\newcommand{\oiii}{\hbox{[O\,{\scriptsize III}]}}
\usepackage{graphbox}

\shorttitle{VODKA: Methodology}
\shortauthors{Hwang et al.}

\begin{document}

\title{Varstrometry for Off-nucleus and Dual sub-Kpc AGN (VODKA): Methodology and Initial Results with Gaia DR2}


\author[0000-0003-4250-4437]{Hsiang-Chih Hwang}
\affiliation{Department of Physics and Astronomy, Johns Hopkins University}

\author[0000-0003-1659-7035]{Yue Shen}
\altaffiliation{Alfred P. Sloan Research Fellow}
\affiliation{Department of Astronomy, University of Illinois at Urbana-Champaign, Urbana, IL 61801, USA}
\affiliation{National Center for Supercomputing Applications, University of Illinois at Urbana-Champaign, Urbana, IL 61801, USA}

\author{Nadia Zakamska}
\affiliation{Department of Physics and Astronomy, Johns Hopkins University}

\author[0000-0003-0049-5210]{Xin Liu}
\affiliation{Department of Astronomy, University of Illinois at Urbana-Champaign, Urbana, IL 61801, USA}
\affiliation{National Center for Supercomputing Applications, University of Illinois at Urbana-Champaign, Urbana, IL 61801, USA}

\begin{abstract}
Gaia's milli-arcsec (mas) astrometric precision allows systematic identification of optically-selected sub-kpc dual active galactic nuclei (AGN), off-nucleus AGN, and small-scale lensed quasars by `varstrometry' -- where variability-induced astrometric jitter, i.e., temporal displacements of photocenter in unresolved sources, can be reasonably well detected or constrained. This approach extends systematic searches for small-scale ($\gtrsim$ mas) dual and off-nucleus AGN to poorly explored regime between $\sim 10$ pc and $\sim 1$ kpc, with Gaia's full sky coverage and depth to $G\sim 21$. We outline the general principles of this method and calculate the expected astrometric signals from the full time series of photocenter measurements and light curves. We demonstrate the feasibility of varstrometry by using Gaia DR2 data on a sample of variable pre-main sequence stars with known close companions. We find that extended host galaxies have a significant impact on the accuracy of astrometric and photometric variability in Gaia DR2, a situation to be improved in future Gaia releases. Using spectroscopically confirmed SDSS quasars, we present several examples of candidate sub-kpc off-nucleus or dual AGN selected from Gaia DR2. We discuss the merits and limitations of this method and follow-up strategy for promising candidates. We highlight Gaia's potential of systematically discovering and characterizing the sub-kpc off-nucleus and dual AGN population in the entire optical sky. 


\end{abstract}


\section{Introduction}\label{sec:introduction}


The search for and characterization of the binary supermassive black hole population are important both for understanding galaxy formation and for the prospects of low-frequency gravitational wave detection \citep[e.g.,][]{haehnelt94,volonteri03,Jaffe2003,hughes09,Centrella2010,Bogdanovic2015}. Following the merger of two galaxies, the two BHs within each of the galaxies may eventually evolve into a bound binary via dynamical friction and interactions with gas and stars \citep[e.g.,][]{begelman80,gould00,milosavljevic01,blaes02,yu02,Khan2013,DEGN}. Studying SMBH pairs at different evolutionary stages, e.g., from tens of kpc separations at the beginning of the merger to $\lesssim$10 pc scales when the two BHs are gravitationally bound to each other, is important for understanding the impact of galaxy mergers on BH fueling and the dynamical evolution of binary SMBHs \citep[e.g.,][]{colpi09,Dotti2012,Volonteri2016}. 

\begin{figure}
  \includegraphics[width=0.48\textwidth]{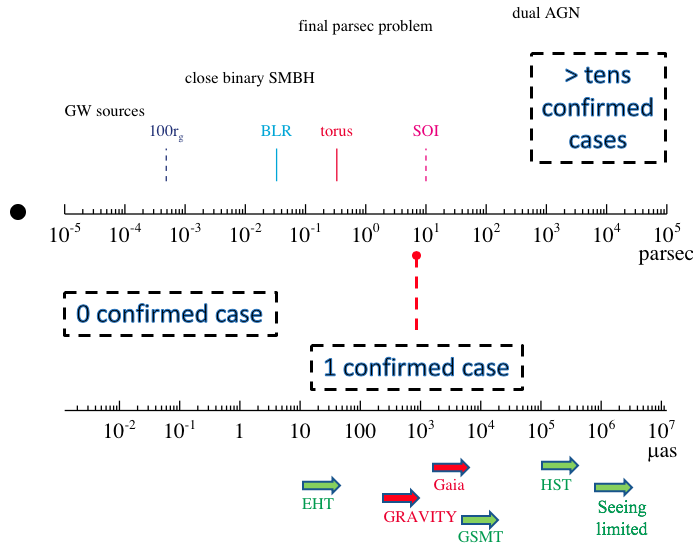}
  \caption{Physical and angular scales around a $10^8\,M_\odot$ SMBH accreting at a typical Eddington ratio of 0.1 at $z=1$. The four vertical line segments mark the location of 100 gravitational radii ($r_g$), the typical distance of the broad-line region (BLR), the typical distance of the dust torus, and the sphere of influence (SOI) of the BH. There are more than a few tens confirmed dual AGN with separations greater than $\sim 1$ kpc. There is only one confirmed $\ll$ kpc binary SMBH with a separation of 7 pc \citep{rodriguez06}. The minimal spatial scale from different facilities are indicated in the bottom arrows. Direct imaging facilities are indicated in green and astrometry facilities are indicated in red. Only the Event Horizon Telescope \citep[EHT,][]{EHT2_2019} with direct radio interferometry imaging and the GRAVITY instrument on VLT \citep{Sturm_etal_2018} with astrometry from IR interferometry can probe smaller scales than Gaia. However, neither EHT nor GRAVITY can survey the entire optical sky and probe to the depth of Gaia.  \label{fig:cartoon}}
  \end{figure}

Beyond $\sim$kpc scales, pairs of SMBHs within merging galaxies can be systematically identified (Fig.\ \ref{fig:cartoon}), if both are accreting as active galactic nuclei (AGN). One possible observational signature of such pairs is double-lined kinematics of the narrow-line regions surrounding each of the BH in their galaxy cores. However, in most cases such structure in the narrow emission lines is due to the complex kinematics (e.g., rotation and outflows) of the narrow line region \citep[e.g.,][]{Shen2011a,Zakamska_Greene_2014,Yuan_etal_2016,Muller-Sanchez_etal_2016}. Therefore the dual AGN nature needs to be confirmed with spatially-resolved imaging and/or spectroscopy \citep[e.g.,][]{Liu2010a,Shen2011a,Fu2012,Fu_etal_2015,Fu_etal_2015b,Muller-Sanchez_etal_2016}. Another method of identifying kpc-scale dual AGN is via spatially resolved X-ray observations \citep[e.g.,][]{Komossa_etal_2003,Comerford2015}. In addition to dual AGN, single off-nucleus AGN, i.e., an inspiraling SMBH pair in which the other member is inactive or obscured \citep[e.g.,][]{barth08,Comerford_Greene_2014,Barrows2017,Tremmel2018}, or spatially-offset recoiling SMBHs \citep[i.e., BHs ``kicked'' from the anisotropic gravitational wave emission following the BH coalescence due to momentum conservation; e.g., ][]{Baker2006,bogdanovic07,civano12,Blecha2016}, can also be systematically searched for with these techniques. All these methods are in general sensitive to scales above $\sim$kpc, with only one or two serendipitous discoveries of low-redshift dual AGN at $\sim 500$ pc projected separations \citep[e.g.,][]{Goulding_etal_2019}. 


However, it becomes challenging to identify AGN pairs at sub-kpc scales before the two SMBHs become gravitionally bound due to the stringent spatial resolution requirement. One spatially-resolved method is to search for closely separated flat-spectrum sources with high-resolution Very Long Baseline Interferometry (VLBI) imaging \citep[e.g.,][]{burke11}, which could be due to compact jets around each of the black holes. There is only one confirmed AGN pair at $\sim$7 pc separation serendipitously discovered with VLBI imaging \citep{rodriguez06}.

The scales of tens to hundreds of parsec are of particular interest to binary SMBH evolution. These scales correspond to the late stage of galaxy merger, while the two BHs are on their way to become a bound binary. The observed frequency of AGN pairs below kpc scales is essential for testing physical recipes for AGN fueling in galaxy merger simulations and for constraining how fast the BH pair can form a bound BH binary \citep[e.g.,][]{yu11,Steinborn2015,Dosopoulou2017,Kelley2017,Tremmel2018}. In addition, the sub-kpc SMBH pair frequency may provide important clues on the physical nature of dark matter particles. For example, in fuzzy dark matter \citep[][]{Hu2000}, a form of dark matter that consists of extremely light scalar particles with masses on the order of $\sim$10$^{-22}$ eV, SMBH pairs would never get much closer than $\lesssim$1 kpc because fuzzy dark matter fluctuations may inhibit the orbital decay and inspiral at kpc scales \citep{Hui2017}.

In this paper, we present a new astrometric technique, motivated by a long history of utilizing astrometry to achieve super-diffraction-limit applications in astronomy \citep[e.g.,][]{Bailey_1998,Shen2012a,Liuyuan2015,Liuyuan2016,Stern_etal_2015,Sturm_etal_2018}. This technique can be used to systematically identify sub-kpc off-nucleus and dual AGN and to quantify their frequency using data from the astrometric space telescope, Gaia \citep{Gaia2016}. 

The working principle of this technique is very simple. Consider an unresolved AGN pair (and at least one of them is an unobscured Type 1 AGN) at $\lesssim$kpc separations (1 kpc corresponds to $\sim$0\farcs2 at $z{>}0.5$) for which Gaia will measure a single-source photocenter\footnote{At high redshift, luminous AGN or quasars often dominate over their host galaxy in the observed-frame optical.}. At parsec to sub-kpc separations, the positions of the two AGN are essentially fixed given the long ($\gg 100$ yr) dynamical timescale of the pair. Since essentially all AGN vary stochastically on all timescales with a typical variability amplitude of $\sim$0.1-0.2 mag on days to years timescales in the restframe UV-optical \citep[e.g.,][]{Sesar2007}, the non-synchronous variation in flux will introduce an astrometric shift in the combined photocenter measured at different observing epochs \citep[][]{Liuyuan2015,Liuyuan2016}, leading to photocenter variations at the $>{\rm mas}$ level detectable by Gaia. This astrometric shift provides a lower limit of the projected physical separation of the AGN pair. However, if the separation is greater than $\sim 1-2$ kpc ($\sim 0\farcs3$), Gaia will likely already resolve the system into two sources given its spatial resolution, scanning direction, and source identification procedure \citep[e.g.,][]{Lemon2017,Ducourant2018}. Therefore, with detected photocenter variations for unresolved sources by Gaia, we can identify AGN pairs with projected separations of $\sim$5 pc to $\sim$1 kpc. Other than direct imaging with VLBI \citep[e.g.,][]{burke11}, whose feasibility, however, relies on the radio brightness of AGN and is limited by observing time, the Gaia astrometric method represents the only currently viable alternative to constrain the sub-kpc AGN pair population in a systematic way, and probes two orders of magnitude smaller angular scales than with HST (e.g., see Fig.\ \ref{fig:cartoon}). The capability of measuring the photocenter to a precision that can be orders of magnitude better than the image resolution underlies this method \citep[e.g.,][]{Lindegren_1978}.

The validity of this method relies upon two factors: an asymmetric intrinsic variability pattern that will cause astrometric jitter in the photocenter of the unresolved source, and sufficient astrometry precision. It is thus distinct from other processes that could also cause photocenter variations not induced by intrinsic variability (such as binary orbital motion, e.g., \citealt{Bansal2017}, and astrometric microlensing, e.g., \citealt{Belokurov2002}). For this reason, here we dub it {\tt varstrometry} (variability$+$astrometry) for the ease of reference and for its general application. A similar idea was applied to unresolved stellar binaries \citep[e.g.,][]{Wielen1996,Makarov2016}, which was referred to as the ``variability-induced motion'' method. 

The paper is organized as follows. We first describe the general principles of this astrometric technique and some simulations in \S \ref{sec:method}, followed by our initial results on the analysis of data from Gaia Data Release 2 \citep[DR2;][]{GaiaCollaboration2018} in \S \ref{sec:gaia} and \S\ref{sec:qso}. We then discuss the implications of our results in \S \ref{sec:disc}. We summarize our findings and conclude in \S \ref{sec:con}. We remind the reader that we are primarily interested in the sub-kpc regime, where only Gaia can provide a systematic search and statistical constraints of such small-scale pairs with its superb astrometric precision, all-sky coverage and decent optical depth. The smallest pair separation that Gaia can potentially recover is $\sim 5-10$ pc. Throughout this paper we adopt a flat $\Lambda$CDM cosmology with $\Omega_0=0.3$ ($\Omega_\Lambda=0.7$) and $H_0=70\,{\rm km\,s^{-1}Mpc^{-1}}$. 





\section{Methods}\label{sec:method}

\begin{figure*}
 \includegraphics[width=0.48\textwidth]{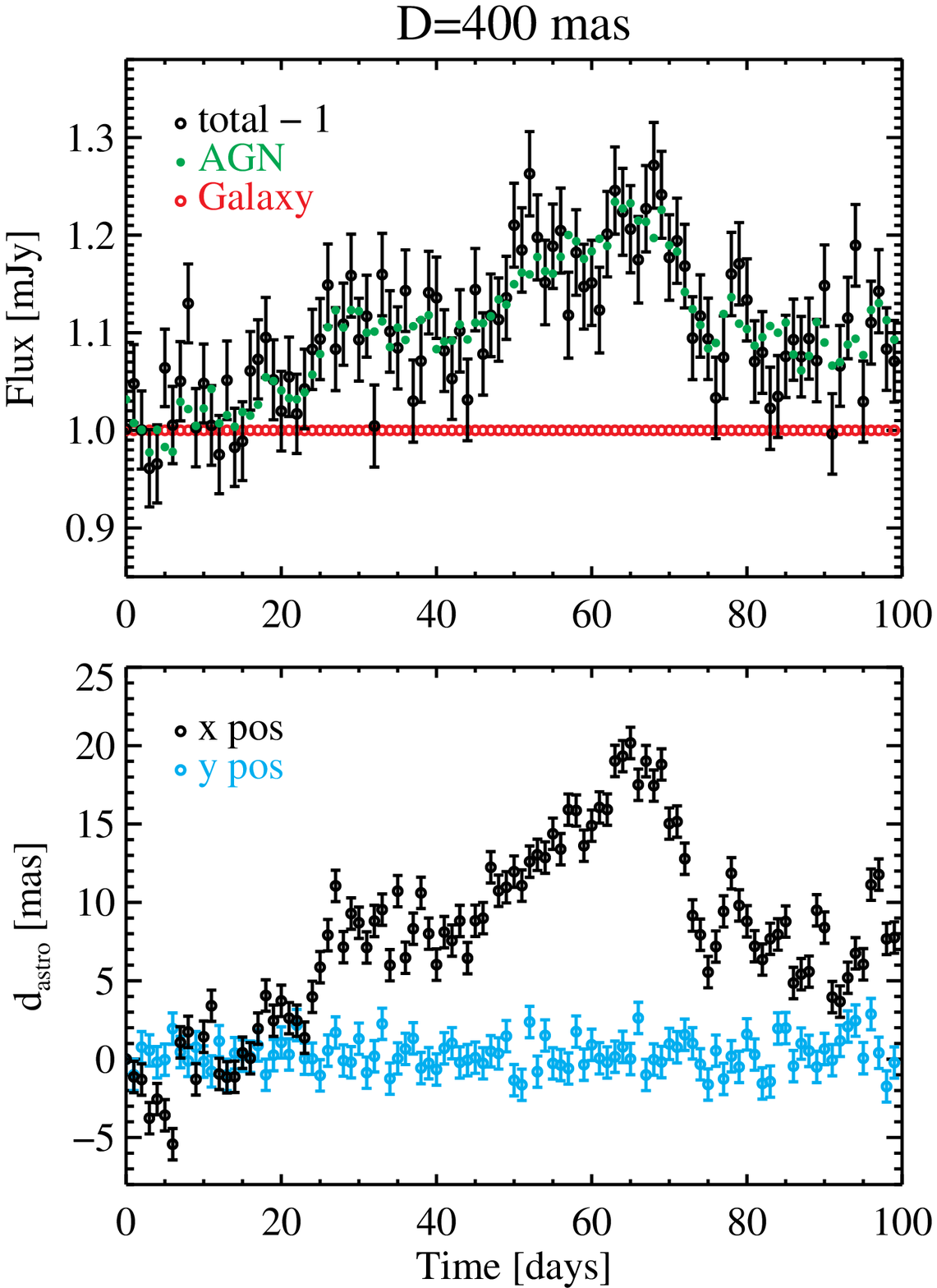}
 \includegraphics[width=0.48\textwidth]{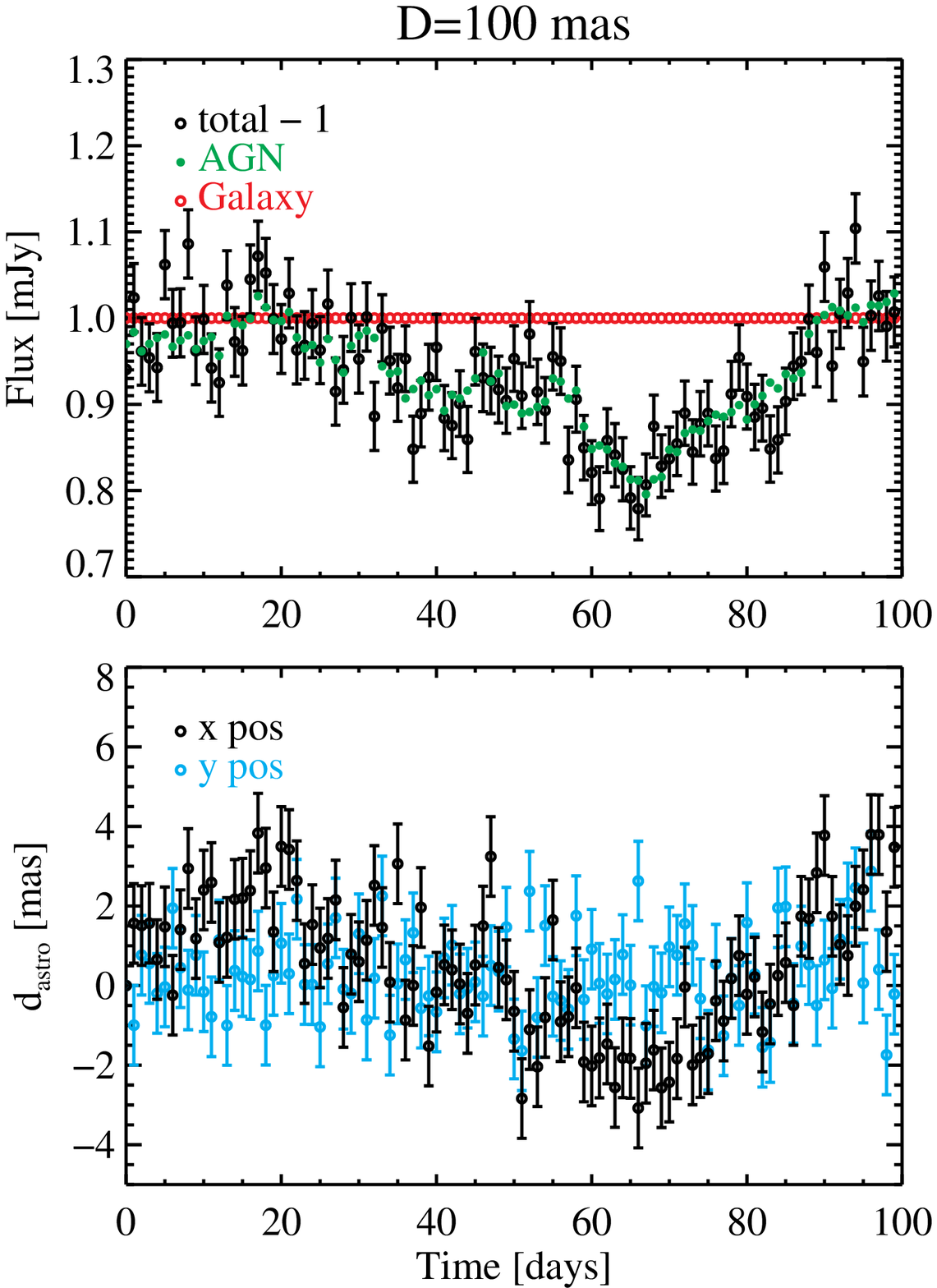}
 \caption{Examples of simulated photocenter and flux variations for an AGN offset from its host galaxy. In this mock observation, we generate light profiles for the AGN assuming a point source, and for the galaxy assuming a S\'ersic profile (with a S\'ersic index $n=4$ and an effective radius of 10 pixels). The pixel scale of the mock imaging is 0.1". We assume a double-Gaussian PSF with the core Gaussian $\sigma=2$ pixels and the second Gaussian $\sigma=4$ pixels. The AGN and galaxy pair is along the $x$ direction with a separation of 4 pixels (400 mas, left) and 1 pixel (100 mas, right), hence the system is marginally resolved or unresolved by Gaia. The total constant flux from the galaxy is 1 mJy (16.4 AB magnitude). The AGN light curves are generated using a Damped Random Walk model using typical parameters from \citet{MacLeod2012} and a long-term mean flux of 1 mJy. The flux and photocenter measurements are sampled daily during a 100-day period, with assumed measurement uncertainties of 2\% for the total flux, and 1 mas for the photocenter positions. \label{fig:sim1}}
 \end{figure*}

In this section we present the {\tt varstrometry} technique. Although this technique generally can be applied to any astrometric data, for demonstration purposes we tailor the discussion to Gaia, which currently provides the best optical astrometric measurements over the entire sky and to a depth ($G\sim21$) that is suitable for the systematic searches for dual and offset AGN. 

\subsection{The case of dual AGN} 

In a sub-kpc pair of SMBHs, if both BHs are actively accreting and appear as unobscured broad-line AGN, we expect that the photocenter of the unresolved system (two AGN plus the host galaxy) varies due to the stochastic variability of both AGN. If the host galaxy light is negligible, the root-mean-square (RMS) dispersion in the photocenter position, $\sigma_{\rm astro}$, depends on the pair separation $D$ and on the flux contrast between the two AGN, as well as on their photometric variability amplitude. By Taylor-expanding to the leading order in the fractional flux variation and by assuming statistically independent flux variability from each AGN, we find that we expect
\begin{equation}\label{eqn:rms1}
    \sigma_{\rm astro}\approx \frac{D}{(\bar{f_1}+\bar{f_2})^2}\sqrt{\bar{f_1}^2\langle\Delta f_2^2\rangle + \bar{f_2}^2\langle\Delta f_1^2\rangle}\ ,
\end{equation}
where $\bar{f}$ is the mean flux of the system and $\Delta f$ is the flux variability. Thus for an equal-flux pair ($\bar{f_1}=\bar{f_2}$), $\sigma_{\rm astro}$ is linearly proportional to both $D$ and the total RMS photometric variability measured from the unresolved system:
\begin{equation}\label{eqn:rms}
    \sigma_{\rm astro}\approx\frac{D}{2}\frac{\sqrt{\langle\Delta f^2\rangle}}{\bar{f}}\ ,
\end{equation}
in the limit of $\sqrt{\langle\Delta f^2\rangle}\ll \bar{f}$. Thus for a 10\% RMS total flux variability, the astrometric signal is 5\% of the pair separation. If we assume the same fractional variability for both AGN, Eqn.\ (\ref{eqn:rms}) can be generalized to non-equal-flux pairs:
\begin{equation}\label{eqn:rms2}
    \sigma_{\rm astro}\approx\displaystyle D\sqrt{\frac{2q^2}{(1+q)^2(1+q^2)}}\frac{\sqrt{\langle\Delta f^2\rangle}}{\bar{f}}\ ,
\end{equation}
where $q\equiv \bar{f_2}/\bar{f_1}$. As expected, a larger flux contrast of the two members or lower total photometric variability diminishes the expected astrometric signal. 

Importantly, the photocenter variations are expected to be bound, aperiodic, and along the direction of the binary, and the largest shifts are expected to be associated with the less frequent, large-amplitude photometric variations. This additional information can be used to confirm the nature of any detected astrometric signals. 

To illustrate the relation between the photometric and astrometric variability, we consider the ideal case where the host light is negligible and the unweighted geometric center of the dual AGN is at the origin. Then the photocenter of the dual AGN system is the ``center of flux''. The displacement of photocenter $d_{\rm astro}$ from the origin is determined by the pair separation $D$ and the instantaneous flux contrast $q^{\prime}\equiv f_2/f_1$: 
\begin{equation}\label{eqn:dphoto}
    d_{\rm astro}=\frac{D}{2}\frac{q^{\prime}-1}{q^{\prime}+1}\ .
\end{equation}
Therefore larger flux changes in one or two of the AGN induce larger photocenter shifts and dominate the astrometric signals. 

Equation (\ref{eqn:dphoto}) also applies to the case (see \S\ref{sec:offagn}) where one member of the pair is the host galaxy and the other member is an AGN, regardless of whether or not the host galaxy is extended.  

\subsection{The case of single off-nucleus AGN}\label{sec:offagn}

The vastrometry technique can also be applied to systems where only one member of the SMBH pair is an unobscured AGN, with the other member being obscured or inactive. Such a system appears as an off-nucleus AGN on sub-kpc scales. An observationally similar, but physically different case is that of a recoiling active SMBH after the merger of the two SMBHs. We expect to see photocenter variations in such systems if the host galaxy contributes significant light and if the variability-induced astrometric signal is large enough. 

Similar to the dual AGN case, the resulting photocenter variations are linear, bound, and aperiodic. The relation between astrometric RMS variability and total photometric RMS variability of the unresolved system is (following Eqn.\ \ref{eqn:rms1}): 
\begin{equation}\label{eqn:rms3}
    \sigma_{\rm astro}\approx D\frac{q}{1+q}\frac{\sqrt{\langle\Delta f^2\rangle}}{\bar{f}}\ ,
\end{equation}
where $q=\bar{f_2}/\bar{f_1}$ and $\bar{f_2}$ corresponds to the constant galaxy flux, and $\langle\Delta f^2\rangle=\langle\Delta f_1^2\rangle+\langle\Delta f_2^2\rangle=\langle\Delta f_1^2\rangle$. In addition, we expect a perfect correlation (in the absence of measurement errors) between the instantaneous photocenter shift and the photometric flux (cf., Eqn.\ \ref{eqn:dphoto}), which can be measured with time series of photocenter measurements and photometric light curves. 

It is straightforward to simulate the expected photocenter shifts given the parameters that describe the pair configuration (e.g., projected separation and position angle), flux contrast, variability characteristics, as well as host contamination. Figure \ref{fig:sim1} demonstrates the expected astrometric signals for an AGN+galaxy pair. In this mock observation we have used observing parameters that approximately resemble the Gaia data although the exact details are different. A much higher cadence is used to demonstrate the continuous variability in flux and photocenter. 

For this simulation, the AGN variability amplitude is typical of the observed values, and the expected photocenter shifts and their strong correlation with the total flux are visually apparent in Figure \ref{fig:sim1} and detectable with Gaia. The resulting photocenter shifts are also consistent with the predictions from  Eqn.\ (\ref{eqn:dphoto}). Since the photocenter shift depends linearly on the pair separation, we expect that Gaia should be able to detect this astrometric signal even when the pair separation $D$ is several times smaller than that assumed in the simulations. The chance of detecting the photocenter shifts increases when large-amplitude flux variations are captured during the observing period. For example, if the AGN initially has equal flux as the galaxy but during the observing period it varied by a factor of two, then Eqn.\ (\ref{eqn:dphoto}) indicates a photocenter shift of $D/6$. Better sampling of the time series of photocenter measurements also helps the correlation analysis with the photometric light curve, even if individual astrometric offsets are marginally detected. 

\begin{figure}
 \includegraphics[width=0.48\textwidth]{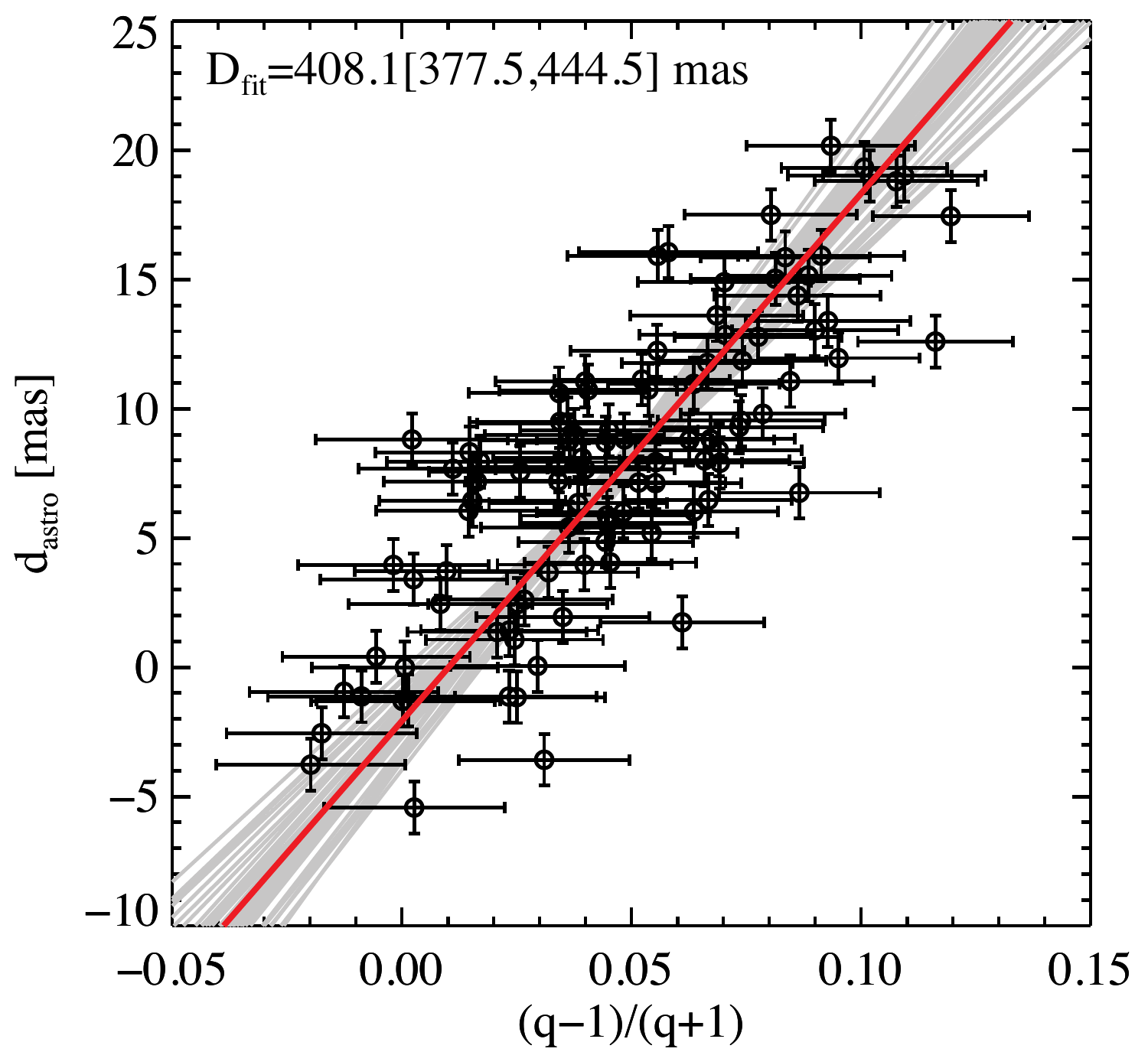}
 \includegraphics[width=0.48\textwidth]{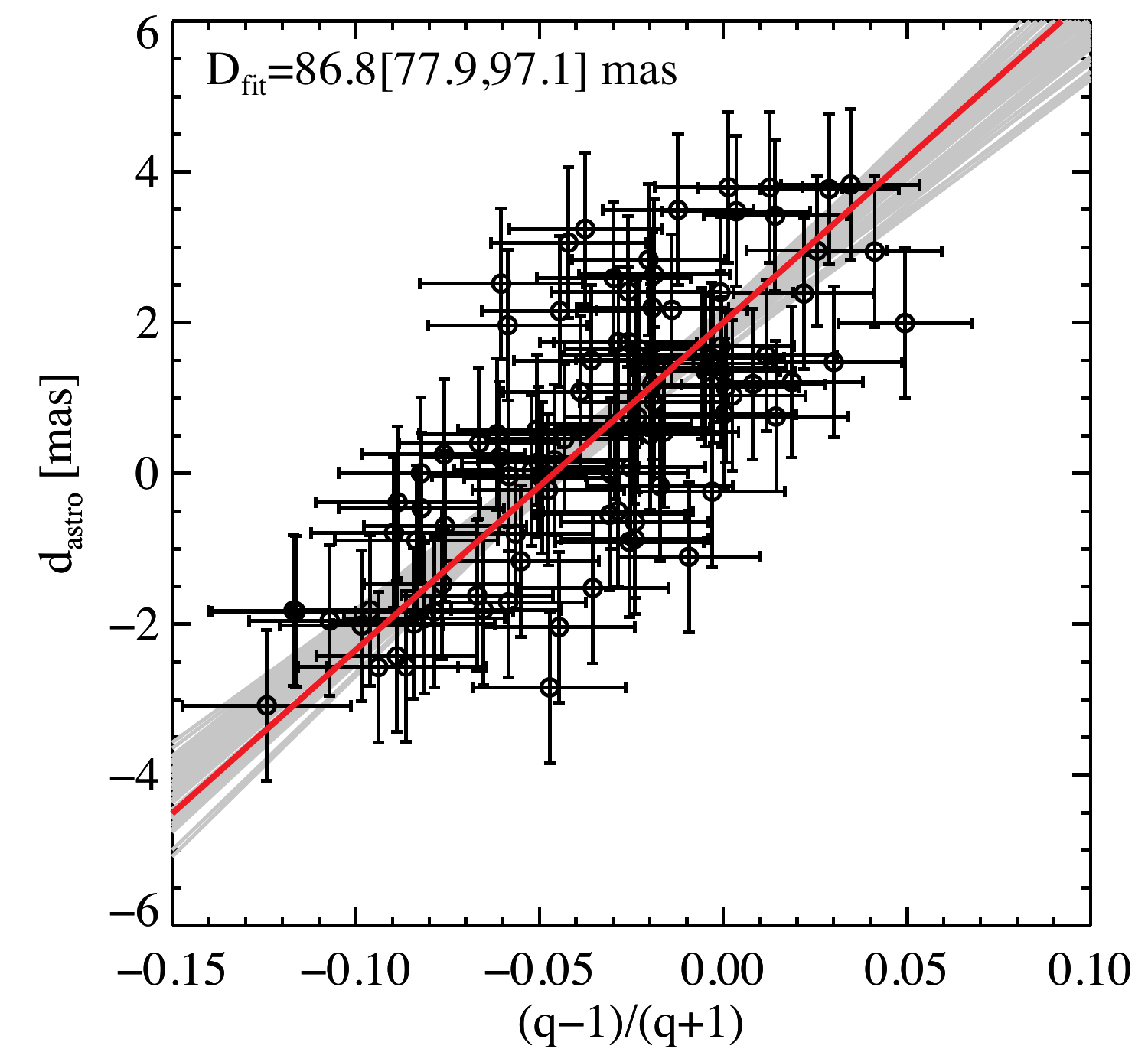}
 \caption{Linear regression between the photocenter shift and the flux ratio measured from each epoch, using the Bayesian regression method in \citet{Kelly_2007}. The two panels correspond to the two examples shown in Fig.\ \ref{fig:sim1} with typical AGN variability and pair separations of 400 and 100 mas, respectively. The gray lines are random draws from the posterior distribution of the regression and the red line indicates the median result. The slope of the regression corresponds to half of the pair separation (cf. Eqn.\ \ref{eqn:dphoto}), and the median value and the 16th/84th percentiles of the fitted pair separation are indicated at the top of the panel. {For simplicity we did not consider the covariance between the photocenter and flux measurements in the regression fits (i.e., the uncertainties along both axes are treated as independent). } \label{fig:sim2}}
 \end{figure}

In the case of an offset AGN+galaxy pair, if the constant galaxy flux can be measured through spectral or imaging decomposition, then the observed flux variation and photocenter jitter can be used to derive the projected pair separation $D$ using Eqn.\ (\ref{eqn:dphoto}) with linear regression analysis. Fig.\ \ref{fig:sim2} presents the linear regression between the photocenter measurements and the flux ratios derived for each epoch, for the same examples shown in Fig. \ref{fig:sim1}. In these two examples with typical AGN variability, we can reasonably measure the pair separation. Even for cases with smaller pair separation and/or lower variability, the non-detection of astrometric shifts can still be used to place upper limits on the pair separation, whereas if the AGN varies more than average during the observing period, we can measure smaller pair separations or place tighter upper limits. In a follow-up paper \citep{vodka2}, we use this method to constrain the population of off-nucleus single AGN.  


\section{Application of Varstrometry to Gaia DR2}\label{sec:gaia}


To utilize the full power of the {\tt varstrometry} technique requires the time series of photocenter measurements and photometric light curves from future releases of Gaia. However, even with the cataloged information in Gaia DR2, i.e., no time series, we can still use proxies for photometric and astrometric RMS variability to test our approach and select initial candidates for follow-up observations. We also examine potential systematics in Gaia DR2 that may impact the reliability of {\tt varstrometry}. 

In \S\ref{sec:proxy} we describe the use of Gaia DR2 quantities to substitute for the astrometric and photometric RMS. In \S\ref{sec:host} we investigate the systematics on Gaia astrometry and photometry in extended sources. In \S\ref{sec:test} we use examples of known binary systems to validate our technique. We defer our analysis on quasars to \S\ref{sec:qso}. Given Gaia's resolution, scanning strategy and window sizes in source identification, dual or off-nucleus AGN on $\gtrsim 0.3$" (or $\gtrsim 1$ kpc) scales are likely resolved into multiple sources in Gaia. While not the focus of this paper, we provide a brief discussion on these potential $\gtrsim 1$ kpc off-nucleus or dual AGN systems in \S\ref{sec:quasar-multi}.




\subsection{Photometric and astrometric RMS variability in Gaia DR2}\label{sec:proxy}

Gaia is an optical all-sky survey which is obtaining photometry, positions, parallaxes, and proper motions for stars with magnitudes down to $\sim21$\,mag and radial velocities for select bright stars. In Gaia DR2 released on 25 April 2018 \citep{GaiaCollaboration2018}, broad-filter G-band magnitudes, blue-band BP magnitudes, red-band RP magnitudes, positions, parallaxes, and proper motions are available for $\sim1.33$ billion objects (5-parameter sources). An additional 0.36 billion objects have available G-band magnitudes and positions (2-parameter sources). Gaia DR2 is based on data collected between 25 July 2014 and 23 May 2016. In addition to Galactic sources, Gaia DR2 contains a large number of extragalactic sources. For example, about half a million WISE-selected quasars are cataloged and are used to define the celestial reference frame in Gaia \citep{Mignard2018, Lindegren2018}. 

While Gaia DR2 has not released the full time series for every source, it does provide an indicator for photometric variability. The photometric errors of the reported mean flux in Gaia DR2 are calculated by
\begin{equation}
{\tt phot\_g\_mean\_flux\_error} = \sigma_G / \sqrt{{\tt phot\_g\_n\_obs}},
\end{equation}
where $\sigma_G$ is the standard deviation of the G-band fluxes in the time series. {\tt phot\_g\_mean\_flux\_error} and {\tt phot\_g\_n\_obs} are the cataloged G-band mean flux error and the number of observations in G-band, respectively. When a star passes through Gaia's focal plane, there are 9 CCDs to measure its photometry in G-band and the corresponding astrometry, and $\sigma_G$ is the standard deviation of all CCD-level fluxes in the time series. The photometric precision of individual CCD measurements can be as low as 2 mmag with $\sim10$\,mmag systematic uncertainties \citep{Evans2018}. As a result, $\sigma_G$ contains information about variability on a wide range of timescales, from seconds of Gaia's individual CCD measurements, hours of the Gaia satellite's spinning period, to weeks and years of Gaia's scanning law. For non-variable sources, $\sigma_G$ represents the measurement (statistical and systematic) uncertainties in flux; for variable sources, $\sigma_G$ has contributions from both the intrinsic variability and measurement uncertainties. Indeed, we find that for matched $G$ magnitudes, quasars have systematically larger $\sigma_G$ than stars, due to their intrinsic variability. 

\begin{figure}
\centering
 \includegraphics[width=0.5\textwidth]{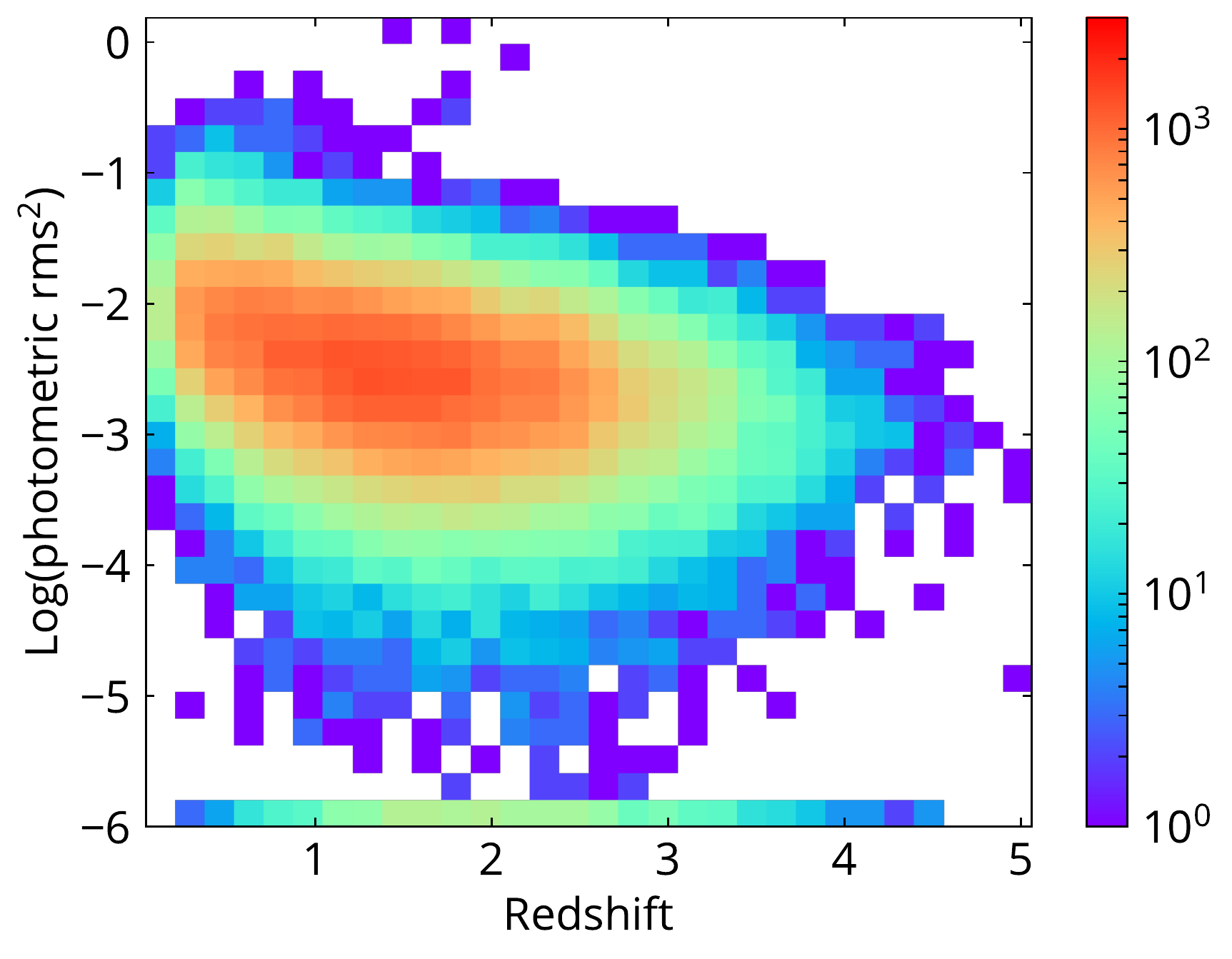}
 \caption{Distribution of SDSS quasars in the fractional photometric RMS and redshift plane, color-coded by the number of objects. The photometric RMS is estimated using the Gaia DR2 flux errors as described in \S\ref{sec:proxy}. The trend of decreasing photometric variability with redshift is mainly caused by the $(1+z)$ time dilation and the general trend that quasars are more variable on longer timescales and at higher luminosities \citep[e.g.,][]{VandenBerk2004,Sesar2007}. This result demonstrates that our proxy for the photometric RMS is reasonable. }  \label{fig:var-z}
\end{figure} 
 
Using {\tt phot\_g\_mean\_flux\_error} and {\tt phot\_g\_n\_obs}, we compute $\sigma_G$ and the fractional variability $f_{G, raw} = \sigma_G/F_G$ for all sources, where $F_{G}$ is the G-band mean flux. To obtain the intrinsic variability estimate, we correct $f_{G, raw}$ for magnitude-dependent instrumental errors \citep{Evans2018}. The instrumental fractional variability (including statistical errors from photon noise), $f_{G,inst}$, is computed from the running medians of $f_{G,raw}$ for a sample of $10$ million nearby stars ($<500$\,pc) across all observed G-band magnitudes. Using running modes instead of medians gives a difference of only $\sim 10^{-3}$ in fractional variability, which is not critical for our purpose since the AGN fractional variability is almost always much higher. The intrinsic fractional variability is then computed as $f_{G}^2 = f_{G, raw}^2 - f_{G, inst}^2$. In this definition, $f_{G}^2$ may be negative if the source does not have significant variability compared to the running median at its G-band magnitude; and we set $f_{G}=0$ in such cases. For quasars, most of which are fainter than 16 mag, the instrumental correction is $f_{G, inst}\sim0.005$ at G$=16$ to $\sim0.08$ at G$=20$. In the following sections, we limit the sample to G-band magnitudes brighter than 19.5\,mag, where the correction is $6$\% in fractional variability so we can investigate intrinsic variability at the few percent level. A similar method has been used to obtain variability information from Gaia dataset to identify RR Lyrae stars \citep{Belokurov2017} and eclipsing binaries (Hwang et~al. in prep).

Fig.~\ref{fig:var-z} displays the fractional photometric RMS with redshift for the spectroscopic SDSS quasar sample (see \S\ref{sec:qso}). The trend of decreasing photometric RMS with redshift is mainly caused by the $(1+z)$ time dilation and the general trend that quasars are more variable on longer timescales and at higher luminosities \citep[e.g.,][]{VandenBerk2004,Sesar2007}. This result demonstrates that our proxy for the photometric RMS is reasonable.


Gaia DR2 catalog provides several astrometric quality indicators, including {\tt astrometric\_excess\_noise} (in units of mas), {\tt astrometric\_sigma5d\_max} (in mas), {\tt astrometric\_chi2\_al} (unitless),  {\tt astrometric\_gof\_al} (unitless), {\tt astrometric\_excess\_noise\_sig} (unitless), and the unit weight error introduced in \cite{Lindegren2018}. Every indicator has its pros and cons, and we refer the reader to \cite{Lindegren2012, Lindegren2018} for more detailed discussion on their properties. 

In the following sections, we present the results using {\tt astrometric\_excess\_noise}. Ideally, if a source is well described by the model (i.e., a single, point-like source in Gaia DR2), then the excess noise is 0. Excess noise becomes non-zero when the source has unmodelled astrophysical behaviors (e.g., in binaries) or when the unmodelled instrumental noise is present \citep{Lindegren2012}. Mathematically, {\tt astrometric\_excess\_noise} is the extra error term added in quadrature to the measurement uncertainty of pixel coordinates in deriving the astrometric solution \citep[][]{Lindegren2012}, and thus it describes the combination of intrinsic astrometric RMS and any residual systematic effects. The astrometric jitter induced by {\tt varstrometry}, $\sigma_{\rm astro}$, is the intrinsic astrometric RMS we want to estimate, and so the {\tt astrometric\_excess\_noise} quantity provides an upper limit on the intrinsic astrometric RMS. Additional benefits of using {\tt astrometric\_excess\_noise} over other astrometric quantities include: 1) it is in units of mas and has a direct physical interpretation; 2) it is available regardless of whether the 5-parameter astrometric solution is successful or not. Therefore we use {\tt astrometric\_excess\_noise} as a proxy (upper limit) for the intrinsic astrometric RMS in photocenter. We also test with other astrometric quality indicators, and the main conclusions of this paper remain unchanged.

Faint sources have larger astrometric errors. The standard deviation of along-scan astrometric measurements is $>5$\,mas for sources fainter than 20\,mag in G-band and is $<0.5$\,mas for G-band magnitudes between 7 and 16.5\,mag \citep{Lindegren2018}. We focus on the sample brighter than 19.5\,mag in G-band where the standard deviation of along-scan astrometric measurements is $<3$\,mas. 

Gaia DR2 has a degree-of-freedom bug when calculating {\tt astrometric\_excess\_noise} for sources brighter than G$\sim17$\,mag \citep{Lindegren2018}. This bug also propagates to other astrometric quality indicators. Including the effects of the bug, for sources brighter than $13$\,mag in G-band, {\tt astrometric\_excess\_noise} is underestimated, while for sources with 13-17\,mag in G-band, {\tt astrometric\_excess\_noise} is overestimated. This bug is corrected in a magnitude-averaged way in Gaia DR2, instead of recomputing the relevant values for every source because the bug was found very late during the data processing of Gaia DR2. The majority of our inactive galaxy and quasar samples are fainter than 16\,mag, and our results are thus not significantly affected by this detail even if the correction for the bug is not perfect. 


\begin{figure*}
\centering
 \includegraphics[width=0.45\textwidth]{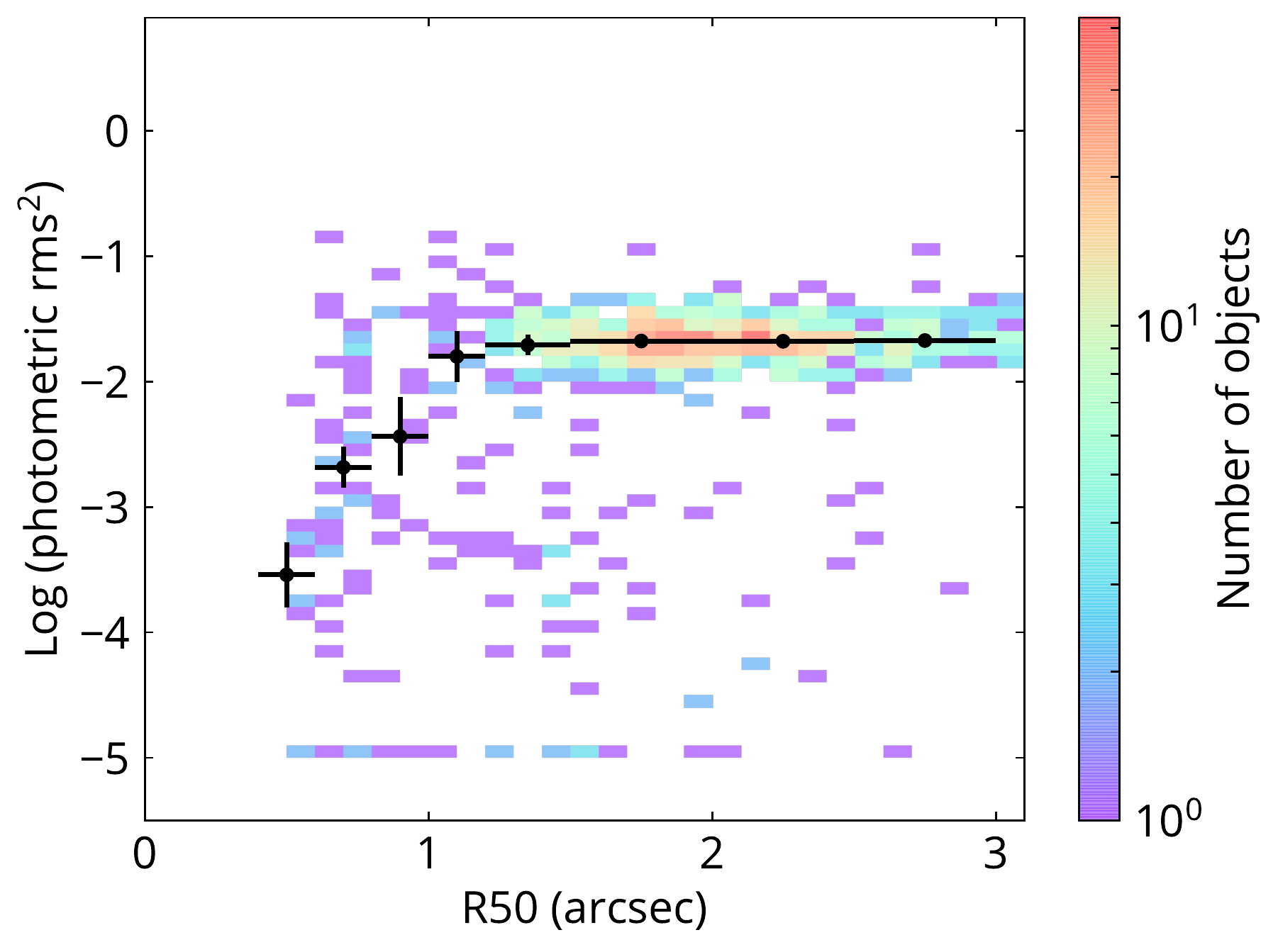}
 \includegraphics[width=0.45\textwidth]{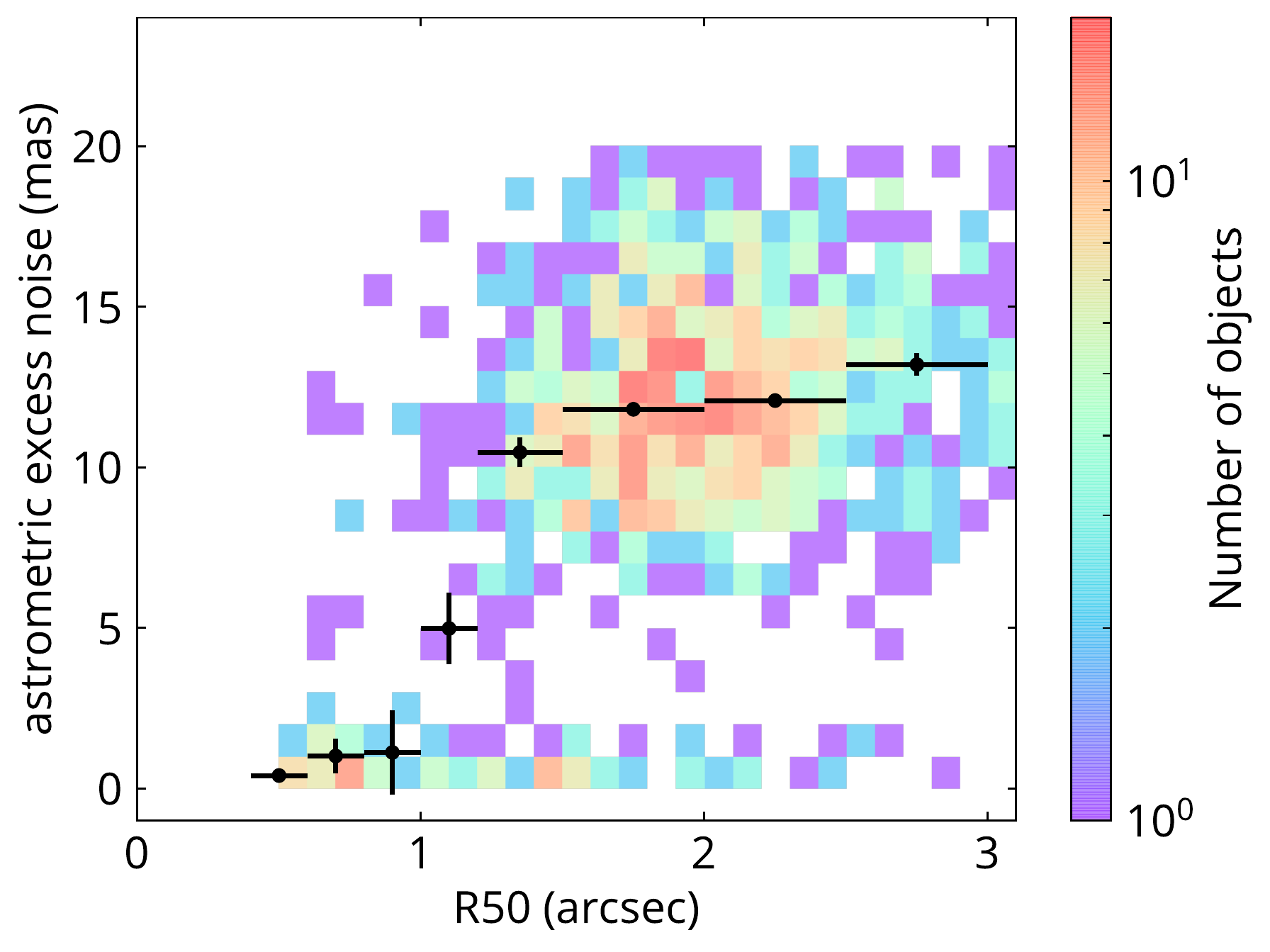}
 \caption{The fractional photometric RMS variability (left) and the astrometric excess noise (right) with respect to the Petrosian radius $R_{50}$ in r-band for star-forming galaxies. The black markers show the running median in each bin, and the x-axis bars indicate the range of the bin and the y-axis error bars are the standard deviation of mean. In Gaia DR2, extended galaxies with $R_{50}>1$\arcsec\ have $\sim10$\% photometric uncertainty and $\sim10$\,mas astrometric excess noise.} \label{fig:galaxy-r50}
\end{figure*}

\subsection{Systematics from extended sources}\label{sec:host}

With {\tt varstrometry} we are concerned with distant quasars that are unresolved by Gaia. Although unobscured AGN should present themselves as point sources at Gaia resolution, AGN are hosted by galaxies which are not point sources. Given the scanning law and processing details of Gaia astrometry and photometry, extended host galaxies (which are treated as single point sources in Gaia DR2) may induce additional systematic uncertainties in the astrometric and photometric measurements. 

We use inactive galaxies to investigate how extended host morphology affects the astrometric and photometric RMS measurements in Gaia DR2. We select star-forming (e.g., emission-line) galaxies from the Portsmouth SDSS DR14 value added catalog \citep{Thomas2013}. We use emission line ratios to remove AGN \citep{Baldwin1981, Kewley2001, Kauffmann2003}. After cross-matching with Gaia DR2, we end up with a sample of 7692 star-forming galaxies at redshifts $<0.6$. Inactive galaxies at higher redshifts are too faint for Gaia. To quantify the extended structure of the host galaxies, we query the Petrosian half-light radius ($R_{50}$) in $r$-band from SDSS DR14.

Fig.~\ref{fig:galaxy-r50} shows the fractional RMS photometric variability and {\tt astrometric\_excess\_noise} from Gaia DR2 as a function of $R_{50}$ for the star-forming galaxy sample. Sources with $R_{50}>1$\arcsec\ have significantly higher photometric variability and astrometric noise. Specifically, when $R_{50}>1$\arcsec, the measured photometric RMS is $\sim$15\%, even though star-forming galaxies are not expected to vary on the timescales of Gaia observations. Therefore, we interpret this photometric RMS as the systematic uncertainty for spatially resolved sources. This uncertainty remains roughly constant for $R_{50}$ between 1\arcsec\ and 3\arcsec. The Gaia-measured {\tt astrometric\_excess\_noise} also rises to $\sim 10$ mas when $R_{50}>1$\arcsec, which we attribute to the same systematic effect in extended sources.

Nevertheless, these levels of impact on the photometric and astrometric RMS estimates are still acceptable, especially for the astrometric RMS estimates. For statistical constraints on the off-nucleus and dual AGN population \citep[e.g.,][]{vodka2}, these systematics will not cause significant issues because a $\sim 10$ mas systematic uncertainty in the astrometric RMS still implies a stringent upper limit of the inferred pair separation. For individual cases, however, we caution that a large astrometric RMS signal may be due to extended host morphology.




Fig.~\ref{fig:galaxy-r50} shows that the impact of host galaxies can be much reduced if we limit the sample to $R_{50}<1$\arcsec. Furthermore, we expect the host-galaxy-induced systematics to be much smaller at high redshift because the host-nucleus flux contrast is much reduced in the observed-frame optical.




\begin{figure*}
\centering
 \includegraphics[width=0.43\textwidth]{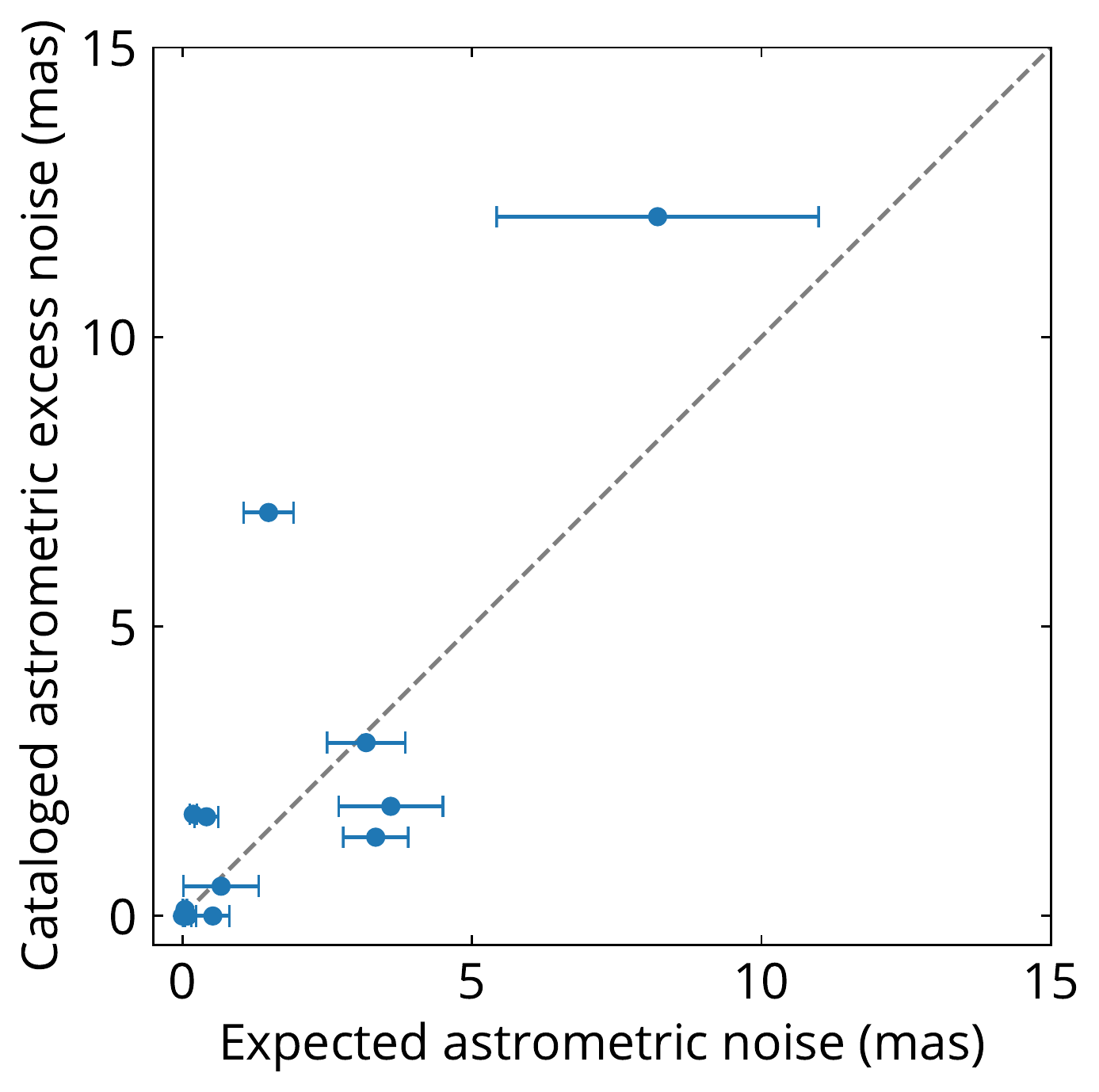}
 \includegraphics[width=0.43\textwidth]{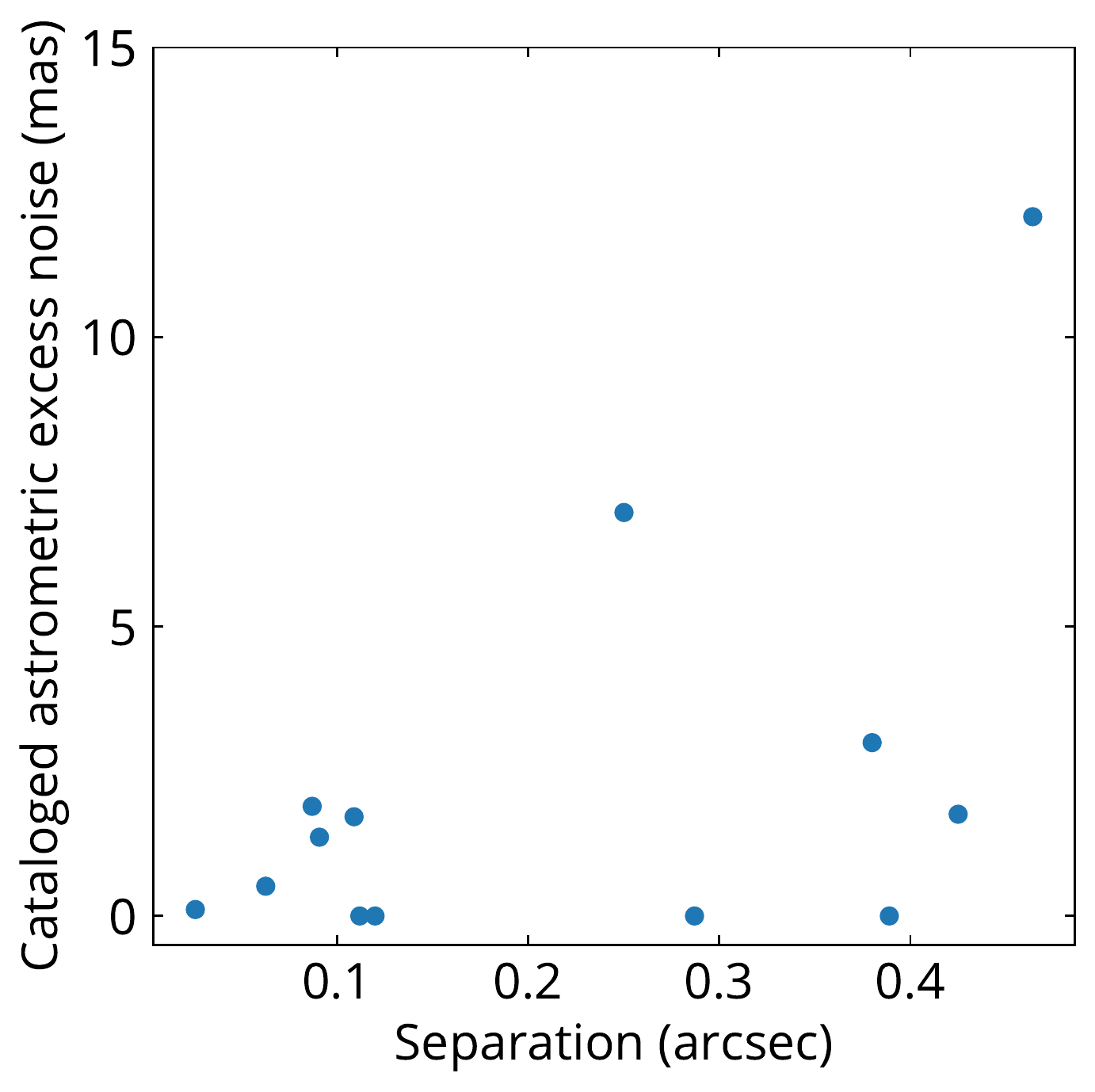}
 \caption{The comparison between the cataloged {\tt astrometric\_excess\_noise} and the expected astrometric noise from Eqn~\ref{eqn:rms2} (left) and the separation from companions (right) for unresolved pre-main sequence binaries in the Taurus-Auriga star-forming region. The cataloged {\tt astrometric\_excess\_noise} is broadly consistent with the expected values, while the separation only sets the upper limit of {\tt astrometric\_excess\_noise}. It is in agreement with {\tt varstrometry} where the astrometric noise is dependent on both angular separation and photometric variability. }
\label{fig:binary-test}
\end{figure*}

\subsection{Testing varstrometry with Galactic sources} \label{sec:test}

To test the validity of the {\tt varstrometry} technique with Gaia DR2, we require known binary (or off-center) systems that are unresolved in Gaia (i.e., the spatial offset must be less than $\sim $0\farcs{3}), variable in the optical, and with expected astrometric RMS greater than the Gaia precision of $\sim 1$\,mas. The preferred test systems would be binary stars, with no complication from extended emission as in galaxies. 

There are not many confirmed stellar binary systems that satisfy these requirements simultaneously. The orbital separations of most eclipsing binaries, cataclysmic variables, and triple star systems are too small. Some spectroscopic binaries might have larger orbits, but they are not necessarily variable. Furthermore, members of the binary need to have comparable fluxes to induce significant astrometric RMS. The last requirement rules out most of pulsating stars in binary systems because the pulsating member typically dominates the flux. 

One suitable test sample for {\tt varstrometry} is pre-main sequence stars that have close companions because pre-main sequence stars are known to be variable. We compile such a sample from \cite{Ghez1997} and \cite{Nguyen2012} and the references therein, where the close companion is resolved by other facilities (e.g., HST) but the system remains unresolved in Gaia DR2. In some cases, sources remain unresolved in Gaia DR2 even if their separations are $>0$\farcs{2}. It was pointed out in \cite{Arenou2018} that Gaia DR2 has a deficiency in binaries where angular separations are between $0$\farcs{12} -- $0$\farcs{5} ($0$\farcs{12} is Gaia's theoretical spatial resolution), which may be due to the combination of Gaia's scanning law, limitations of resolving small-separation pairs with high flux contrast, and the quality of astrometric solutions. We then exclude Gaia-unresolved systems where the separation is $>0$\farcs{5} because for such wide-separation pairs the light distribution may significantly deviate from the PSF and {\tt varstrometry} may not be the dominant source of astrometric jitter. We further require that {\tt visibility\_periods\_used}$\ge 9$ so that there are sufficient numbers of Gaia observations for astrometric and photometric measurements. This results in 13 systems with companions separated by $0$\farcs{03}-$0$\farcs{46}. The properties of these 13 systems are listed in Table~\ref{tab:pMS}. All of them are in the Taurus-Auriga star-forming region which is about 140\,pc away. Using a nearby sample is important because otherwise it would be difficult to constrain the intrinsic astrometric jitter with noisy measurements of astrophysical proper and parallax motions.

About half (6/13) of the pre-main sequence stars in our sample have intrinsic photometric variability $>4$\% in Gaia G-band. In addition, 9 out of 13 have flux ratios between 0.1 and 10. The ranges of photometric variability and flux ratios make these pre-main sequence binaries a suitable test sample for {\tt varstrometry}. 

\begin{table*}
\caption{Properties of pre-main sequence stars with close companions.}
\center
\begin{tabular}{cccccccc}
\hline \hline
NAME & AEN\tablenotemark{a} & AENS\tablenotemark{b} & separation & flux ratio\tablenotemark{c} & $\log(f_G^2)$ & expected AN\tablenotemark{d} & ref \tablenotemark{e} \\
 & ($\mathrm{mas}$) &  & (arcsec) &  &  & (mas) &  \\
\hline
DF Tau & 1.89 & 1277.43 & 0.0871 & 1.33 & -2.14 & 3.60 & 4 \\
V773 Tau & 0.51 & 129.46 & 0.0628 & 3.5 & -2.91 & 0.67 & 4 \\
V410 Tau & 0.00 & 0.00 & 0.2871 & 265.0 & -2.48 & 0.09 & 4 \\
GG Tau & 6.97 & 23723.72 & 0.2502 & 9.33 & -2.72 & 1.49 & 4 \\
RX J0430.8+2113 & 0.00 & 0.00 & 0.389 & $\lesssim$0.01 & -2.92 & 0.00 & 5,6 \\
FF Tau & 0.11 & 4.67 & 0.026 & 0.1 & -3.77 & 0.04 & 3,6 \\
V807 Tau & 2.99 & 2922.86 & 0.39 & 0.12 & -2.51 & 3.17 & 1,6 \\
DI Tau & 0.00 & 0.00 & 0.12 & 0.01 & -3.55 & 0.03 & 2,3,6 \\
RX J0437.2+3108 & 1.71 & 1054.44 & 0.109 & 0.15 & -3.36 & 0.42 & 5,6 \\
V827 Tau & 1.36 & 921.19 & 0.0909 & 0.31 & -1.88 & 3.34 & 6 \\
HD286179 & 0.00 & 1.09 & 0.112 & 0.22 & -3.45 & 0.52 & 5,6 \\
RX J0452.9+1920 & 1.76 & 1785.35 & 0.425 & 0.01 & -3.01 & 0.19 & 5,6 \\
RX J0438.2+2023 & 12.08 & 49719.17 & 0.464 & 0.87 & -2.90 & 8.21 & 5,6 \\
\hline \hline
\multicolumn{8}{l}{
    \begin{minipage}{5.5in}
    $^a${\tt astrometric\_excess\_noise} from Gaia DR2. 
    $^b$ {\tt astrometric\_excess\_noise\_sig} from Gaia DR2. $^c$\,For sources from \cite{Ghez1997}, we adopt the flux ratios in F675W. For other sources, we adopt the flux ratios in R-band derived in \cite{Nguyen2012}. $^d$\,Expected astrometric noise computed from Equation~(\ref{eqn:rms2}). $^e$\,(1) \cite{Leinert1993}; (2) \cite{Ghez1993}; (3) \cite{Simon1995}; (4) \cite{Ghez1997}; (5) \cite{Kohler1998}; (6) \cite{Nguyen2012}.
    \end{minipage}
}\\
\end{tabular}
\label{tab:pMS}
\end{table*}


Fig.~\ref{fig:binary-test} shows the comparison between the astrometric RMS expected from Equation~(\ref{eqn:rms2}) and the {\tt astrometric\_excess\_noise} from the Gaia DR2 catalog. The uncertainty in the expected RMS is computed using the uncertainties of flux ratios and separations from the references listed in Table~\ref{tab:pMS}, and assuming that the uncertainty of the fractional photometric RMS from Gaia is $1$\%. For cases where the errors of separations and flux ratios are not available, we assume an uncertainty of 5 mas for separations and a 10\% relative uncertainty for flux ratios. The flux ratios can change over time, so the flux ratios measured from the literature may be different from those in Gaia DR2, but it remains a reasonable first-order estimate. Overall, Fig.~\ref{fig:binary-test} reveals a reasonably good correlation between the expected and measured intrinsic astrometric jitter for the pre-main sequence binary sample, considering the many uncertainties and assumptions in this comparison. 

While Gaia DR2 provides a dimensionless quantity {\tt astrometric\_excess\_noise\_sig} as an indicator of whether the non-zero astrometric excess noise is significant \citep{Lindegren2012}, we cannot easily convert it to an uncertainty estimate for {\tt astrometric\_excess\_noise}. For sources where {\tt astrometric\_excess\_noise}$>1$\,mas, they all have very high {\tt astrometric\_excess\_noise\_sig}, ranging from $\sim900$ to $\sim50000$, meaning that their astrometric measurements are inconsistent with the single-star model.



We compare {\tt astrometric\_excess\_noise} and the angular separation of the binaries in the right panel of Fig.~\ref{fig:binary-test}. One may argue that the significant detection of {\tt astrometric\_excess\_noise} in Fig.~\ref{fig:binary-test} (left panel) are due to the large companion separation that makes the light profile deviate from a single point spread function. However, even for binaries with separations as small as $\sim $0.1\arcsec\ Gaia detects significant intrinsic astrometric jitter, consistent with the expectation from {\tt varstrometry}. 



Even without a good knowledge of the intrinsic flux ratio between the binary components, Fig.~\ref{fig:binary-test} broadly supports that we are seeing the intrinsic astrometric jitter caused by {\tt varstrometry} in unresolved binaries. This test also validates the use of Equation~(\ref{eqn:rms2}) to explore {\tt varstrometry} in Gaia DR2 even when the full light curve and single-transit astrometric measurements are not available. Furthermore, it demonstrates the more general application of {\tt varstrometry}, e.g., the search for Galactic unresolved variable binary/multiple stars, including pre-main sequence binaries and triple systems with eclipsing inner binaries.

\subsection{Testing varstrometry with extragalactic sources}

We now examine the few confirmed dual or off-nucleus AGN systems reported in the literature with Gaia data. The pair separations in these systems satisfy the requirements for {\tt varstrometry} with Gaia DR2.  

CXOC\,J100043.1+020637 (also known as CID-42) is a double-nucleus AGN and one of the nuclei is an unobscured Type-1 AGN \citep{Civano2010}, so there is expected optical variability. The separation of the two nuclei is $\sim$0\farcs{4} ($\sim 2.5$ kpc). Gaia DR2 does not resolve the system into two objects, and the single Gaia source does have high {\tt astrometric\_excess\_noise} of 13\,mas. It has a fractional photometric variability of $f_{G}=20\%$. Adopting a host-to-AGN flux ratio of $q=2/3$ from \citet[][]{Civano2010}, we estimate an expected astrometric RMS of $\sim 32$ mas using Eqn.\ (\ref{eqn:rms3}), which is larger than the reported {\tt astrometric\_excess\_noise}. However, given the low redshift of CID-42, the Gaia photometric RMS may be overestimated due to the extended host. Indeed, the light curves from the Dark Energy Survey Y3 data \citep{Diehl_etal_2016} covering roughly the same period as Gaia DR2 indicate an intrinsic photometric RMS of only $\sim 8\%$ in $g$ and $i$ bands (Y. Chen, private communication), which would result in an expected astrometric RMS of $\sim 13$ mas, almost in perfect agreement with Gaia {\tt astrometric\_excess\_noise}. Of course, we are unable to rule out that the Gaia DR2 astrometric RMS measurement is also affected by the extended host galaxy because of its low redshift of $z=0.359$.

\begin{figure*}
\centering
 \includegraphics[width=0.47\textwidth]{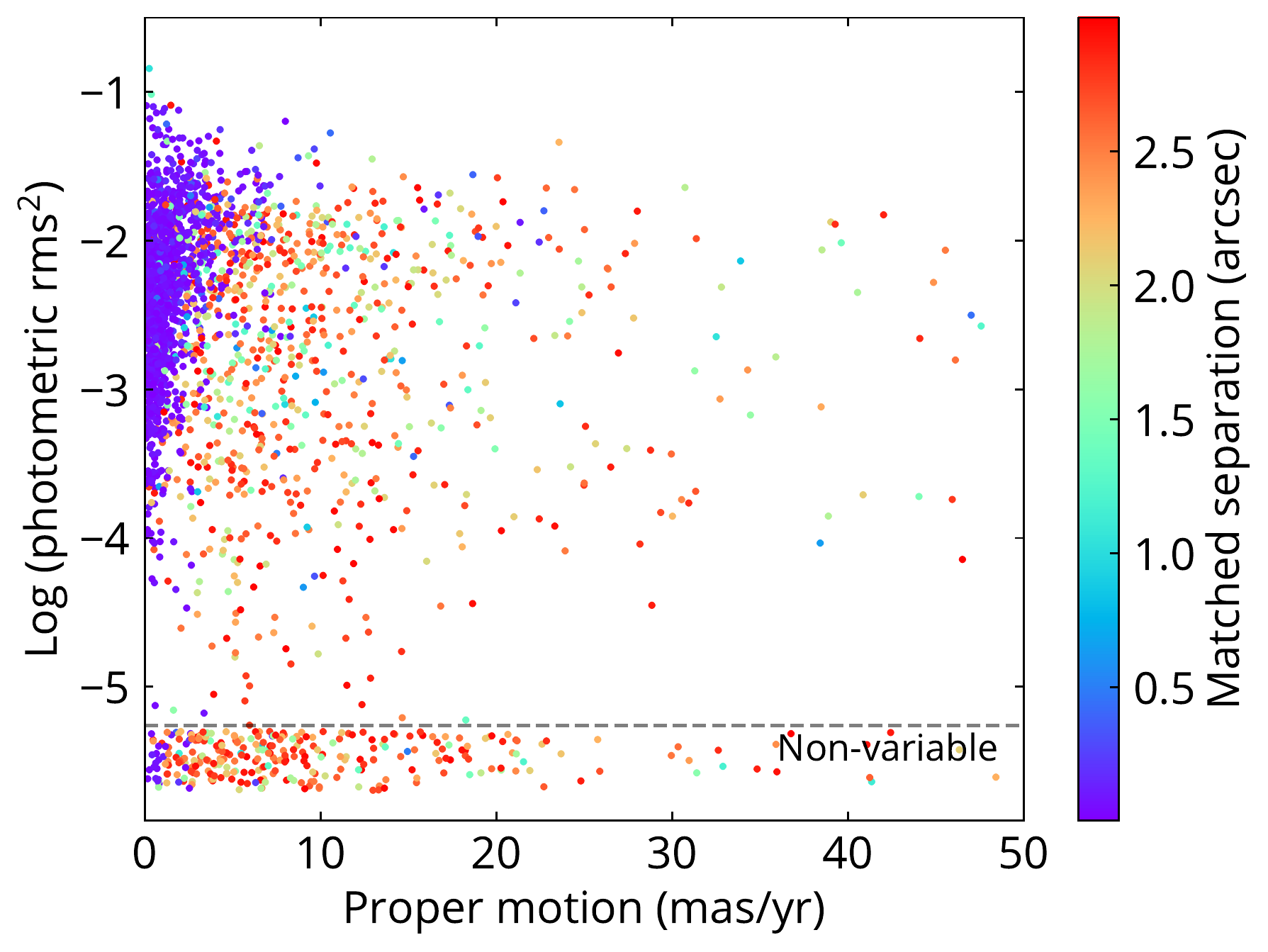}
 \includegraphics[width=0.47\textwidth]{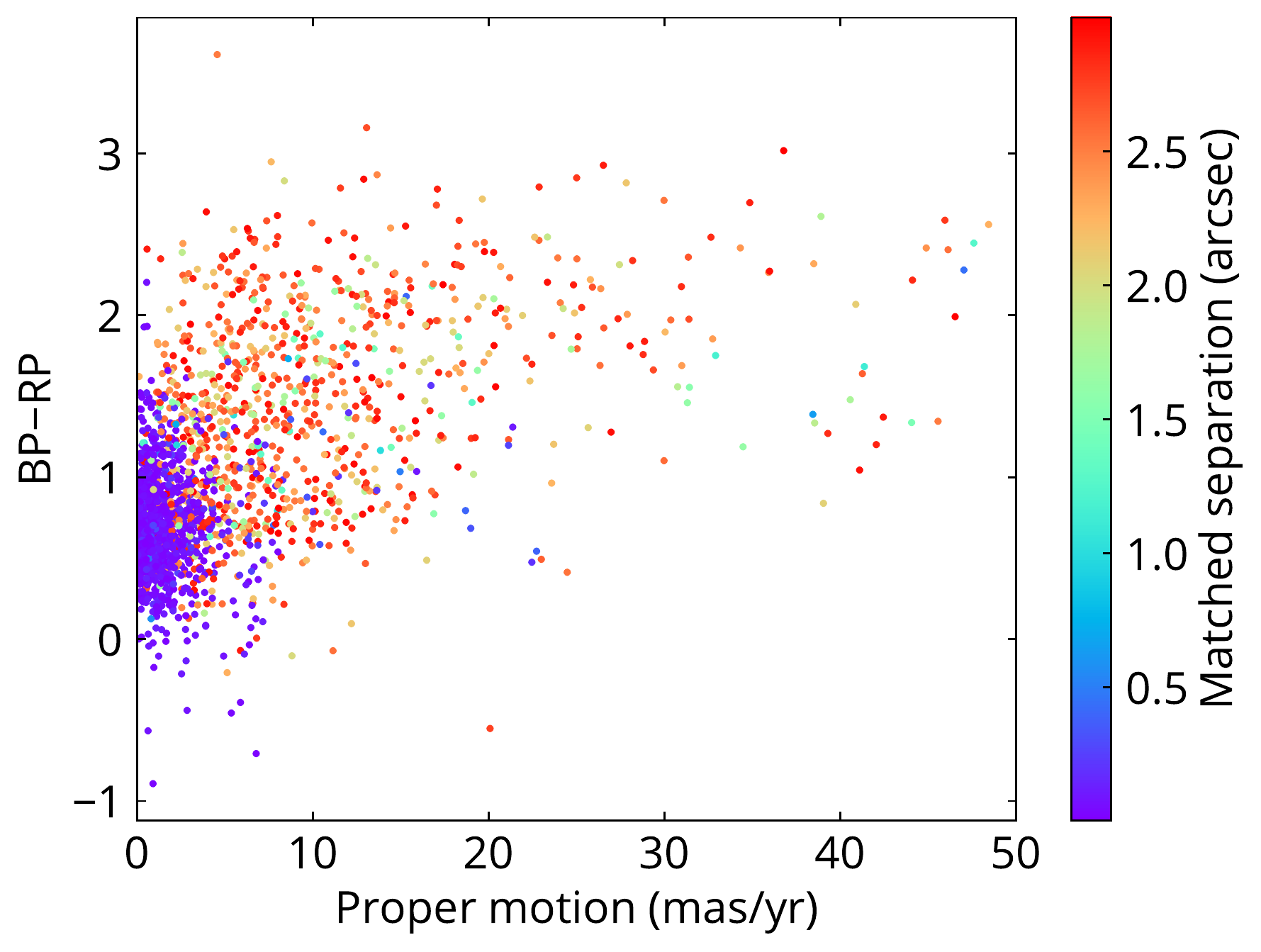}
 \caption{Left: fractional photometric variability versus total proper motion for SDSS quasars that have two Gaia matches within 3\arcsec. For sources where the photometric RMS is below the instrumental level, we place them below the dashed line with some small random vertical offsets. Gaia matches with separations $<0$\farcs{3} have small proper motions and significant variability, while sources with larger matched separations have higher proper motions, tend to be less variable, and have redder colors, suggesting that these larger-separation matches are likely to be foreground stars. Nevertheless, there are many systems where the multiple Gaia matches within 3\arcsec\ are consistent with AGN in terms of photometric variability, proper motion and color. These are promising dual AGN candidates on $\gtrsim$kpc scales that are already resolved by Gaia.}
\label{fig:var-pm}
\end{figure*}

The radio galaxy 0402+379 has a 7-pc AGN binary reported in \cite{rodriguez06}, but its host galaxy is too extended and smooth in optical and therefore it is not considered as a point-like source cataloged in Gaia DR2. 

SDSSJ092455.24+051052.0 is a dual AGN with a projected angular separation of 0\farcs{4} \citep{Liu2018}. While it does have high astrometric excess noise of $15$\,mas, it may be due to its extended host galaxy because its redshift is only $z=0.1495$ and it is a Type-2 AGN where strong photometric variability is not expected.

\section{Varstrometry of Quasars in Gaia DR2}\label{sec:qso}

We now proceed to investigate the photometric and astrometric RMS properties of quasars using Gaia DR2. We emphasize that in this initial study we will not be able to confirm any sub-kpc dual/off-nucleus AGN or small-scale lenses. Instead, we mainly use this exercise to understand potential systematics and to formulate our follow-up strategy for potential candidates (\S\ref{sec:strategy}).  

Our main quasar sample includes $\sim 500,000$ spectroscopically confirmed quasars from the SDSS DR7 quasar catalog \citep{Shen2011} and the SDSS DR14 quasar catalog \citep{Paris2018}. In \S\ref{sec:wiseqso}, we also consider a photometric quasar sample selected using data from the Wide-field Infrared Survey Explorer \citep[WISE,][]{Wright_etal_2010} survey and presented in \citet{Secrest2015}, to increase the overlap with the Gaia sky coverage.

\subsection{Quasars with multiple Gaia source detections}
\label{sec:quasar-multi}


We cross-match the SDSS quasars with Gaia DR2 with a matching radius of 3\arcsec, resulting in $\sim$350,000 quasars with Gaia matches out of the $\sim$500,000 parent quasars. The remaining SDSS quasars are too faint for Gaia DR2's detection limit ($\sim21$\,mag in G band). During the cross-match, we keep all Gaia sources within 3\arcsec\ (i.e., there could be multiple Gaia sources within 3\arcsec) because Gaia may resolve dual/off-nucleus AGN and lensed quasars separated by $\gtrsim 0.2$". 

We first investigate the multiple-quasar systems where each member is spectroscopically confirmed as a quasar in SDSS. There are 16 such systems in total: 11 of them are lensed/dual quasar candidates where the redshifts of the members are very close ($\Delta z<0.1$); 3 have members at different redshifts that are projected quasar pairs; one system contains two quasars, SDSSJ123401.31+063214.9 and SDSSJ123401.24+063212.1, with nearly identical redshifts but different colors. The last system, SDSSJ114653.06+164425.3, turns out to be an SDSS imaging processing problem and it is in fact a single quasar. In these cases, quasars may have Gaia cross-matches of the other member, so we manually identify the correct Gaia matches for individual members. 

\begin{figure*}
\centering
  \includegraphics[align=c,height=2.in]{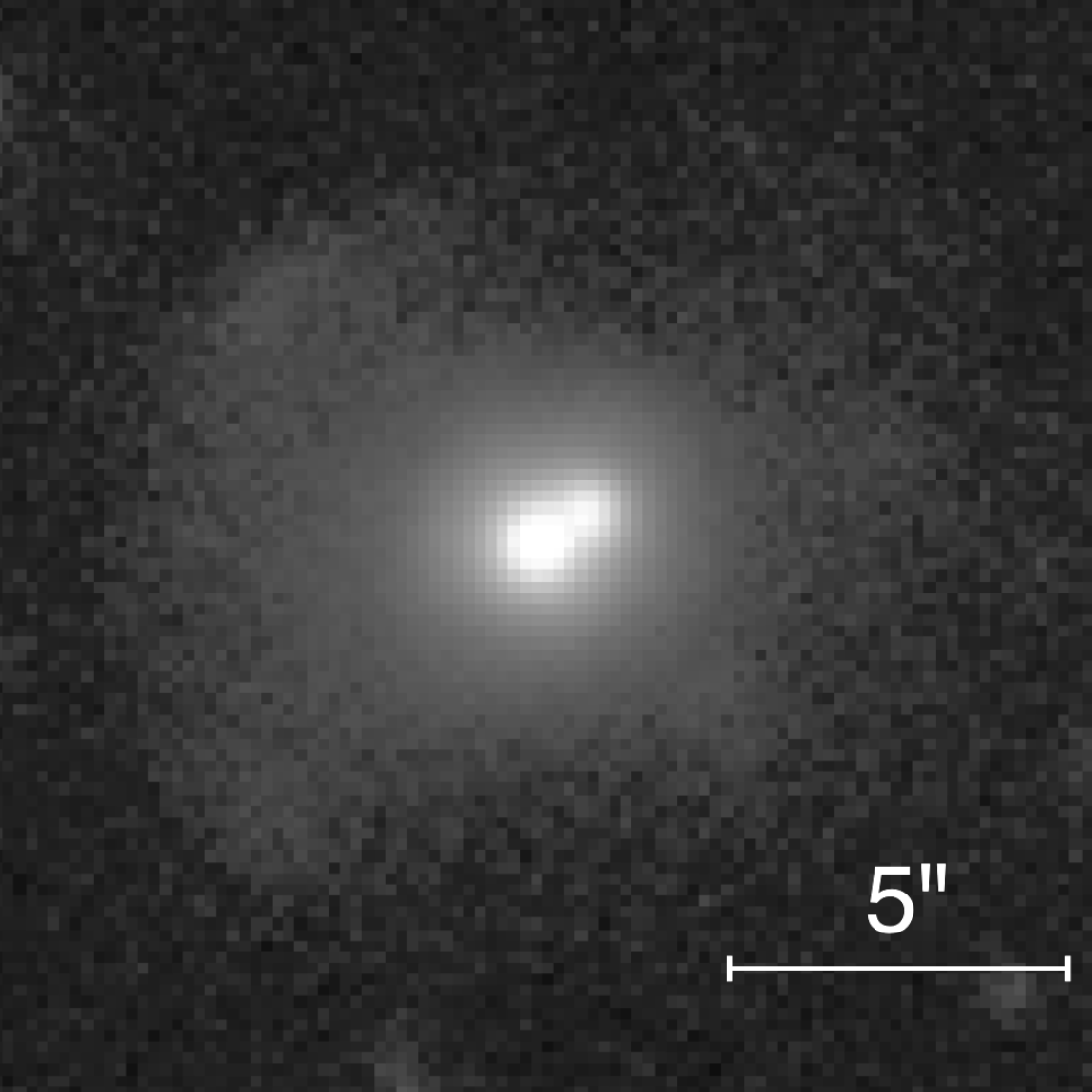}
  \hspace*{.3in}
 \includegraphics[align=c,height=2.7in]{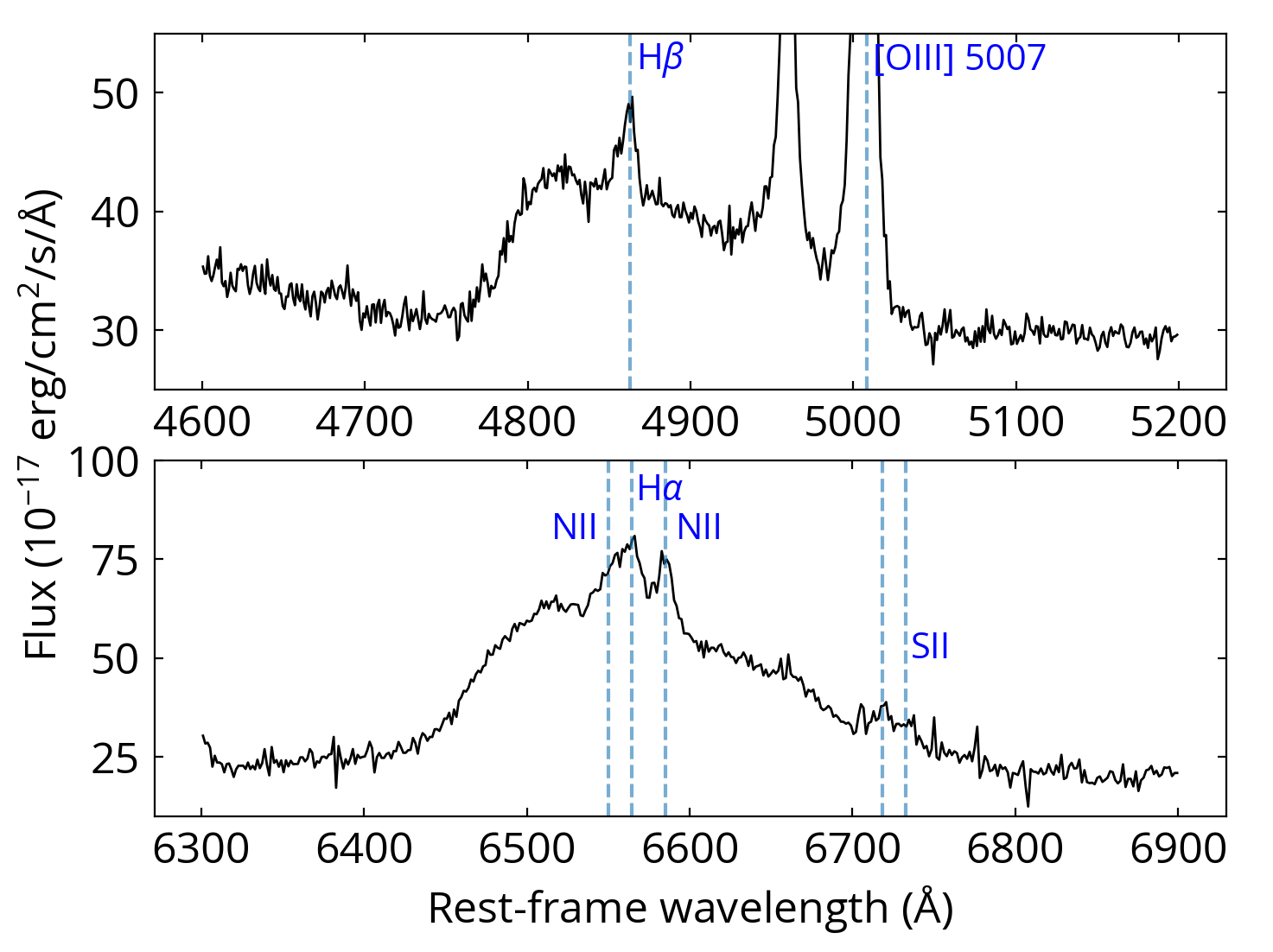}
 \caption{The HSC $i$-band image (left panel; north/east is up/left) and SDSS spectrum (right) of SDSSJ000710.01+005329.0, a quasar at $z=0.32$. Gaia DR2 detects two sources separated by $0.79$\arcsec\ in the source, coincident with the two cores visible in the HSC image. There is also a faint extended shell-like structure indicative of an evolved merger. Its SDSS spectrum shows disk-emitter-like features in the broad Balmer lines. This is likely a dual AGN with a projected separation of 3.7\,kpc. } 
 \label{fig:quasar0007}
\end{figure*}

Next we investigate single SDSS quasars with multiple matches in the Gaia database. Out of the $\sim$350,000 quasars with at least one Gaia matches within 3\arcsec, one quasar has 4 matched Gaia sources, five have 3 Gaia sources, and 1600 have 2 Gaia sources within 3\arcsec. In these multiply-matched cases, the closest match is usually the correct match to the quasar while others are close companions. The one having 4 Gaia sources is a well-known quadruply imaged, lensed quasar PG1115+080 (SDSSJ111816.94+074558.2) at $z=1.73$ \citep[e.g.,][]{Weymann1980, Young1981, Kristian1993}. Out of the five sources having 3 Gaia matches, 4 have resolved members with significantly different optical colors in Pan-STARRS \citep[][]{Chambers2016} images, while the last one (SDSSJ141546.24+112943.4) is a known quadruply lensed quasar \citep{Magain1988}. 

For the 1600 sources having two Gaia matched sources, they can be either dual AGN or off-nucleus AGN on $\gtrsim$ kpc scales, gravitational lensed quasars, projected quasar pairs, or quasars close in projection to foreground stars. Fig.\ \ref{fig:var-pm} shows the distributions of the Gaia matched sources for these 1600 quasars in the proper motion, photometric variability and color space, color-coded by the distance from the SDSS position. In Gaia DR2, BP and RP bands are measured from the sum of fluxes within a $3.5\times2.1$\,arcsec$^2$ window (no deblending treatment), so the BP$-$RP color in Fig.~\ref{fig:var-pm} represents the average color within such a window around the source. On average, Gaia sources with larger matched separations have higher proper motions, tend to be less variable, and have redder colors. Therefore we conclude that most of these large-separation sources are Milky Way stars. Nevertheless, there are many sources that are consistent with AGN in terms of photometric variability, proper motion and color. These are promising dual AGN candidates on $\gtrsim $kpc scales already resolved by Gaia DR2.


Fig.~\ref{fig:quasar0007} presents a $3.7$ kpc dual AGN candidate SDSSJ000710.01+005329.0 at $z=0.32$ that has two Gaia source detections separated by $0.79$\arcsec. The $i$-band image (left panel) from the Hyper Supreme-Cam (HSC) Survey DR2 \citep{Aihara_etal_2019} clearly shows two resolved cores and a more extended faint structure indicative of an evolved merger. Its SDSS spectrum (right panel) shows disk-emitter-like features in the broad Balmer lines \citep{Eracleous1994,Strateva2003}. The northwest Gaia source has {\tt astrometric\_excess\_noise}=0 and a fractional photometric variability of $\log(f_G^2) = -2.8$, and the southeast one has {\tt astrometric\_excess\_noise}=0.77\,mas and a fractional photometric variability of $\log(f_G^2) = -1.6$. Because the source is resolved by Gaia, {\tt varstrometry} should not lead to astrometric RMS, consistent with the $<1$\,mas astrometric excess noise for these two Gaia sources. The northwest Gaia source only has a 2-parameter astrometric solution (i.e., ra and dec), while the southeast one has a 5-parameter solution with proper motions and parallax consistent with zero. This example demonstrates that it is desirable to further investigate quasars that have multiple Gaia detections since they may be promising dual AGN candidates resolved by Gaia. The spatial resolution requirement to resolve these systems is well within the reach of ground-based adaptive optics or HST imaging, as well as radio interferometry. 




\subsection{Quasars with significant Gaia parallax or proper motion}


We now investigate quasars with significant parallaxes and proper motion detections in Gaia DR2. Ideally, quasars should have zero parallaxes and zero proper motions. Quasars with significant parallaxes and proper motions may inform us of sample contamination and systematics, or can be due to interesting physical causes, such as {\tt varstrometry}. We keep in mind that {\tt astrometric\_excess\_noise} is a more generally applicable indicator of astrometric RMS than parallax/proper motion as it is available even for sources with 2-parameter astrometric solution only (i.e., no available parallax or proper motion measurements). However, under certain circumstances, the intrinsic astrometric RMS may be mistakenly reported as a valid astrometric solution for parallax and proper motion measurements, with negligible reported {\tt astrometric\_excess\_noise} from Gaia DR2. The purpose of this subsection is to examine cases where Gaia reports significant parallaxes and proper motion detections for quasars, regardless of their {\tt astrometric\_excess\_noise} values. 

We select quasars that have positive parallaxes by (1) \texttt{parallax\_over\_error}$>5$, i.e., parallax inconsistent with zero at $5\sigma$; (2) the Gaia-SDSS separation $<1$\arcsec\ to reduce contamination from foreground stars (see Fig.~\ref{fig:var-pm}). For each quasar, we consider all Gaia matches satisfying the above criteria to include potential dual AGN resolved by Gaia. This results in 25 SDSS objects, where 15 have single Gaia matches and 10 have two Gaia matches within 1\arcsec. To determine the physical nature of these sources, we investigate their SDSS spectra and their optical colors in SDSS and Pan-STARRS 1. 

For sources with single Gaia matches, we classify them into three categories: (1) genuine quasars where broad emission lines are identifiable in SDSS spectra; (2) sources with nearly featureless continuum in SDSS spectra; (3) magnetized white dwarfs where the line splitting from the Zeeman effect is identifiable. Category (2) can either be weak-line quasars or DC (no absorption lines) white dwarfs \citep{Collinge2005}.

Out of the 15 SDSS objects with significant non-zero parallaxes and single Gaia matches, 9 are in category (1), 4 are in category (2), and 2 are in category (3). We compute the significance of their proper motions and find that sources in category (2) and (3) have non-zero proper motions at $\gtrsim 20\sigma$, while sources in category (1) have significance less than $10\sigma$ (only one is above $5\sigma$). By using their parallaxes to compute the G-band absolute magnitudes, we find that sources in category (2) and (3) are consistent with the location of white dwarfs in the color-absolute magnitude diagram, while sources in category (1) are mostly scattered between the white dwarf track and the main sequence of stars. Therefore, we conclude that sources in category (2) and (3) are mostly (if not all) white dwarfs misclassified as quasars in the SDSS quasar catalog, and category (1) is a clean quasar sample.


For the 10 SDSS objects that have two close Gaia matches within 1\arcsec, only one of the two Gaia matches has a significant non-zero parallax. We investigate their proper motions and optical colors. We find that five Gaia sources with non-zero parallaxes have non-zero proper motions at $>10\sigma$, and red optical colors of BP-RP$>0.9$ whenever applicable. Furthermore, the SDSS spectra of two of these systems show identifiable M-dwarf spectral features (SDSSJ092853.51+570735.3 and SDSSJ014349.15+002128.3). We conclude that these five quasars are in close projection to foreground stars. The other five Gaia matches to SDSS quasars with non-zero parallaxes have bluer optical colors of BP-RP$<0.8$ (similar to Gaia matches with zero parallaxes) and less significant proper motions ($<10\sigma$). They may be multiply lensed quasars, although we cannot rule out the possibility that they are blue foreground stars.

We conclude with a sample of 14 genuine quasars that have non-zero parallaxes, where nine have single Gaia matches and five have two Gaia matches within 1\arcsec. Their redshifts range from 0.44 to 3.5. For the nine quasars with single Gaia matches within 1\arcsec\ the astrometric solution is not affected by crowding. Out of these nine quasars, 6 have \texttt{visibility\_periods\_used}$\ge 9$, so they should have sufficient numbers of observations to derive reliable parallaxes. Three out of nine have {\tt astrometric\_excess\_noise}$>1$\,mas, suggesting these sources may have unusual behaviors in their astrometric measurements. These nine quasars are good candidates for {\tt varstrometry}-selected sub-kpc dual/off-nucleus AGN. 


We also investigate quasars that have non-zero proper motions. We select candidates with (1) total proper motion $5\sigma$ inconsistent with 0, and (2) Gaia-SDSS cross-match separation $<1$\arcsec. The selection results in 51 SDSS quasars. This sample contains 11 systems that have non-zero parallaxes which we have discussed above, but none of them are considered as genuine quasars by our previous classification. Excluding these 11 non-zero parallax systems, we end up with 40 SDSS quasars that have non-zero proper motions, with 19 of them having a single Gaia source and 21 having two Gaia sources within 1\arcsec. 

For quasars that have non-zero proper motions, we check their SDSS spectra, and for those having multiple Gaia detections, we check whether the detected sources have similar colors in order to classify them as either foreground stars or lensing candidates. Out of the 19 sources with single Gaia source detection, the SDSS spectra show that 8 are genuine quasars, 7 have prominent stellar features (F, G, K, M stars) with some trace of AGN features indicating star+quasar superposition, one has peculiar spectral features that might be a carbon star (SDSSJ093306.61+500544.4), and 3 are featureless.


For the 21 quasars with non-zero proper motions and two matched Gaia sources, the SDSS spectra of 10 objects have prominent stellar spectral features and therefore are superpositions with foreground stars. For the remaining 11 quasars, all of them have AGN-dominant spectral features with little trace of foreground stars. Some of them are known lensing systems, for example, SDSSJ091127.61+055054.1 (RXJ0911.4+0551). 





We compile tables, Pan-STARRS $g+z$ optical images, and SDSS spectra for genuine quasars that have significant non-zero parallaxes or proper motions in the Appendix. We exclude white dwarfs misclassified as quasars and quasar-star superpositions. 

It is unclear why some genuine quasars and lensing systems have non-zero parallaxes and/or proper motions. {\tt Varstrometry} could be one physical cause, i.e., the astrometric RMS is mistaken as a valid astrometric solution. Alternatively, the ``detected'' parallaxes and/or proper motions are due to systematics, such as extended hosts. Nevertheless, quasars with significant parallaxes and proper motions are rare, only $\sim0.01\%$ of the entire sample, thus we must be careful with the possibility that these non-zero parallaxes and proper motions are simply statistical outliers. To confirm the nature of these systems we need independent follow-up observations that can potentially resolve their small-scale structure, or wait for future Gaia releases to provide more information and reduced systematics.

\subsection{Varstrometry for quasars}

\begin{figure*}
\centering
 \includegraphics[width=0.9\textwidth]{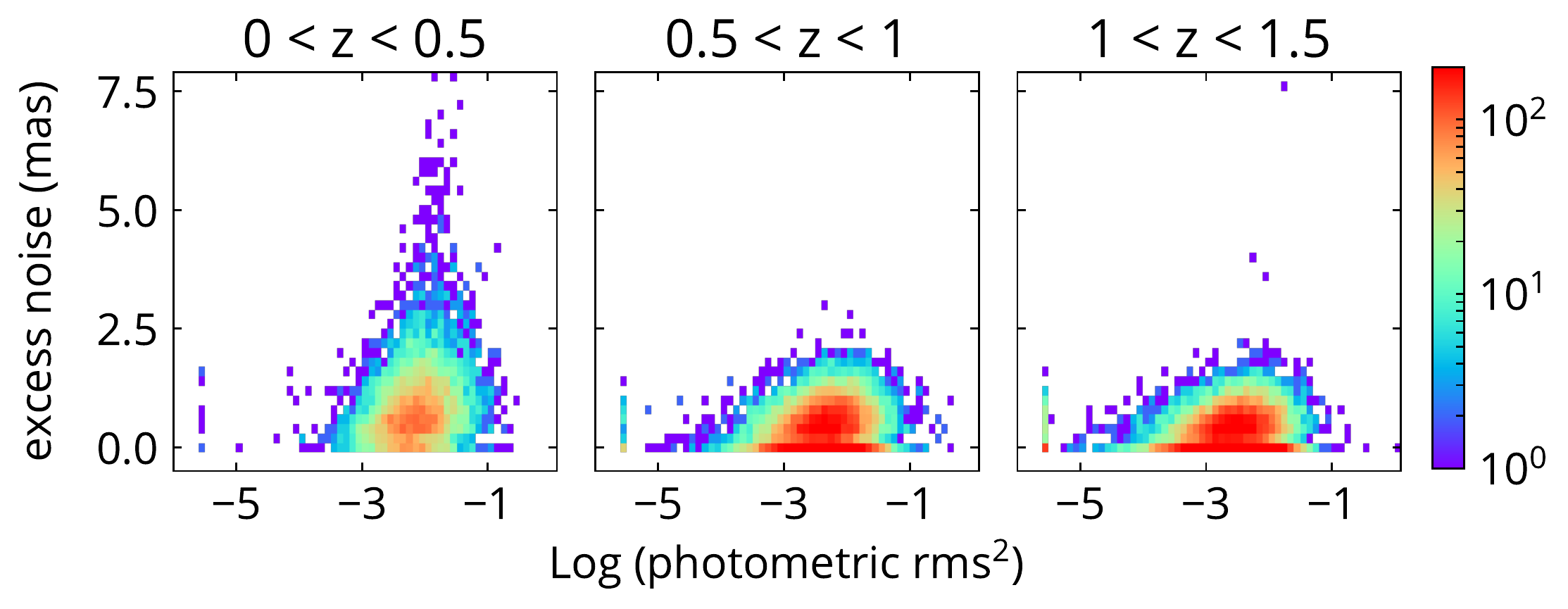}
 \caption{Gaia {\tt astrometric\_excess\_noise} versus photometric variability for spectroscopic SDSS quasars in three redshift ranges. There is a tail extending to high {\tt astrometric\_excess\_noise} in the sample of $z<0.5$. The tail is caused by extended host galaxies at lower redshifts and is not seen at higher redshifts. Quasars with high {\tt astrometric\_excess\_noise} at redshifts $>0.5$ may be the sub-kpc dual/lensed AGN candidates. The mild correlation between astrometric excess noise and variability is likely due to covariance in the measurements, e.g., both RMS values are measured from fluxes in the same bandpass and measuring window. The locus of the distribution suggests that there is a floor level of $\sim 1$ mas for the reliability of {\tt astrometric\_excess\_noise}.} \label{fig:ss-quasar}
\end{figure*}




\begin{figure*}
\centering
 \includegraphics[align=c,height=2.in]{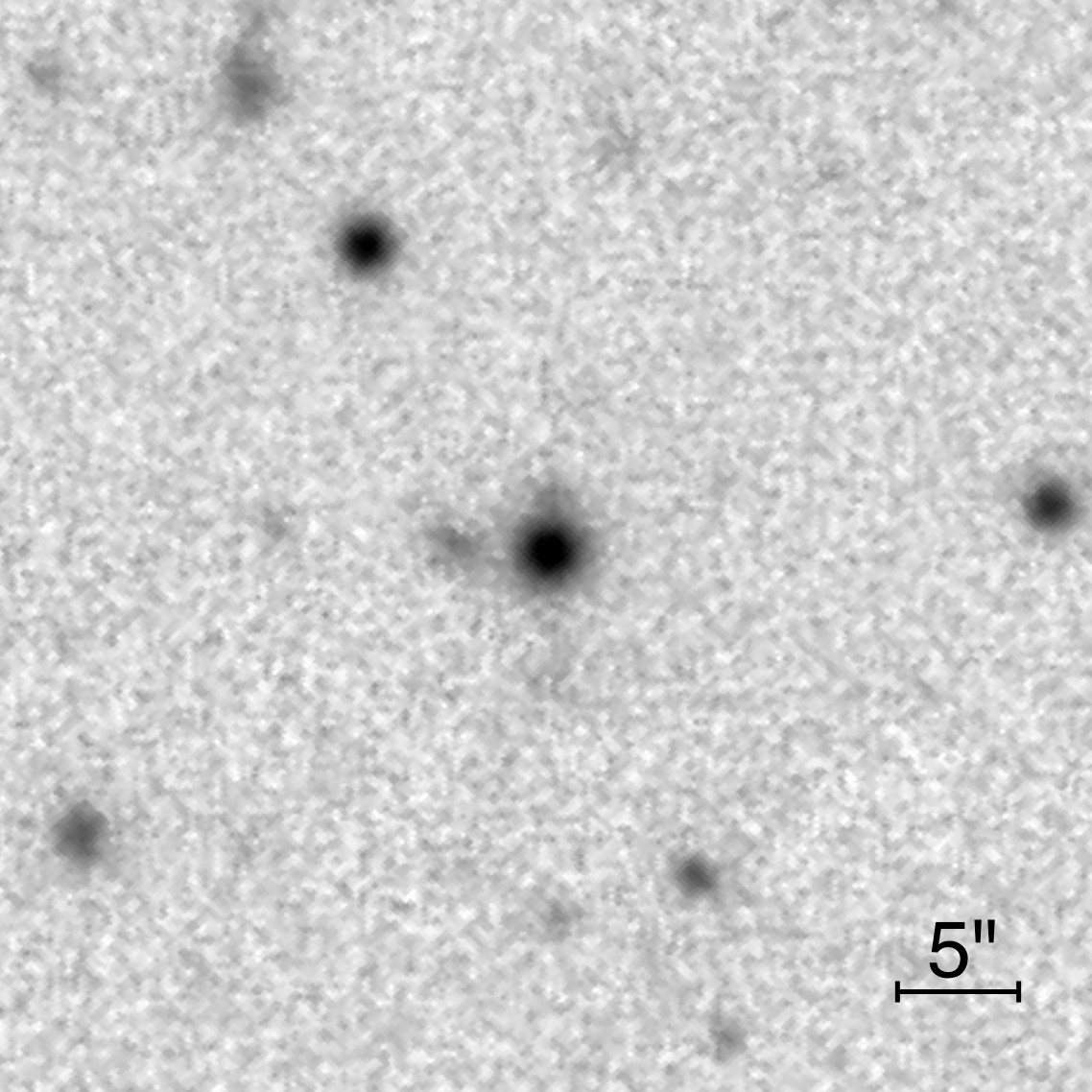}
 \hspace*{.3in}
 \includegraphics[align=c,height=2.7in]{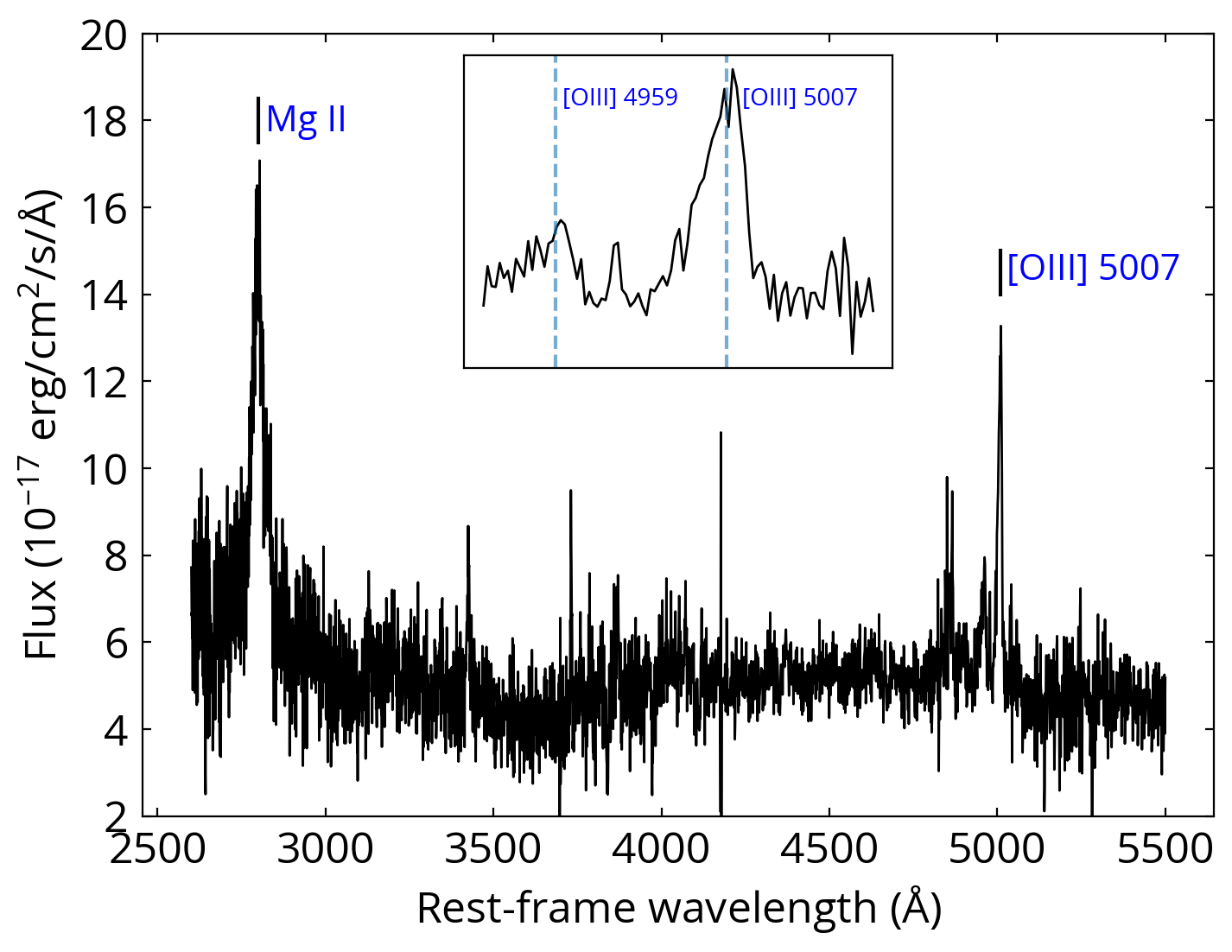}
 \caption{The Pan-STARRS $i$-band optical image (north/east is up/left) and SDSS spectrum of SDSSJ112101.30+080926.3, a quasar (centered in the image) that has an astrometric excess noise of 2.9\,mas at $z=0.51$. The $i$-band image shows some faint companions around the quasar, one to the north and one to the east. Its \oiii$\lambda5007$ narrow emission line shows an asymmetric profile (when zoomed in) that may be caused by multiple kinematic components associated with a dual AGN. } 
 
 \label{fig:quasar1121}
\end{figure*}

\begin{figure*}
\centering
 \includegraphics[align=c,height=2.in]{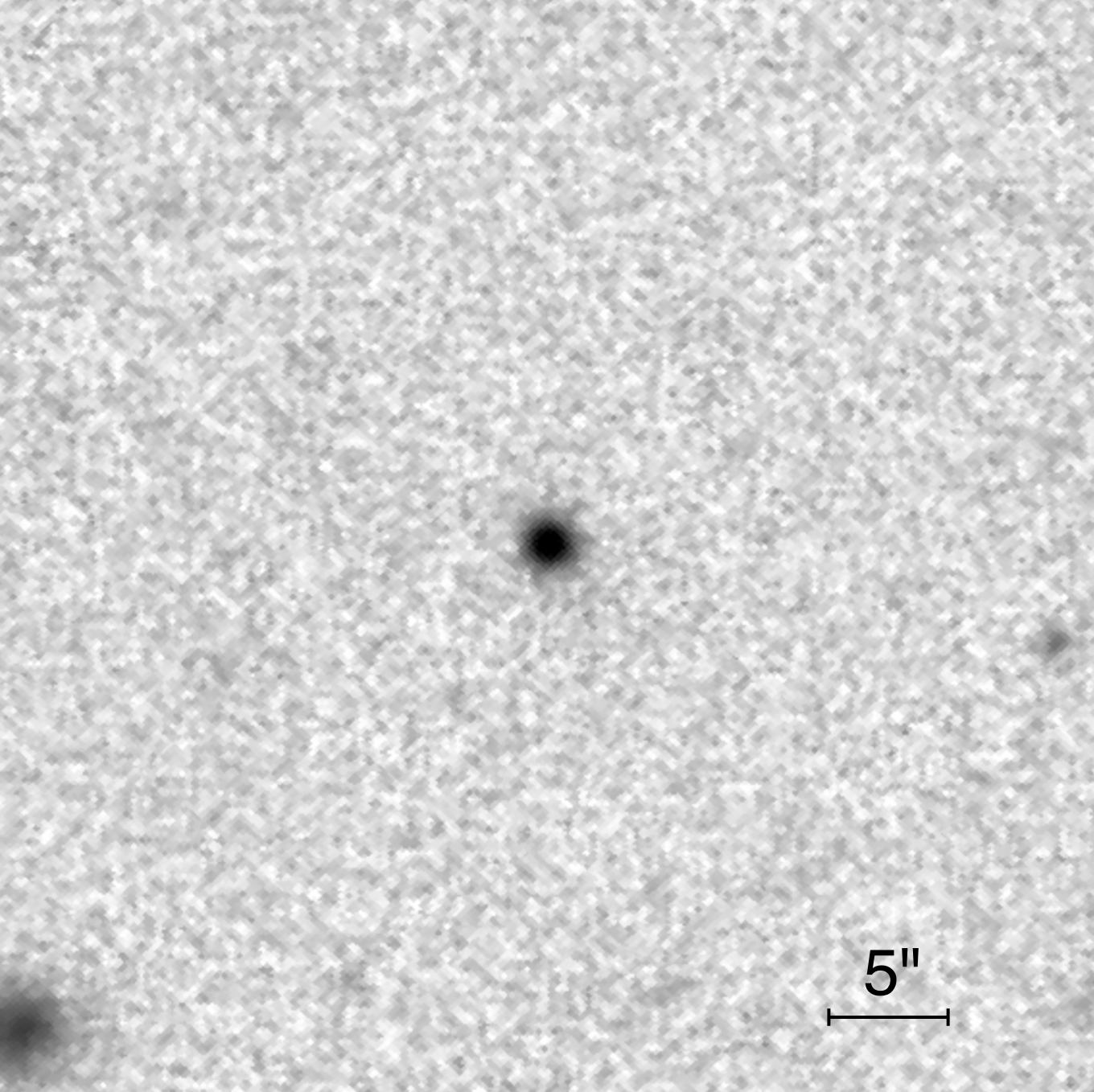}
 \hspace*{.3in}
 \includegraphics[align=c,height=2.7in]{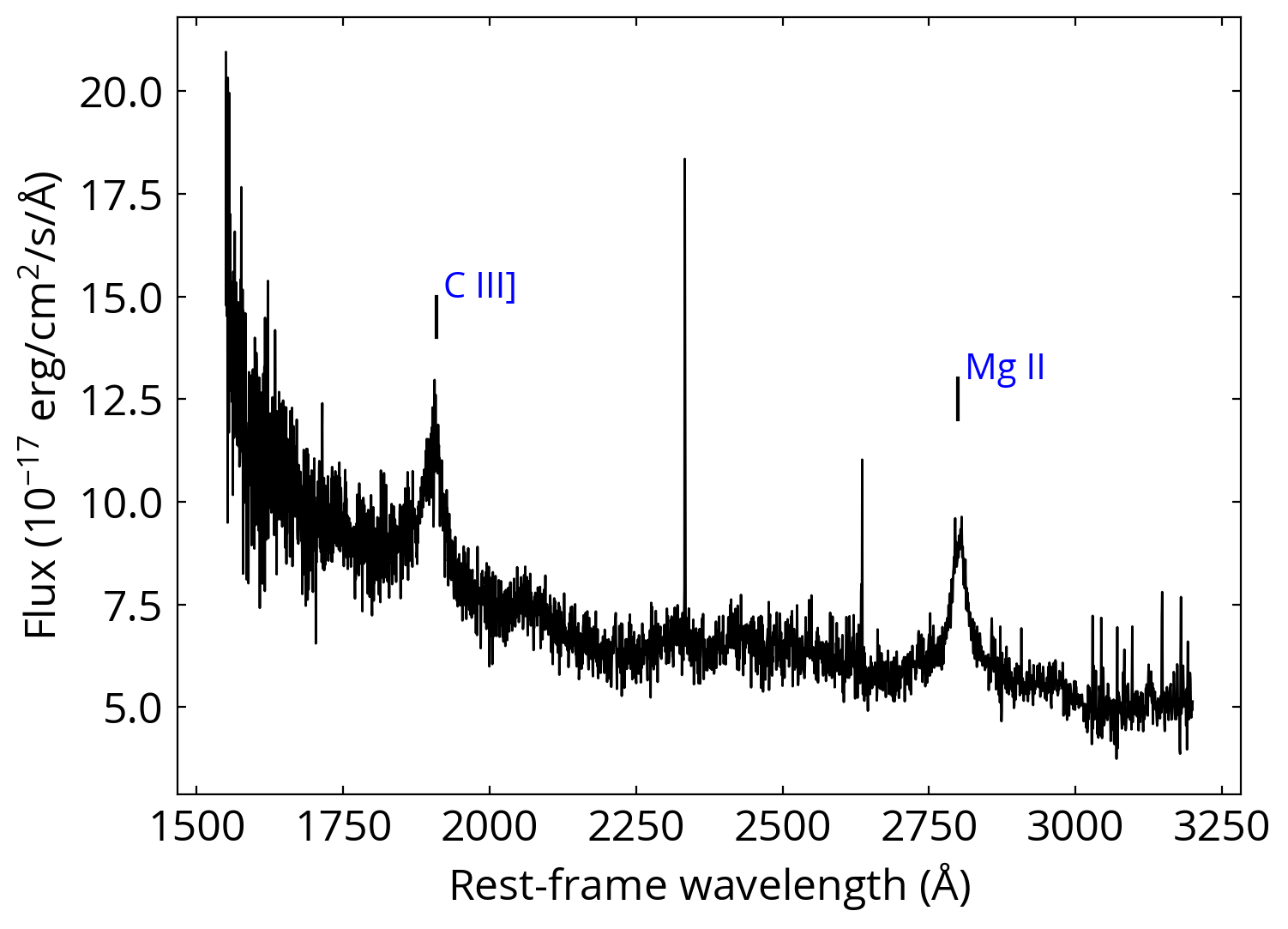}
 \caption{The Pan-STARRS $i$-band optical image (north/east is up/left) and SDSS spectrum (right) of SDSSJ000252.60$-$034345.1, a quasar (centered in the image) that has an {\tt astrometric\_excess\_noise} of 3.7\,mas at $z=1.39$. No source is strongly detected within $20$\,arcsec around the quasar down to Pan-STARRS limiting magnitude of 23.1. } 
 \label{fig:quasar0002}
\end{figure*}

Fig.~\ref{fig:ss-quasar} shows the distribution of SDSS quasars in the {\tt astrometric\_excess\_noise} vs. fractional photometric variability plane, for three redshift ranges. Here we only consider quasars that have a single Gaia matched source within 3\arcsec, are brighter than 19.5\,mag in G band, and have {\tt visibility\_periods\_used}$\ge 9$ so that Gaia DR2 has sufficient number of observations for photometric and astrometric RMS measurements. These additional cuts result in $\sim86,000$ SDSS quasars out of the parent $\sim350,000$ sample with Gaia matches. For sources where the variability is insignificant compared to the instrumental correction (i.e., $f_{G,raw} \leq f_{G,inst}$), we place them in the left-most column of the plots. For the $z<0.5$ sample, there is a long tail at $\log(f_G^2)\sim-2$ extending to high {\tt astrometric\_excess\_noise}. The samples at $0.5<z<1$ and $1<z<1.5$ have very similar distributions in Fig.~\ref{fig:ss-quasar}, and neither of them displays a tail towards high astrometric noise seen in the $z<0.5$ sample. This tail of high {\tt astrometric\_excess\_noise} has values similar to those seen in the inactive galaxy sample (\S\ref{sec:host} and Fig.~\ref{fig:galaxy-r50}), and thus it is mostly due to the extended structure (still treated as a single source in Gaia DR2) of the host galaxies that becomes more prominent at low redshifts. This effect of host galaxies seems insignificant for bright ($G<19.5$\,mag) quasars at $z>0.5$. This trend is expected because (1) SDSS probes more luminous quasars at higher redshifts that dominate the total flux, (2) the angular sizes of the host galaxies are smaller at higher redshifts, and (3) host galaxies are very faint in the rest-frame UV at high redshifts, and contribute much less to the total flux measured by Gaia than at lower redshifts.  


Fig.~\ref{fig:ss-quasar} also reveals a mild correlation between the photometric RMS and {\tt astrometric\_excess\_noise}, which we believe is due to the covariance in the astrometric and photometric RMS estimates from the same data. For example, any residual host galaxy systematics resulting from window sizes and scanning directions of Gaia measurements increases both the photometric RMS and astrometric RMS estimates. The locus of the distribution in Fig.~\ref{fig:ss-quasar} therefore roughly defines a ``floor'' value of reliable {\tt astrometric\_excess\_noise} around $\sim 1$ mas for quasars, below which we do not consider the intrinsic astrometric RMS estimate reliable. 

Among $\sim 37,000$ quasars with redshifts of 0.5--1.5, $6$\% have {\tt astrometric\_excess\_noise} $>1$\,mas, and they tend to have photometric variability at the $\sim10$\% level. This {\tt astrometric\_excess\_noise} is difficult to explain with host galaxies in this redshift range. These quasars may be the sub-kpc dual/off-nucleus AGN and lensed AGN that we are interested in. Confirming (i.e., resolving) them at $<100$ mas imaging resolution may be challenging. But at the very least, these astrometric RMS estimates place stringent limits on the off-nucleus distance of these small-scale pairs \citep[e.g.,][]{vodka2}. 

We inspect the optical spectra and images of quasars at redshifts$>0.5$ with ${\tt astrometric\_excess\_noise}>2.5$\,mas. In the redshift range $0.5<z<1$, two sources, SDSSJ123913.86+281434.1 and SDSSJ112101.30+080926.3, have astrometric excess noise of 3.1\,mas and 2.9\,mas, respectively. Both of them show arguably double-peaked \oiii$\lambda$5007, and SDSSJ112101.30+080926.3 clearly has an asymmetric broad \oiii$\lambda$5007 emission (Fig.~\ref{fig:quasar1121}). The asymmetric broad \oiii\ profile could be due to a galactic wind (e.g. \citealt{Zakamska_Greene_2014}) or an double-peaked \oiii\ from a dual AGN (e.g. \citealt{Liu2018}). Their optical images in Pan-STARRS show that both of them have red faint sources within $\sim3$\arcsec\ that are not detected by Gaia. These red sources may be foreground stars, merger companions of the quasar, or tidal streams. Even though these close red sources are undetected in Gaia, we are unable to rule out the possibility that the {\tt astrometric\_excess\_noise} may be caused by these faint sources. Also, these two quasars happen to have redshifts of 0.52 and 0.51, and it is possible that their host galaxies may contribute to some uncertainties in the astrometric RMS.

There are three quasars with astrometric excess noise $>2.5$\,mas at redshift $1<z<1.5$. Their SDSS spectra clearly show quasar broad-line features, with no trace of stellar spectra from chance superposition with a foreground star. Fig.~\ref{fig:quasar0002} presents one example, SDSSJ000252.60-034345.1, that has an {\tt astrometric\_excess\_noise} of $3.7$\,mas at $z=1.39$. In its $i$-band image from Pan-STARRS 1 (left panel), there is no detected extended host or companion within 15\arcsec\ that can affect the astrometric measurements. If its {\tt astrometric\_excess\_noise} is from an unresolved equal-flux dual AGN, Equation~(\ref{eqn:rms}) implies a projected separation of 81\,mas, or 700\,pc at its redshift.


\subsection{Extension to the WISE quasar sample}\label{sec:wiseqso}

So far, we have used the spectroscopic quasar sample from SDSS to reduce source contamination. To increase the sky overlap with Gaia, we use an all-sky quasar sample selected using WISE data and presented in \citet{Mateos2012} and \citet{Secrest2015}. These WISE-selected photometric quasars are used by Gaia to define the celestial frame \citep{Mignard2018, Lindegren2018}. While most of them should be genuine quasars, not all of them are spectroscopically confirmed and have known redshifts. Similar to the SDSS quasar sample, here we only consider WISE-selected quasars that are brighter than 19.5\,mag in G band and have {\tt visibility\_periods\_used}$>8$.

Here we use a subset of these WISE quasars to demonstrate the varstrometry selection of candidate sub-kpc dual/off-nucleus AGN, and the utility of Gaia to select blazars. We first limit the WISE quasar sample to those that Pan-STARRS 1 flags as point sources in all filters to reduce the systematics caused by extended host galaxies. To include radio information, we cross match the sample with the catalog of Faint Images of the Radio Sky at Twenty Centimeters (FIRST; \citealt{Becker1995, White1997}). 

\begin{figure}
\centering
 \includegraphics[width=0.45\textwidth]{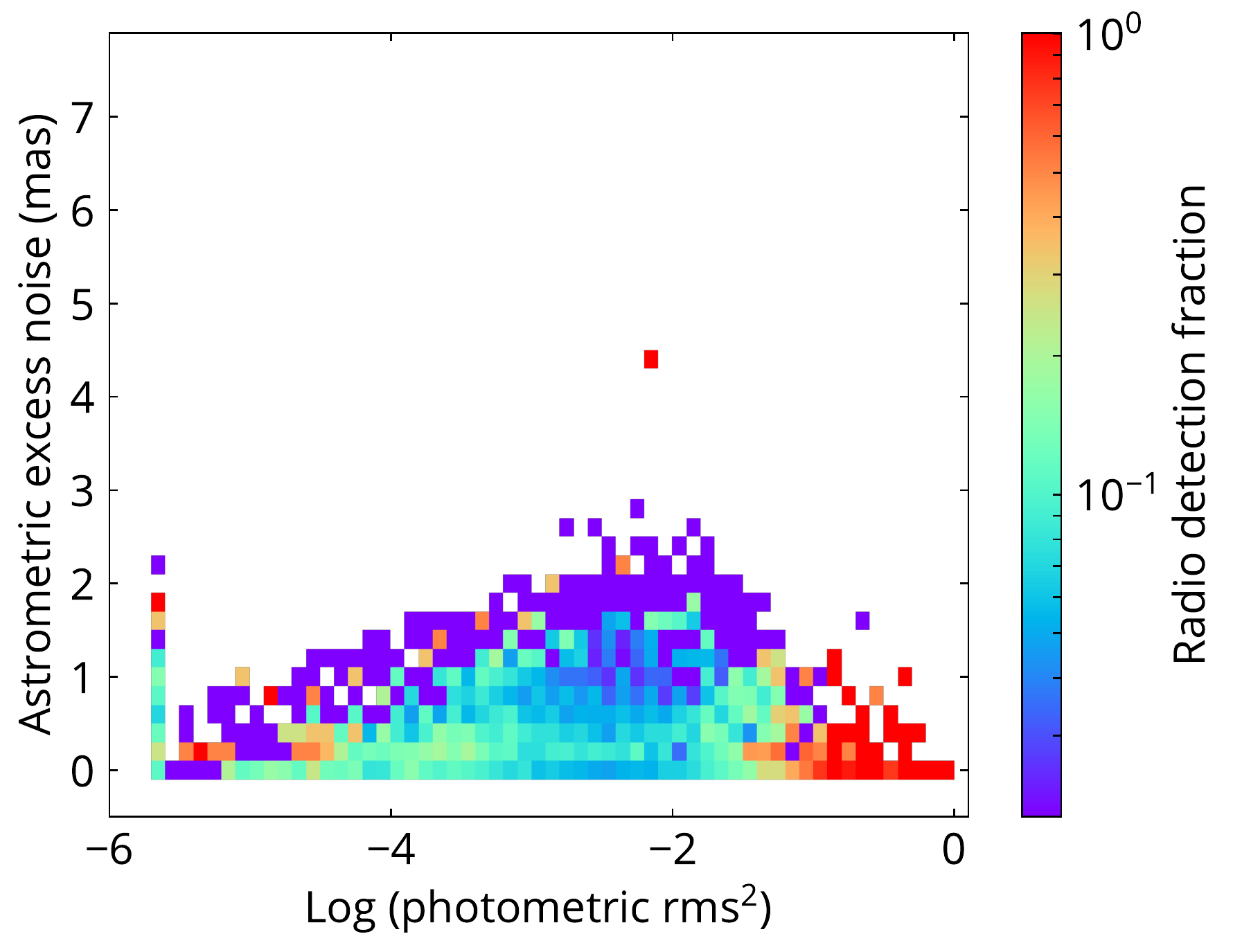}
 \caption{The radio detection fraction for WISE-selected quasars that are classified as point source in Pan-STARRS and are within the FIRST footprint. Compared to Fig.~\ref{fig:ss-quasar}, one significant difference is that the WISE-selected sample has more highly variable (fractional photometric RMS $>30\%$) blazars with a high radio detection fraction but low {\tt astrometric\_excess\_noise}.} \label{fig:ss-wisequasar}
\end{figure}

In Fig.~\ref{fig:ss-wisequasar}, we present the radio-detection fraction in FIRST for the WISE-selected quasar sample. One significant difference between the spectroscopic SDSS quasars in Fig.~\ref{fig:ss-quasar} and WISE-selected quasars in Fig.~\ref{fig:ss-wisequasar} is that WISE-selected quasars retain a larger number of strongly variable sources ($\log(f_G^2)>-1$). Fig.~\ref{fig:ss-wisequasar} shows that these strongly variable sources have very high radio detection fractions but low {\tt astrometric\_excess\_noise}. These strongly variable sources may be blazars or flat-spectrum radio quasars, and the properties of variability-selected extragalactic sources in Gaia are currently being investigated (Isler et~al., in prep.).

A few strongly variable sources ($\log(f_G^2)>-1$) still have {\tt astrometric\_excess\_noise} $>1$\,mas. If the astrometric jitter is caused by the superluminal motion of the jet, it requires a minimal Lorentz factor of 50 to explain a jitter of $\sim1$\,mas, and the jet has to be propagating within an angle of 0.02 radian to the line of sight. In addition to that, the jet has to be extended and optically bright enough to induce significant astrometric jitter, which is difficult. Alternatively, the astrometric RMS can be due to variable sub-kpc jets and knots in the optical, or genuine dual/off-nucleus AGN. We plan to investigate these systems in a follow-up study.

\section{Discussion}\label{sec:disc}


\subsection{Merits and limitations of {\tt varstrometry} }

We have laid out the working principles of {\tt varstrometry} in the context of finding stationary (position-wise), unresolved offset or dual AGN systems where the photocenter varies along with photometric variations. The expected astrometric signal at fixed pair separation is strongest when the members of the unresolved source are highly variable and have comparable fluxes. The most significant advantage of this technique is its ability to explore AGN separations below $\sim 1$ kpc, utilizing the superb astrometric precision from astrometry missions like Gaia. The technique may also have a broader application in general, e.g., to probe Galactic stellar binary systems \citep[e.g.,][]{Makarov2016}. 

Using {\tt varstrometry} and the all-sky Gaia data, we can perform systematic searches of sub-kpc off-nucleus and dual AGN, and compare with predictions from simulations. For example, some galaxy formation simulations predict a large number of wandering active SMBHs in the host galaxy during the inspiraling phase \citep[e.g.,][]{Tremmel2018}, and this population of off-nucleus AGN can be best constrained by this {\tt varstrometry} technique and Gaia data. In a similar spirit we can also confront the observational constraints using this technique with the predicted population of recoiling SMBHs in simulations \citep[e.g.,][]{Blecha2016} to test the kick velocity distribution, correlations with galactic potential, and accretion recipes in these simulations. Another application is strong gravitationally lensed quasars with closely separated multiple images below $\sim$0\farcs2 Gaia resolution: because of the time delays in the intrinsic quasar variability between these gravitationally lensed images \citep{BlandfordNarayan1992}, we will observe similar astrometric jitter in the source photocenter. In the future with much higher astrometric precisions \citep[e.g., $\mu$as;][]{Vallenari2018}, it may even be possible to apply {\tt varstrometry} to the variable broad-line region of quasars \citep[e.g.,][]{Shen2012a} to independently constrain the broad-line region size.

The major limitation of this technique is that it only applies to systems where at least one offset member is an optically unobscured AGN to ensure photometric variability of the system. In addition, for the induced astrometric signature to be detectable given the astrometric precision, it requires a suitable combination of pair separation (or off-nucleus distance), flux ratio, and photometric variability amplitude (Eqns. \ref{eqn:rms2}--\ref{eqn:rms3}).

There are also technical limitations of the application of {\tt varstrometry} that are specific to Gaia. We have shown that extended hosts (albeit still treated as point source in Gaia DR2) may impact the quality of Gaia astrometry and the interpretation of the reported {\tt astrometric\_excess\_noise}. Thus follow-up observations, with sufficient spatial resolution to resolve the tentative sub-kpc pair, are required to confirm Gaia-selected candidates. Gaia DR2 does not release time series of photocenter measurements and light curves, hence we can only use proxies in the Gaia DR2 catalog to approximate the (intrinsic) photometric and astrometric RMS. Ideally we require a full understanding and characterization of the systematics in the measured astrometric RMS, at a level that will make the {\tt varstrometry} technique competitive. For example, if we can control the total systematic astrometric RMS to less than a few mas, we can confidently attribute a large measured astrometric RMS to the {\tt varstrometric} signal. 

Another case where we can confidently confirm an off-nucleus AGN+galaxy system with {\tt varstrometry} alone is where the photocenter variations are aperiodic, bound and linear, and correlates with the photometric light curve (\S\ref{sec:offagn}). Time series from Gaia are required for such analyses.

\subsection{Follow-up strategy}\label{sec:strategy}

To further explore systematics in Gaia astrometry, and to confirm candidate sub-kpc dual or off-nucleus AGN selected from Gaia DR2, we require high-resolution imaging follow-up observations. Additional observations, such as spatially-resolved spectroscopy, are also important to reduce ambiguities. Below we discuss our follow-up strategy.  

Gaia provides many resolved pairs at $\gtrsim0$\farcs{3} scales; most of them are foreground stars but some of them may be off-nucleus or dual AGN on $\sim 1-10$ kpc scales. For AGN with multiple Gaia detections within 3\arcsec ($D\sim 0.3-3$\arcsec), the following candidates are of high priority:

\begin{enumerate}
    \item[$\bullet$] targets with resolved Gaia sources that are consistent with AGN in terms of color, variability and parallax (proper motion) measurements -- to remove most contaminants from foreground stars;
    \item[$\bullet$] targets with radio detection -- to facilitate high-resolution imaging with radio interfreometry. 
\end{enumerate}

For $<1$ kpc candidates selected with {\tt varstrometry} from unresolved Gaia sources ($D\lesssim$0\farcs{3}), the following candidates are of high priority:
\begin{enumerate}
    \item[$\bullet$] targets with less extended morphology from ground-based seeing-limited imaging -- to mitigate Gaia DR2 systematics on extended sources;
    \item[$\bullet$] targets whose expected separation is resolvable, i.e., those with large photometric and astrometric RMS, or significant parallax and/or proper motion detections;
    \item[$\bullet$] targets with significant color variability -- to increase the probability of dual/off-nucleus AGN;
    \item[$\bullet$] targets with radio detection because dual/off-nucleus radio-loud AGN are suitable targets for VLBI imaging.
\end{enumerate}

In addition, for those with spatially-integrated fiber spectroscopy from SDSS (within 3\arcsec\ or 2\arcsec\ fiber diameter), candidates with obvious AGN-star superposition are not considered. These follow-up strategies are designed to improve the overall success rate of confirmation, and also to explore additional systematics from Gaia DR2. 

\section{Summary and Conclusions}\label{sec:con}

In this paper, we have introduced a new technique {\tt varstrometry}, based on the ideas proposed in \citet{Shen2012a} and \citet{Liuyuan2015} where astrometric jitter induced by intrinsic photometric variability in unresolved off-center sources can be used to constrain the pair separation. A particular application of this technique is to identify sub-kpc dual and off-nucleus AGN with mas-precision astrometry. We present the basic principles of this technique, and explore its feasibility and potential systematics using Gaia DR2. Our main results are:

\begin{enumerate}
	\item Candidate sub-kpc dual or off-nucleus AGN can be identified based on Gaia detected photocenter variations that are bound, linear and aperiodic. If there is further a strong correlation between the photocenter shift and the total flux variability, it will signal a single off-nucleus AGN that dominates the variable light. If the host light can be further derived from spectral or imaging decomposition in any epoch, the time series of the photocenter and flux measurements can be used to fit for the pair separation via Eqn.\ (\ref{eqn:dphoto}). 
	
    \item If only RMS measurements of the photocenter and fluxes are available, one can still derive an estimate of the pair separation or off-nucleus distance via Eqn.\ (\ref{eqn:rms2}) and Eqn.\ (\ref{eqn:rms3}) for the dual AGN case and the off-nucleus single AGN case, respectively. 

	\item We use pre-main sequence stars that have Gaia-unresolved companions to test the feasibility of {\tt varstrometry} in Gaia DR2. In general, the observed astrometric RMS is consistent with the expectation from {\tt varstrometry} (Fig.~\ref{fig:binary-test}), supporting that we are indeed seeing the variability-induced astrometric RMS that can be constrained by Gaia.
	
	\item We show that extended host galaxies (albeit still treated as point sources in Gaia DR2) with $R_{50}>1$\arcsec\ have extra systematic photometric RMS of $\sim15$\% and astrometric RMS of $\sim10$\,mas in Gaia DR2 (Fig.~\ref{fig:galaxy-r50}). This may be improved for future Gaia data releases with better treatments for extended sources. In Gaia DR2, the effect of extended host galaxies is important for quasars at $z\lesssim0.5$.
	
	\item We investigate the astrometric and photometric RMS properties of spectroscopically-confirmed quasars from SDSS, and present several example candidates that are potentially genuine sub-kpc dual or off-nucleus AGN. There are at least hundreds of good sub-kpc dual/off-nucleus AGN candidates with {\tt astrometric\_excess\_noise}$\gtrsim 2$ mas where the follow-up efforts should be centered. We also explore the properties of WISE-selected photometric quasars and identified a large number of blazar candidates based on Gaia photometric variability (Fig.~\ref{fig:ss-wisequasar}). 
	
	\item We discuss the merits and main limitations of this technique, and our follow-up strategies for promising candidates (\S\ref{sec:disc}). 
	
	
\end{enumerate}



In future work, we will further explore the utility of the combination of {\tt varstrometry} and Gaia data. For example, we have applied this technique to statistically constrain the sub-kpc off-nucleus AGN population at low redshift \citep[][]{vodka2}. We are planning follow-up observations to observe our most promising candidates selected from Gaia that are potentially genuine sub-kpc dual AGN, off-nucleus AGN, or gravitational lenses.  

With future Gaia releases that provide time series of photocenter and light curves, we will be able to refine the selection by computing RMS directly from time series and by cross-correlating photocenter and flux variability. Extended time baselines from Gaia will also improve the detection of photocenter variations (e.g., more frequent large-amplitude AGN variations on longer timescales) and facilitate the cross-correlation analysis between photocenter positions and light curves. Refined treatments for extended sources will also help mitigate the systematics on Gaia astrometric RMS estimation. In the meantime, our follow-up observations will help provide a better understanding of Gaia systematics, and thus to facilitate the application of {\tt varstrometry}. 

\acknowledgments

We gratefully acknowledge the Heising-Simons Foundation and Research Corporation for Science Advancement for their support to this work. YS acknowledges partial support from an Alfred P. Sloan Research Fellowship and NSF grant AST-1715579. XL acknowledges support from the University of Illinois Campus Research Board.



\appendix

Here we provide summary tables, optical images from Pan-STARRS, and SDSS spectra for genuine quasars that have significant non-zero parallaxes or proper motions from Gaia DR2. We summarize quasars with significant parallax detections in Table \ref{tab:para} and those with significant proper motion detections in Table \ref{tab:prop}. The Gaia coordinates in these tables have a precision of $\sim0.04$\,mas. The red dots in the Pan-STARRS images are all Gaia DR2 detections at the J2015.5 epoch. Images have 10\arcsec\ on each side.

\begin{table*}
\caption{List of genuine quasars that have significant non-zero parallaxes from Gaia DR2. Positions are also from Gaia DR2 and have precisions of $\sim0.04$\,mas. Some SDSS quasars have two Gaia matches within 1\arcsec.}
\center
\begin{tabular}{ccccc}
\hline \hline
SDSS name & RA & DEC & parallax & parallax\_over\_error \\
 & deg & deg & mas &  \\
\hline
010212.54$+$014032.3 & 15.55226876 & 1.67566157 & 3.49 & 5.96 \\
024634.09$-$082536.1 & 41.64186178 & -8.42654860 & nan & nan \\
                   & 41.64211573 & -8.42672206 & 1.63 & 6.80 \\
091402.90$+$084121.6 & 138.51208139 & 8.68936409 & 1.61 & 5.05 \\
102806.07$+$502126.1 & 157.02533446 & 50.35723731 & 2.53 & 7.07 \\
114734.05$+$411928.4 & 176.89189707 & 41.32453769 & 7.18 & 5.59 \\
123204.05$+$375855.8 & 188.01691412 & 37.98213972 & 8.15 & 5.06 \\
132128.67$+$541855.5 & 200.36940864 & 54.31544109 & 4.92 & 6.44 \\
                   & 200.36978657 & 54.31539680 & nan & nan \\
144432.35$+$602939.4 & 221.13473179 & 60.49432884 & 7.35 & 7.59 \\
                   & 221.13508711 & 60.49417643 & 0.92 & 0.80 \\
151651.32$+$064105.5 & 229.21388692 & 6.68486792 & 6.25 & 6.15 \\
153320.05$+$442156.8 & 233.33356955 & 44.36581427 & 3.98 & 6.01 \\
165043.44$+$425149.3 & 252.68110564 & 42.86339288 & 0.43 & 0.86 \\
                   & 252.68102328 & 42.86371533 & 0.91 & 5.40 \\
212315.32$+$095050.6 & 320.81378486 & 9.84730542 & nan & nan \\
                   & 320.81394408 & 9.84747621 & 5.29 & 5.30 \\
223412.95$-$011309.6 & 338.55400872 & -1.21933023 & 13.96 & 9.20 \\
223558.48$-$092811.7 & 338.99368620 & -9.46991281 & 4.09 & 5.82 \\
\hline \hline
\end{tabular}\label{tab:para}
\end{table*}

\begin{table*}
\caption{List of genuine quasars that have significant non-zero proper motions from Gaia DR2. Positions are also from Gaia DR2 and have precisions of $\sim0.04$\,mas. Some SDSS quasars have two Gaia matches within 1\arcsec.}
\center
\begin{tabular}{ccccc}
\hline \hline
SDSS name & RA & DEC & total proper motion & pm\_over\_error \\
 & deg & deg & $\mathrm{mas\,yr^{-1}}$ &  \\
\hline 
003250.12$-$105357.6 & 8.20890136 & -10.89935725 & 22.44 & 14.00 \\
                   & 8.20871290 & -10.89932825 &  &  \\
022404.85$+$014941.9 & 36.02016279 & 1.82833372 & 7.51 & 10.55 \\
022723.99$-$010623.4 & 36.85003493 & -1.10656877 & 30.30 & 27.79 \\
074817.13+191003.0 & 117.07139158 & 19.16751662 & 11.92 & 16.53 \\
075824.27+145752.4 & 119.60109938 & 14.96456712 & 12.55 & 11.92 \\
082155.98+340412.9 & 125.48325872 & 34.07025103 & 3.55 & 5.42 \\
                   & 125.48293423 & 34.07023015 & 7.97 & 22.73 \\
083956.37+433248.6 & 129.98492865 & 43.54683719 & 43.52 & 13.65 \\
085122.37+472249.0 & 132.84303687 & 47.38050201 & 9.24 & 13.68 \\
                   & 132.84332839 & 47.38017838 & 0.20 & 0.67 \\
091127.61+055054.1 & 137.86506252 & 5.84856614 &  &  \\
                   & 137.86513579 & 5.84840999 & 10.39 & 10.82 \\
121044.84+452730.2 & 182.68691512 & 45.45837117 & 11.43 & 13.16 \\
                   & 182.68675057 & 45.45839074 &  &  \\
133243.66+343300.6 & 203.18173257 & 34.55023612 & 0.32 & 0.52 \\
                   & 203.18200803 & 34.55007318 & 22.71 & 28.02 \\
141803.88+183532.4 & 214.51614365 & 18.59229451 & 13.47 & 16.04 \\
150324.78+475829.7 & 225.85327225 & 47.97495113 & 3.05 & 14.40 \\
151020.42+153115.1 & 227.58511015 & 15.52087067 & 8.48 & 20.85 \\
151623.88+310336.2 & 229.09947298 & 31.06009574 & 0.61 & 1.51 \\
                   & 229.09978114 & 31.06007692 & 33.90 & 19.52 \\
153646.86+383035.6 & 234.19514960 & 38.50974338 & 9.03 & 18.29 \\
                   & 234.19529623 & 38.50999595 & 7.75 & 7.58 \\
154817.88+044101.4 & 237.07436189 & 4.68365592 &  &  \\
                   & 237.07462435 & 4.68377879 & 18.97 & 18.62 \\
155259.18+230104.9 & 238.24675652 & 23.01803655 & 7.51 & 20.85 \\
                   & 238.24653013 & 23.01804974 & 1.03 & 1.95 \\
172308.15+524455.5 & 260.78401847 & 52.74866790 & 9.66 & 47.75 \\
                   & 260.78370514 & 52.74890673 & 0.14 & 0.49 \\
\hline \hline
\end{tabular}\label{tab:prop}
\end{table*}
\begin{figure*}
\includegraphics[align=c,width=0.18\textwidth]{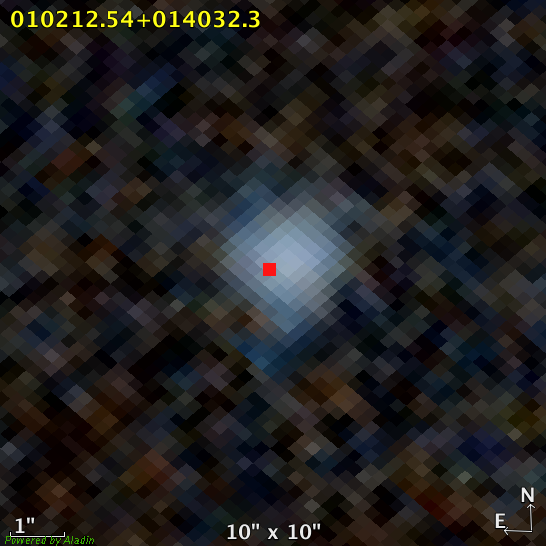}
\includegraphics[align=c,width=0.28\textwidth]{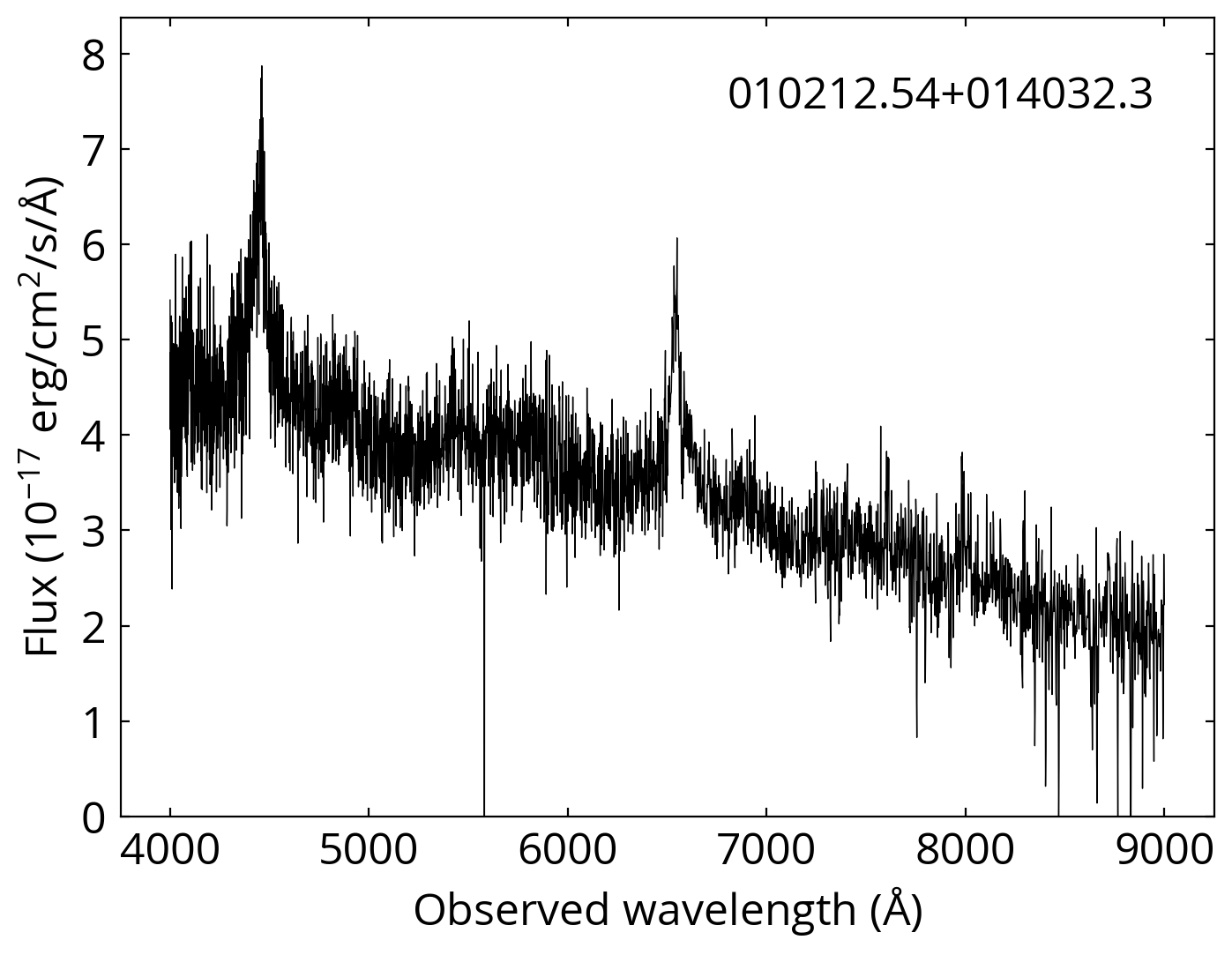}\hspace{0.01\textwidth}
\includegraphics[align=c,width=0.18\textwidth]{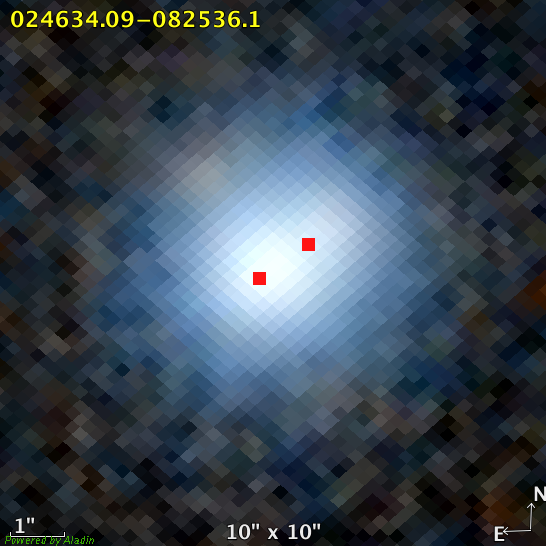}
\includegraphics[align=c,width=0.28\textwidth]{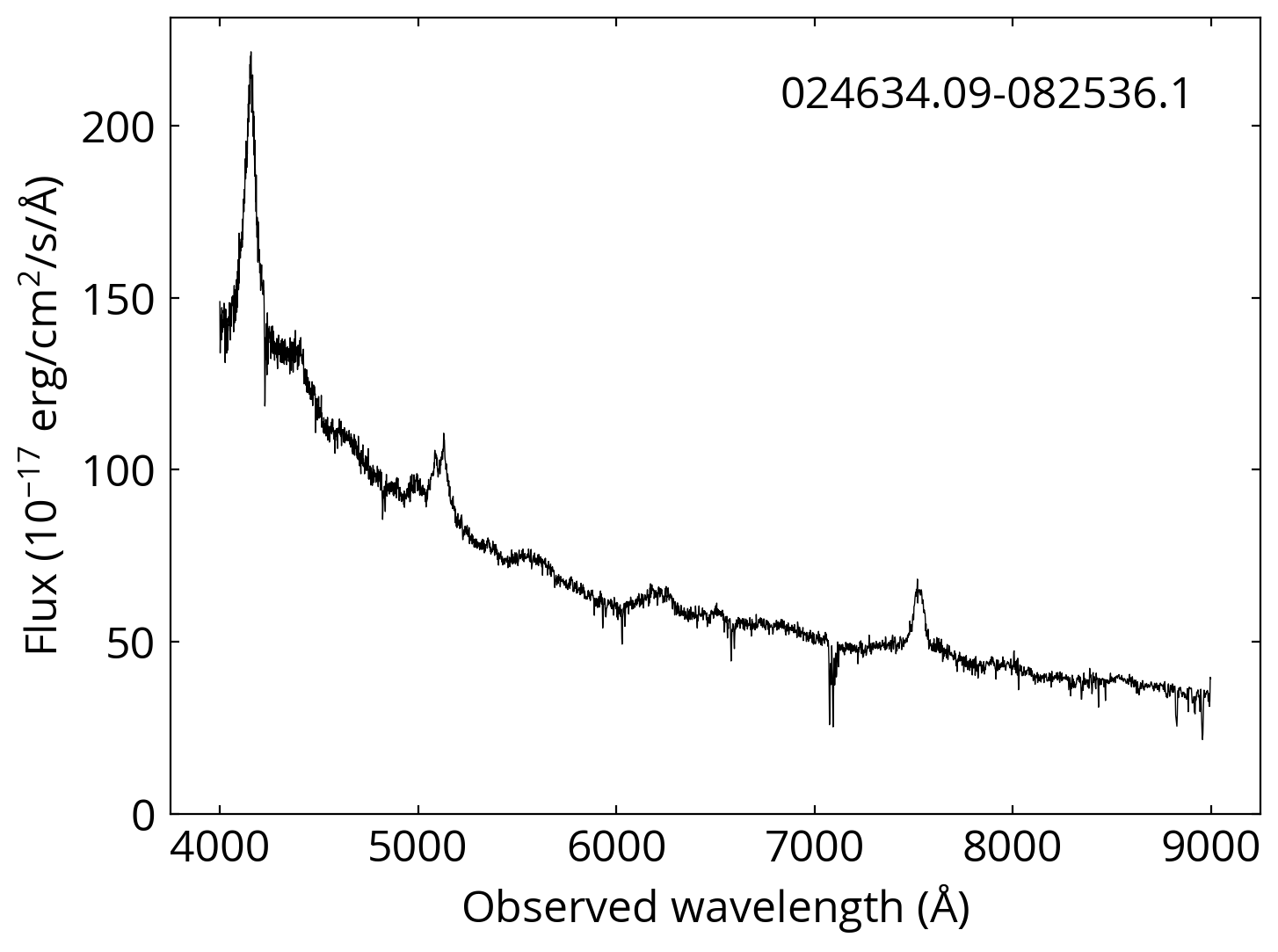}
\includegraphics[align=c,width=0.18\textwidth]{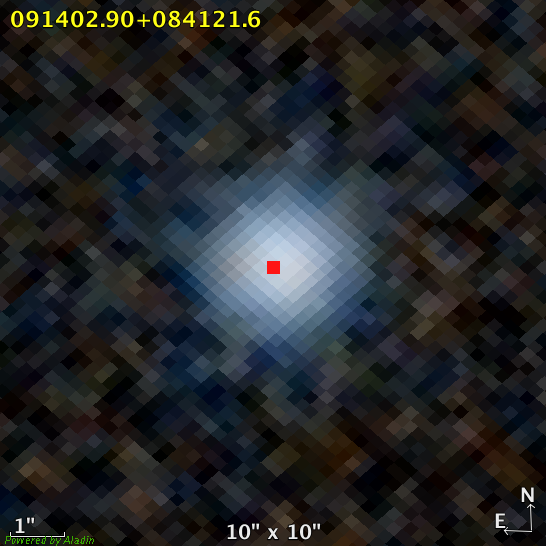}
\includegraphics[align=c,width=0.28\textwidth]{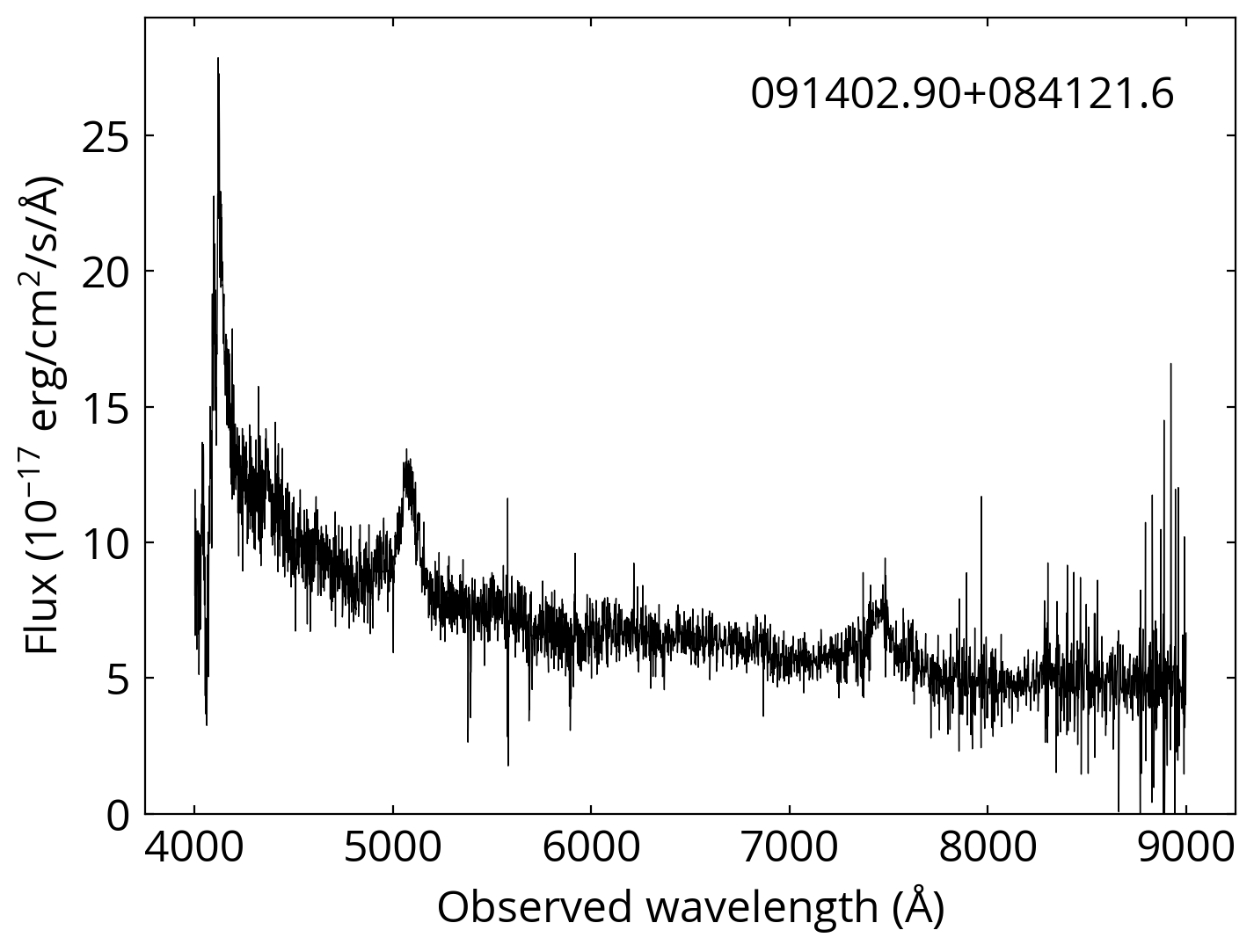}\hspace{0.01\textwidth}
\includegraphics[align=c,width=0.18\textwidth]{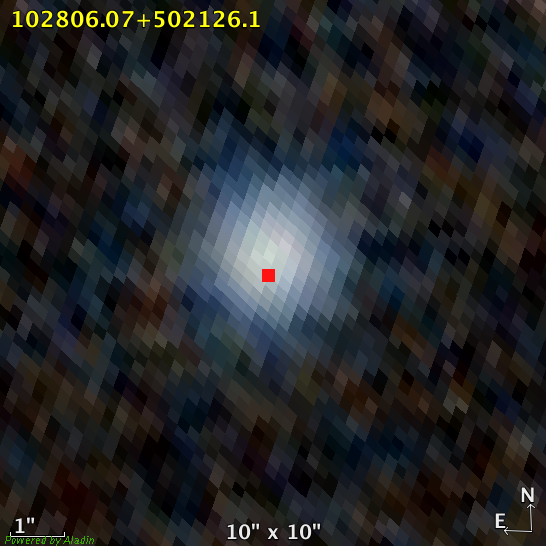}
\includegraphics[align=c,width=0.28\textwidth]{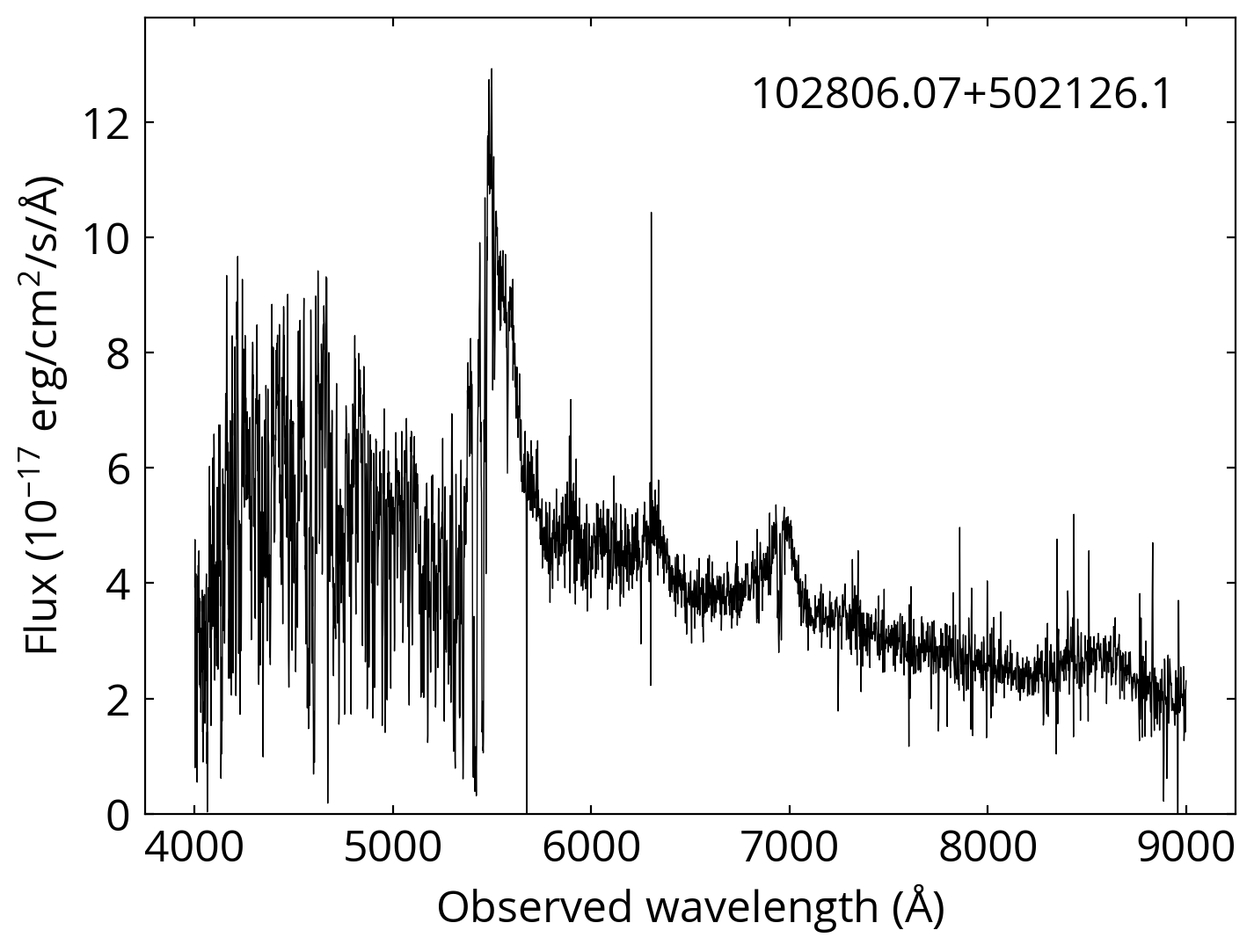}
\includegraphics[align=c,width=0.18\textwidth]{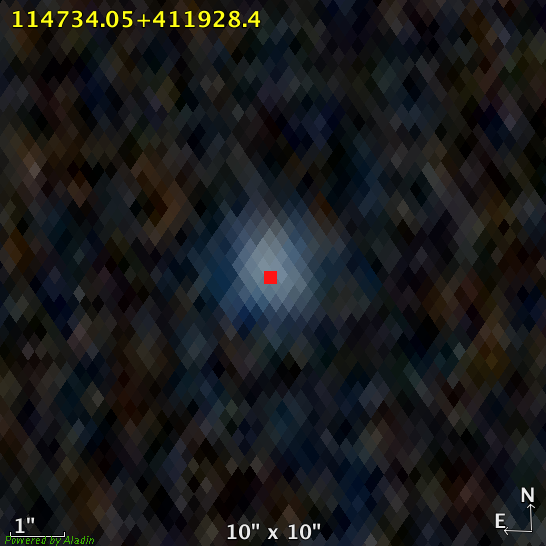}
\includegraphics[align=c,width=0.28\textwidth]{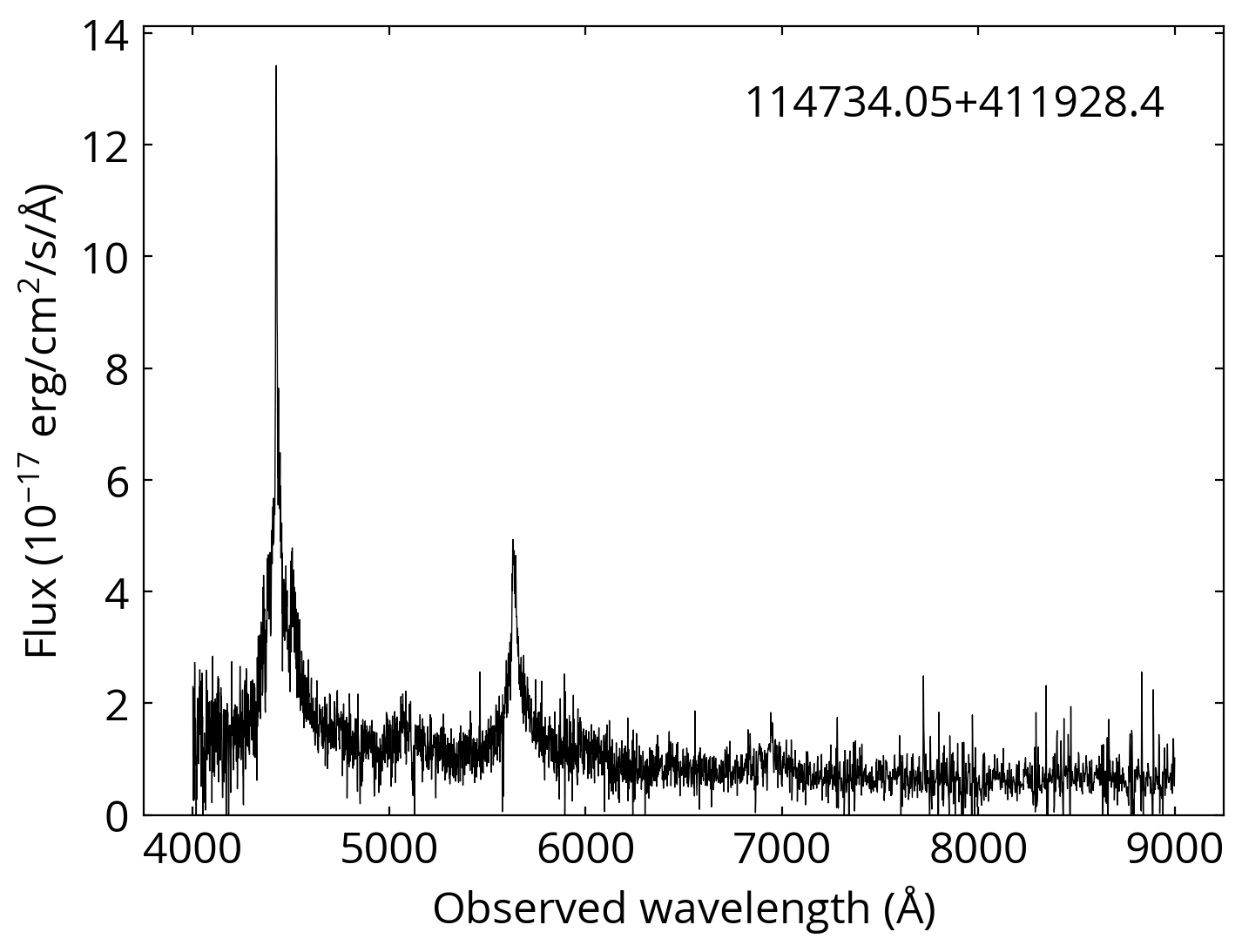}\hspace{0.01\textwidth}
\includegraphics[align=c,width=0.18\textwidth]{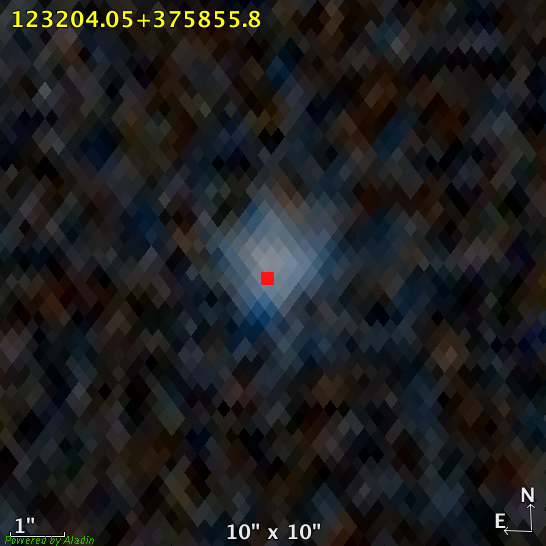}
\includegraphics[align=c,width=0.28\textwidth]{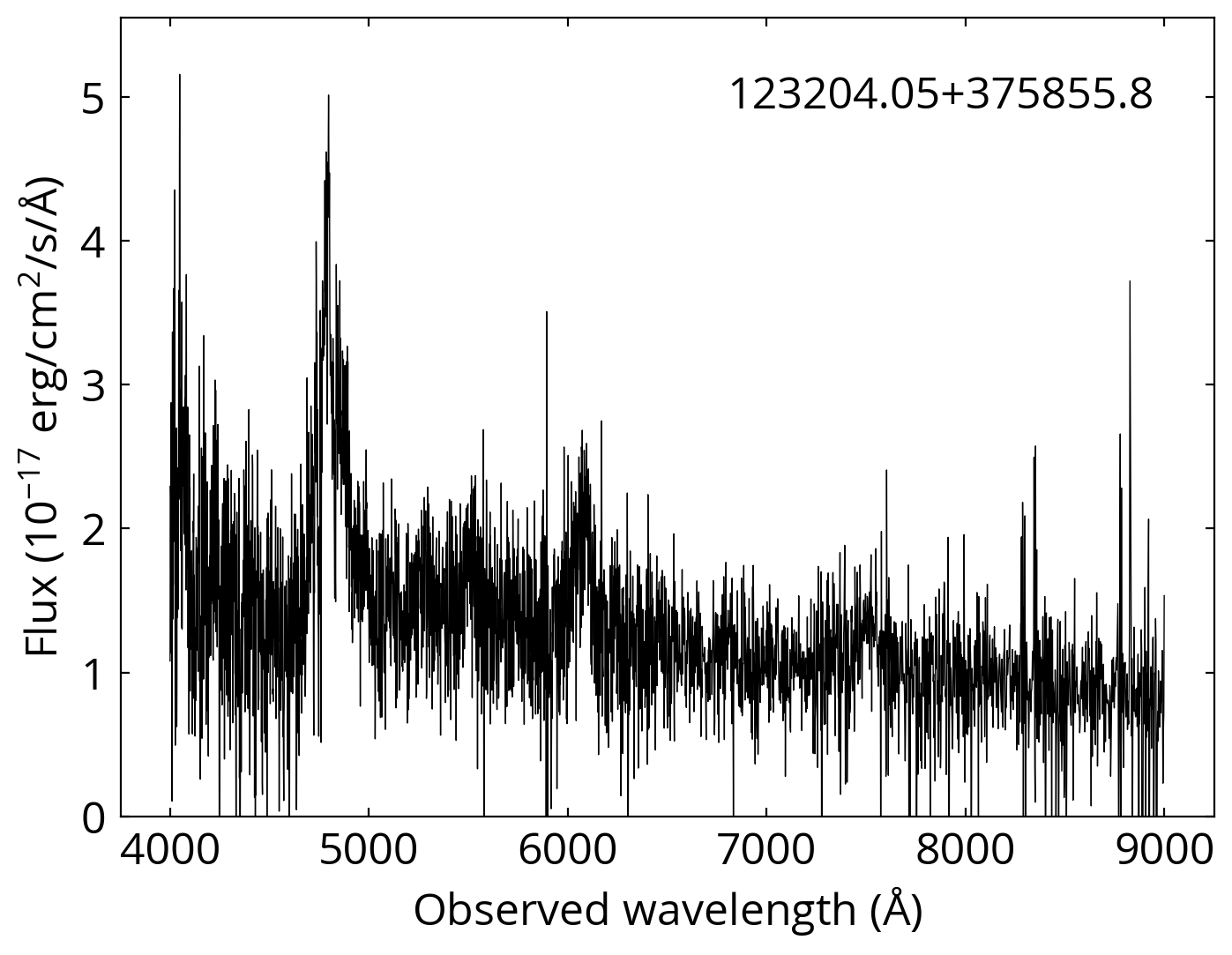}
\includegraphics[align=c,width=0.18\textwidth]{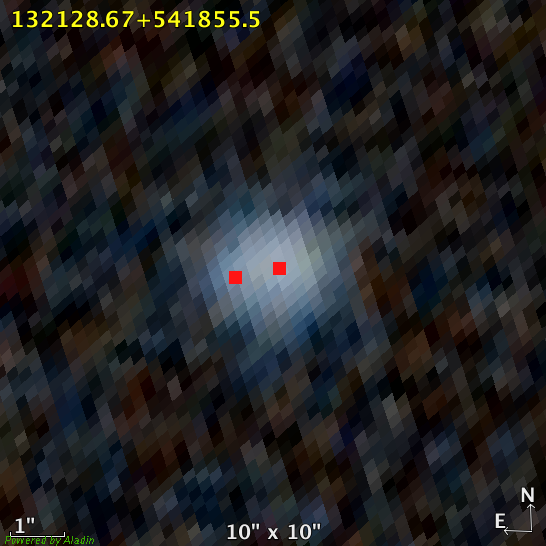}
\includegraphics[align=c,width=0.28\textwidth]{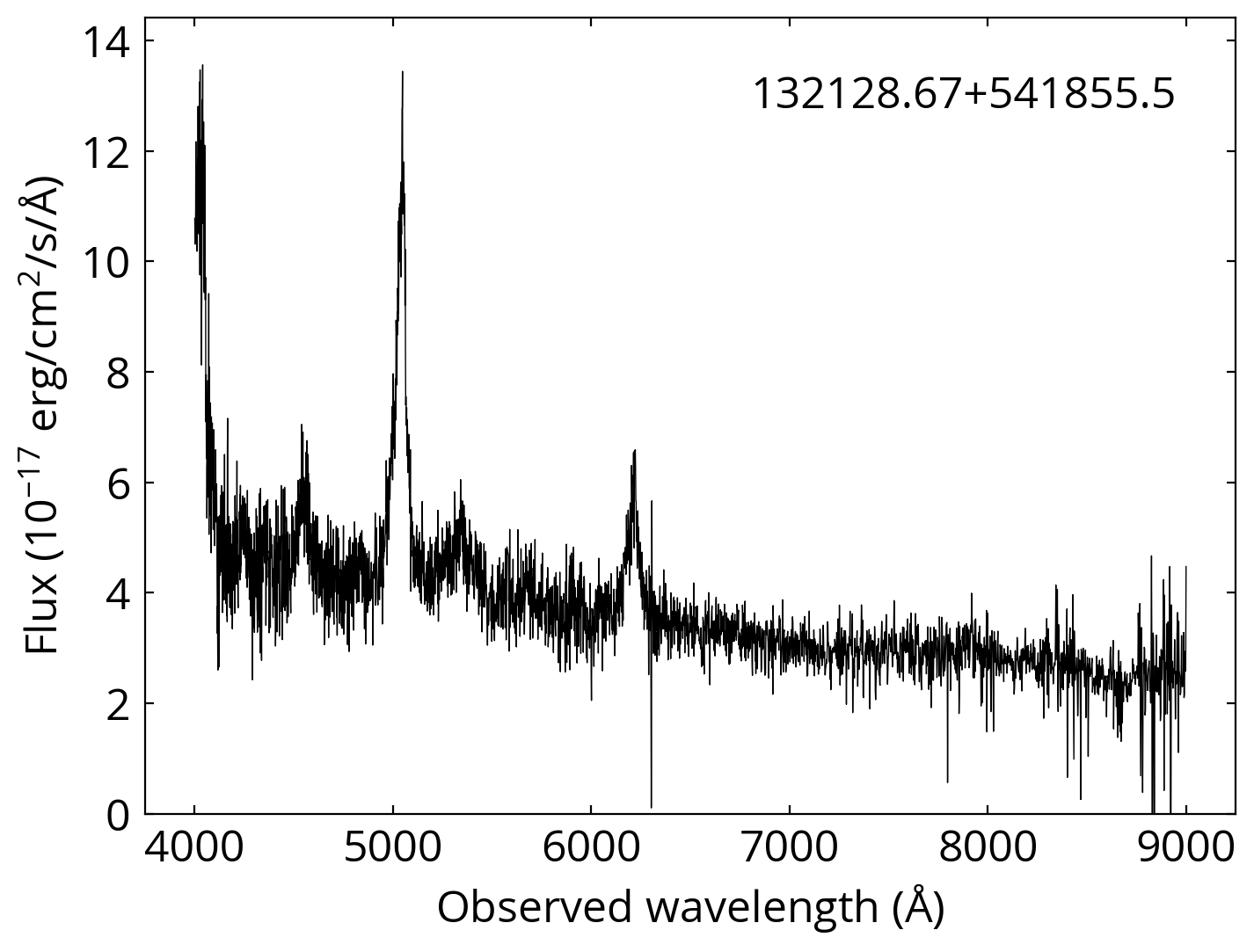}\hspace{0.01\textwidth}
\includegraphics[align=c,width=0.18\textwidth]{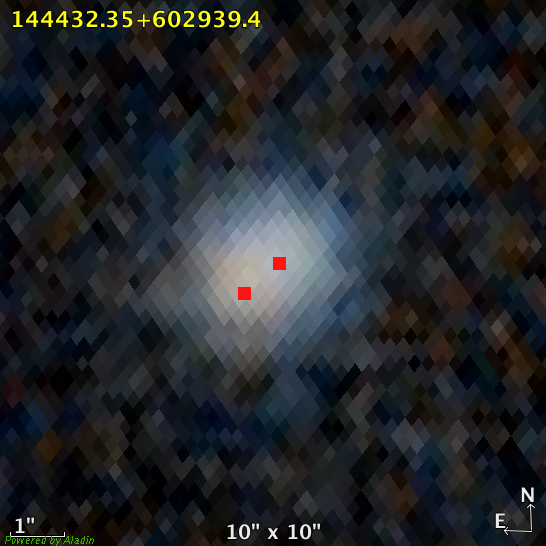}
\includegraphics[align=c,width=0.28\textwidth]{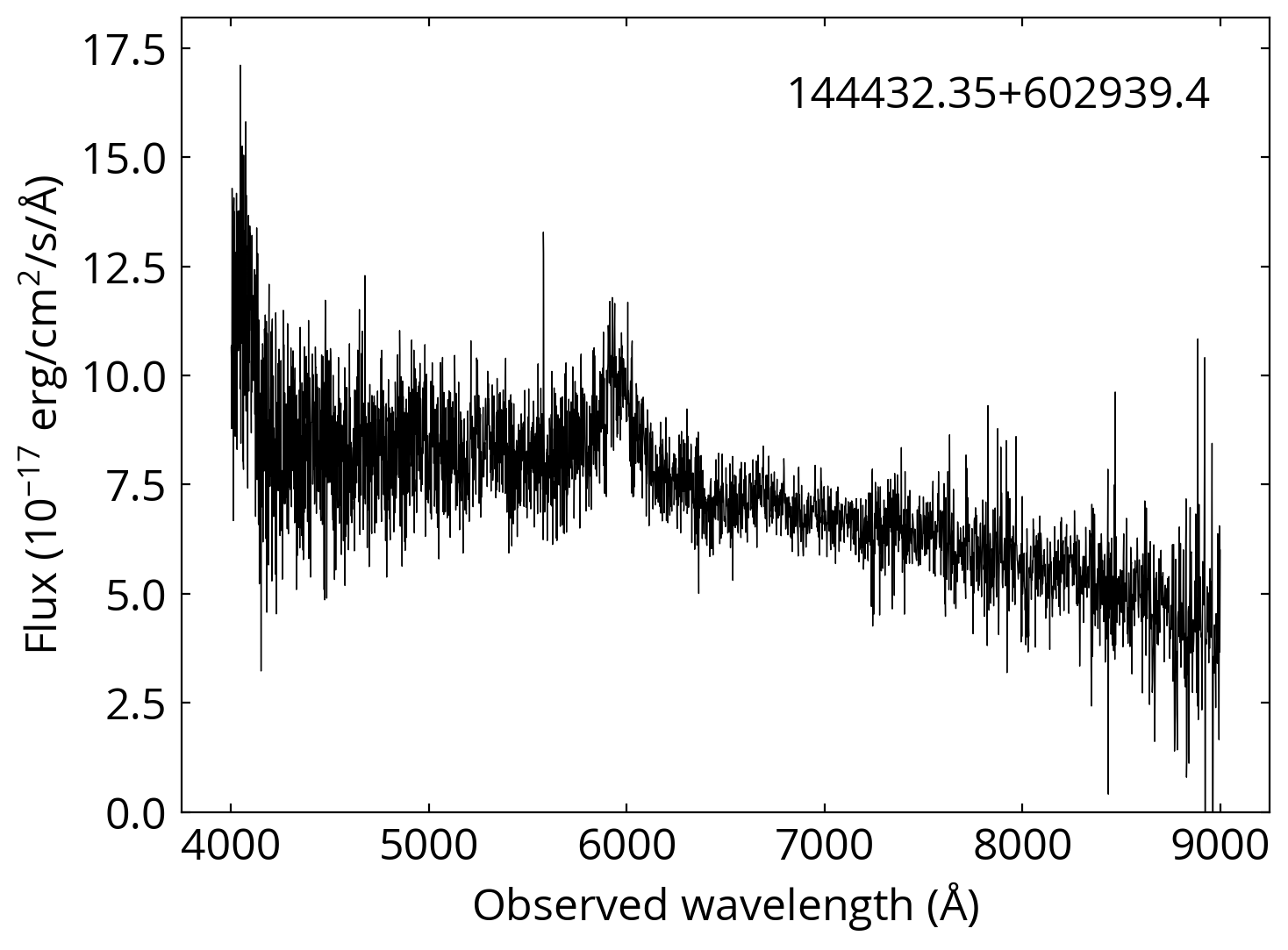}
\includegraphics[align=c,width=0.18\textwidth]{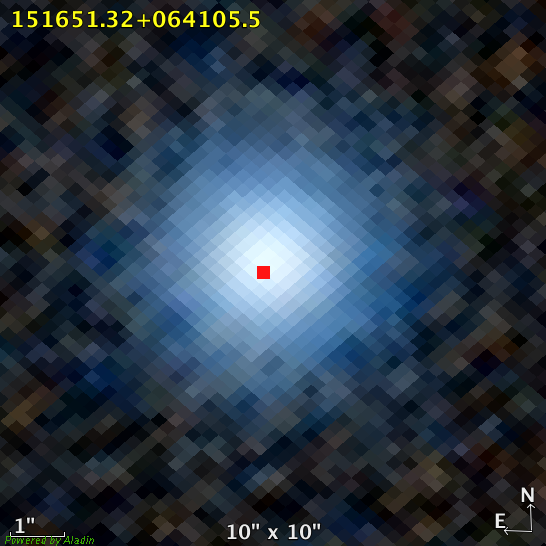}\hspace{0.012\textwidth}
\includegraphics[align=c,width=0.29\textwidth]{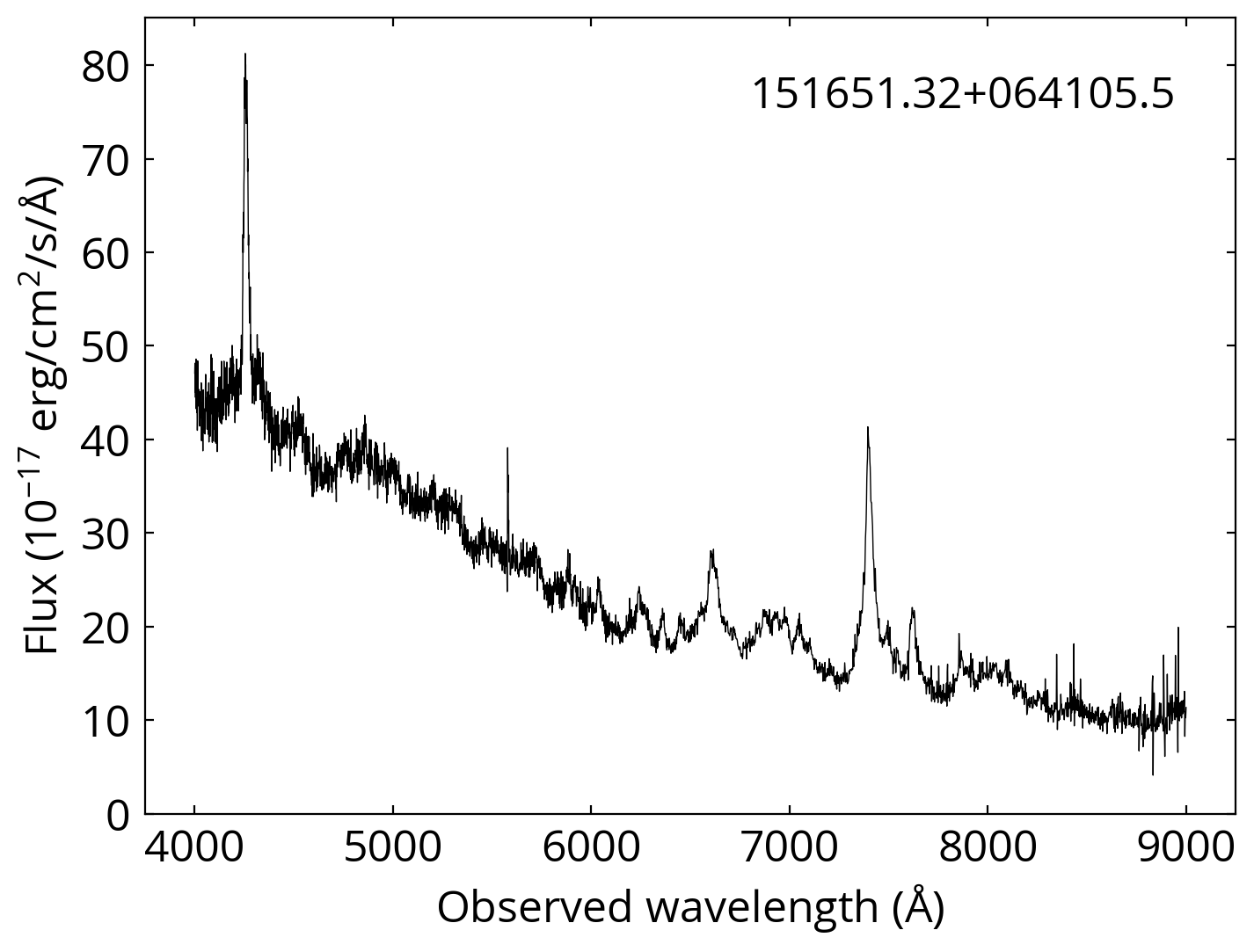}\hspace{0.025\textwidth}
\includegraphics[align=c,width=0.18\textwidth]{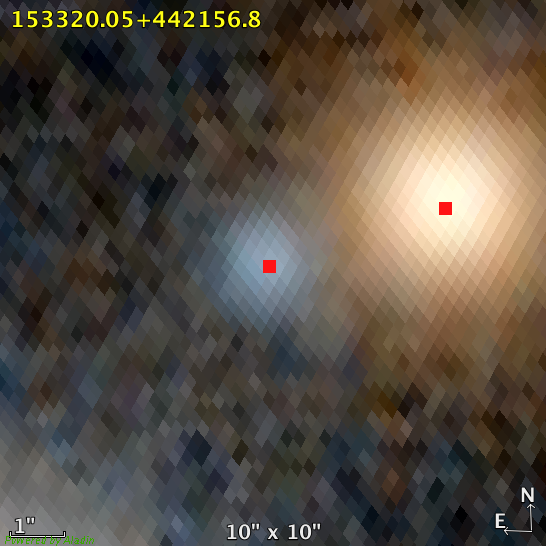}\hspace{0.012\textwidth}
\includegraphics[align=c,width=0.29\textwidth]{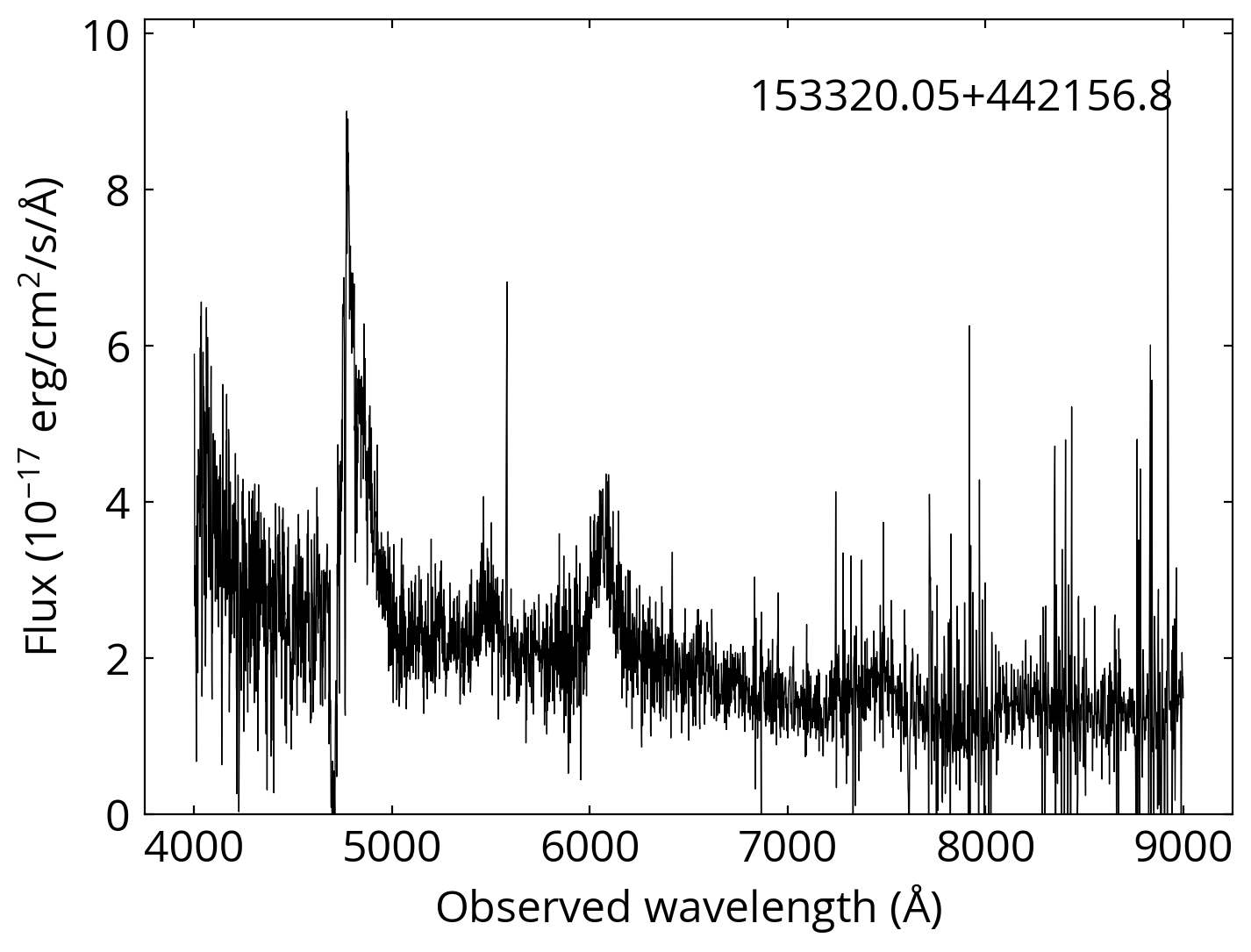}
\caption{Pan-STARRS 1 color-composite images (left) and SDSS spectra (right) for genuine quasars with non-zero parallaxes from Gaia DR2. The red dots in the Pan-STARRS images are the Gaia detections at the J2015.5 epoch. Images have 10\arcsec\ on each side, and north (east) is up (left). }\label{fig:app1}
\end{figure*}

\begin{figure*}
\includegraphics[align=c,width=0.18\textwidth]{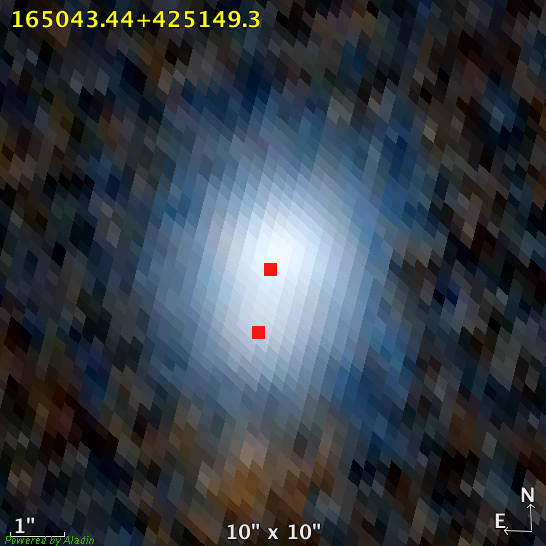}
\includegraphics[align=c,width=0.28\textwidth]{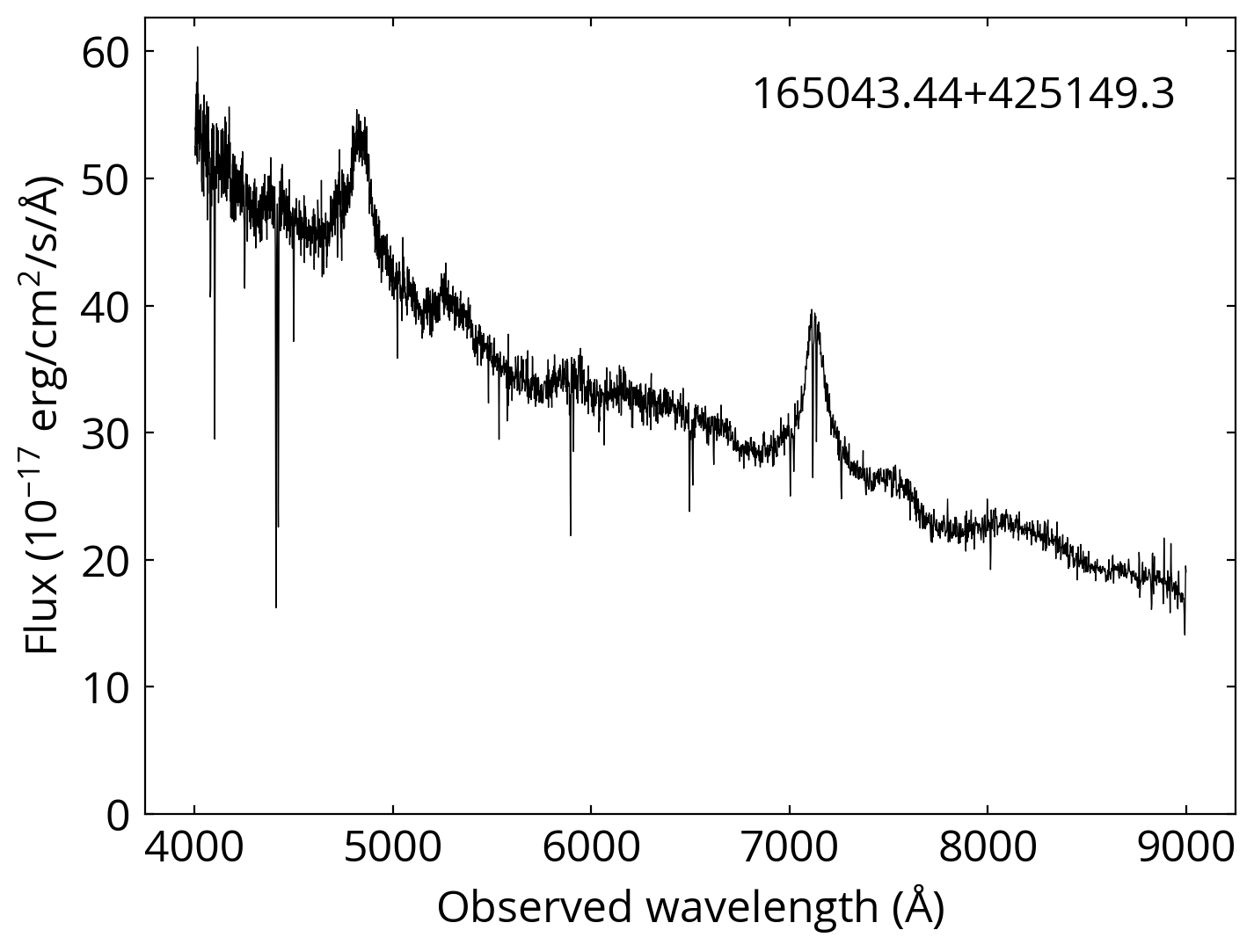}\hspace{0.01\textwidth}
\includegraphics[align=c,width=0.18\textwidth]{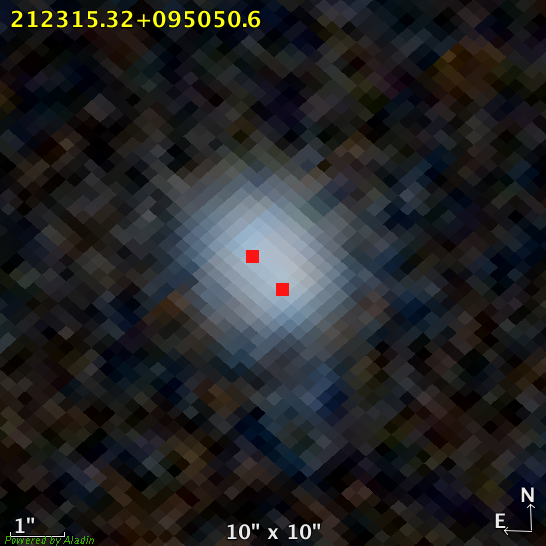}
\includegraphics[align=c,width=0.28\textwidth]{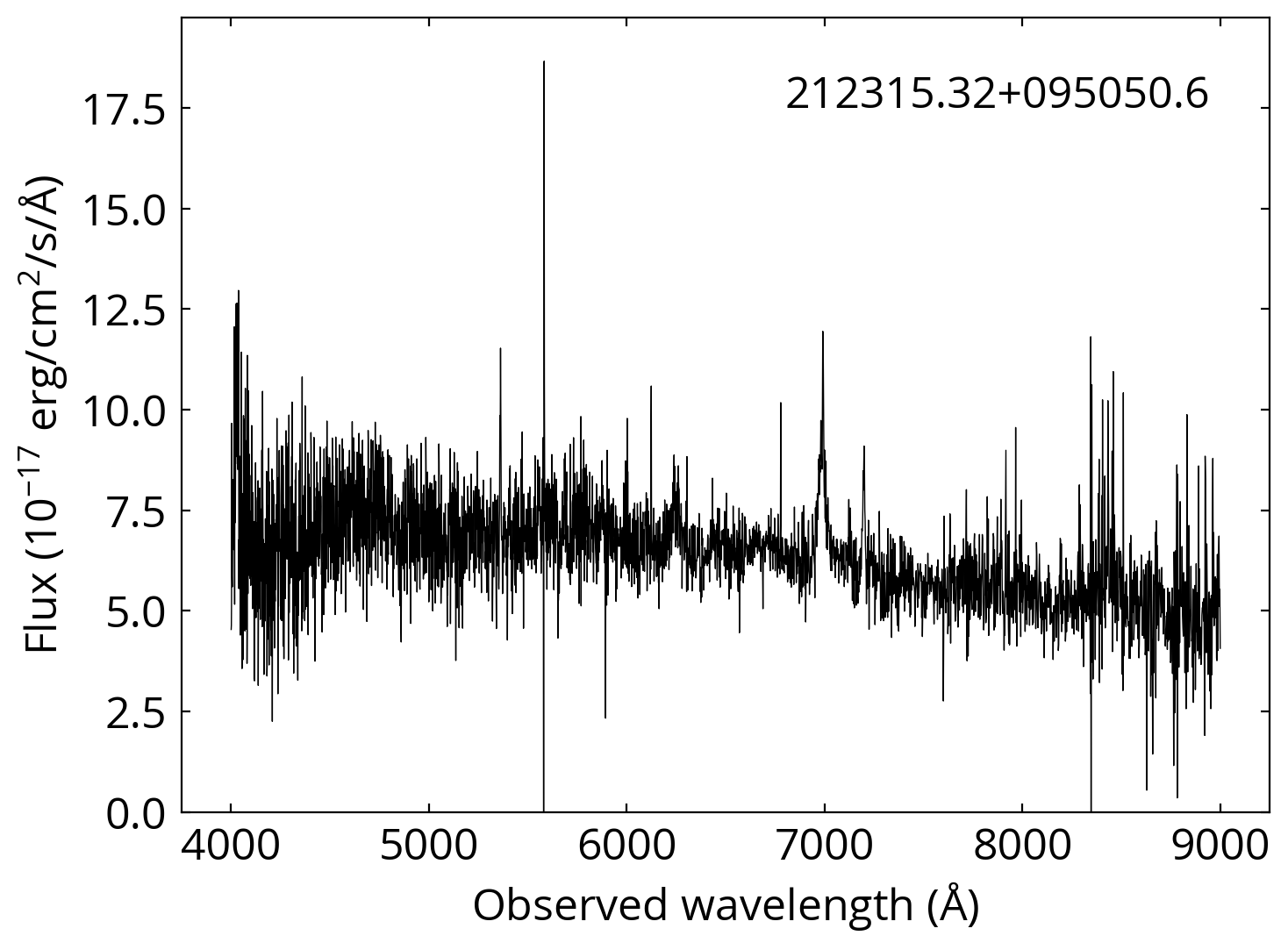}
\includegraphics[align=c,width=0.18\textwidth]{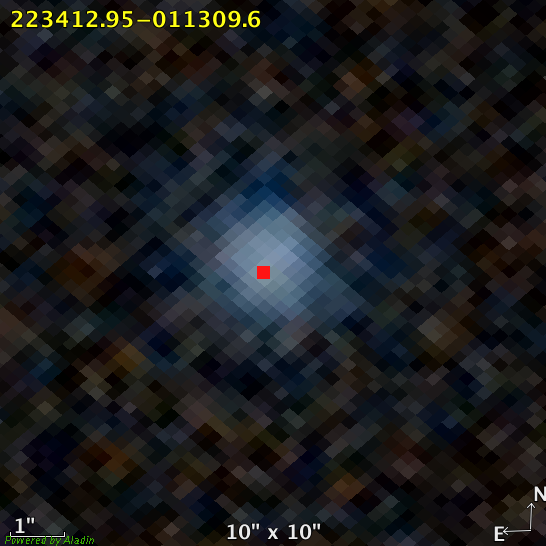}\hspace{0.012\textwidth}
\includegraphics[align=c,width=0.28\textwidth]{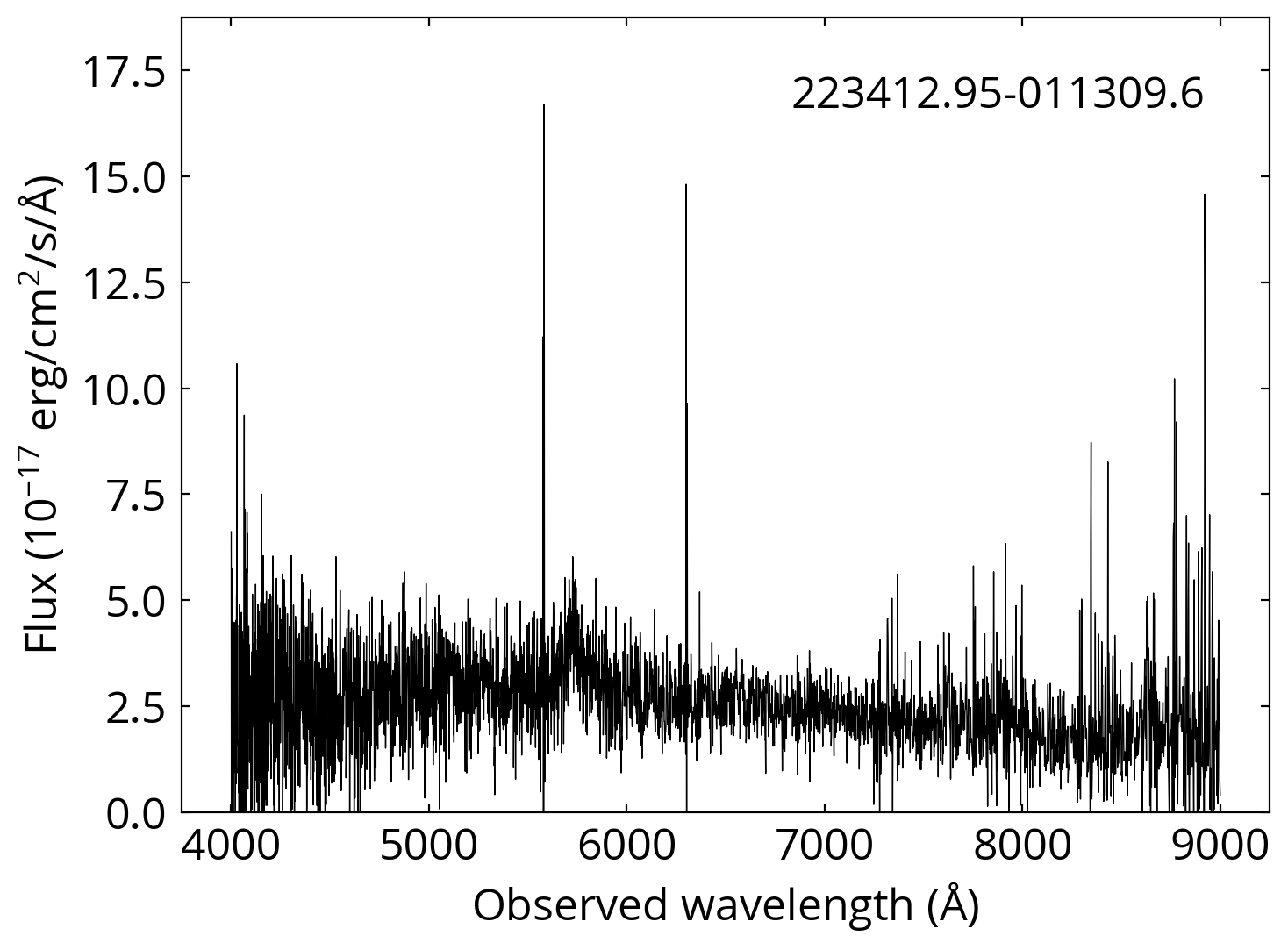}\hspace{0.03\textwidth}
\includegraphics[align=c,width=0.18\textwidth]{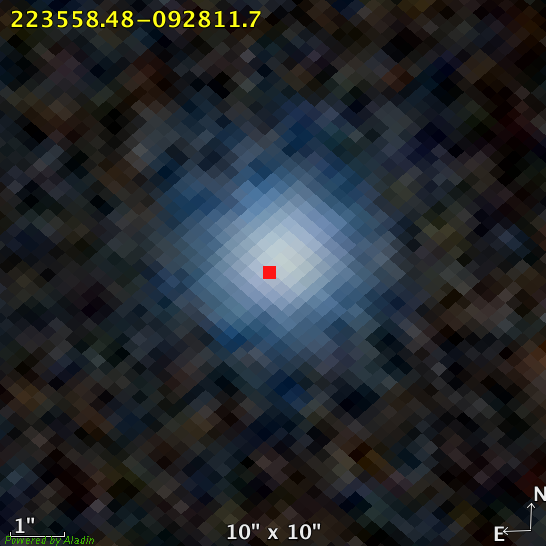}\hspace{0.012\textwidth}
\includegraphics[align=c,width=0.28\textwidth]{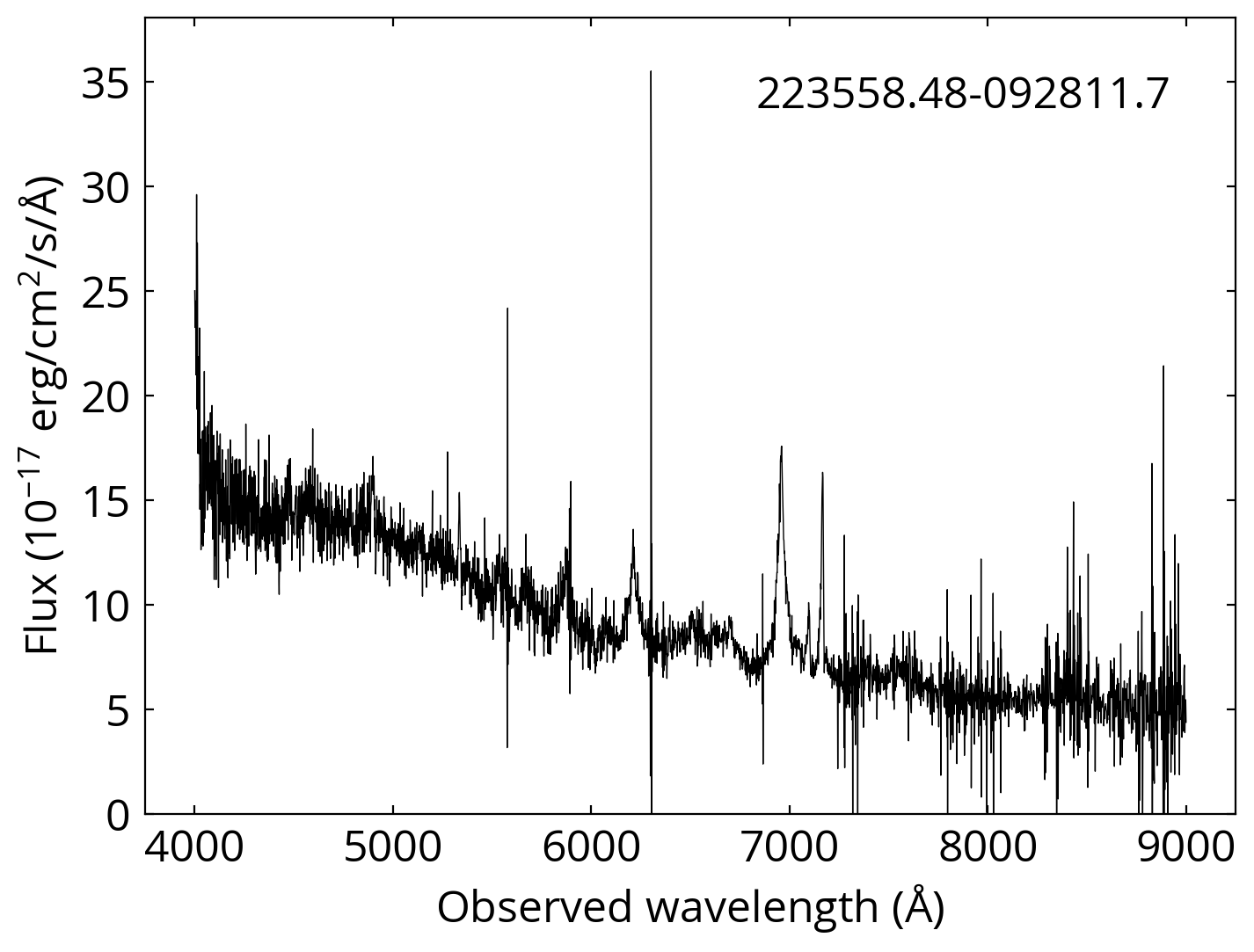}
\caption{Same as Fig.~\ref{fig:app1}, but for a different set of objects. }
\end{figure*}

\begin{figure*}
\includegraphics[align=c,width=0.18\textwidth]{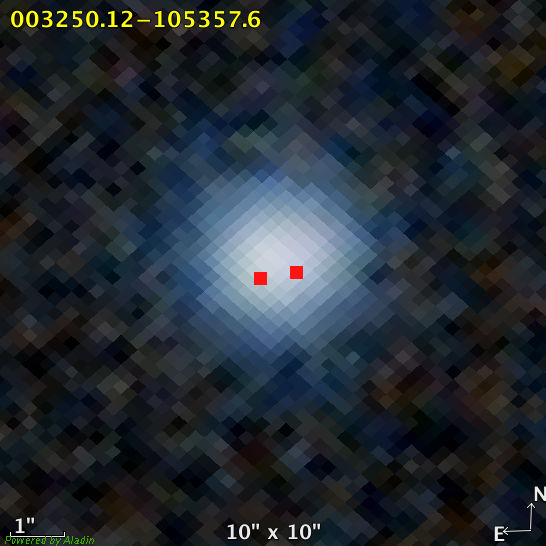}
\includegraphics[align=c,width=0.28\textwidth]{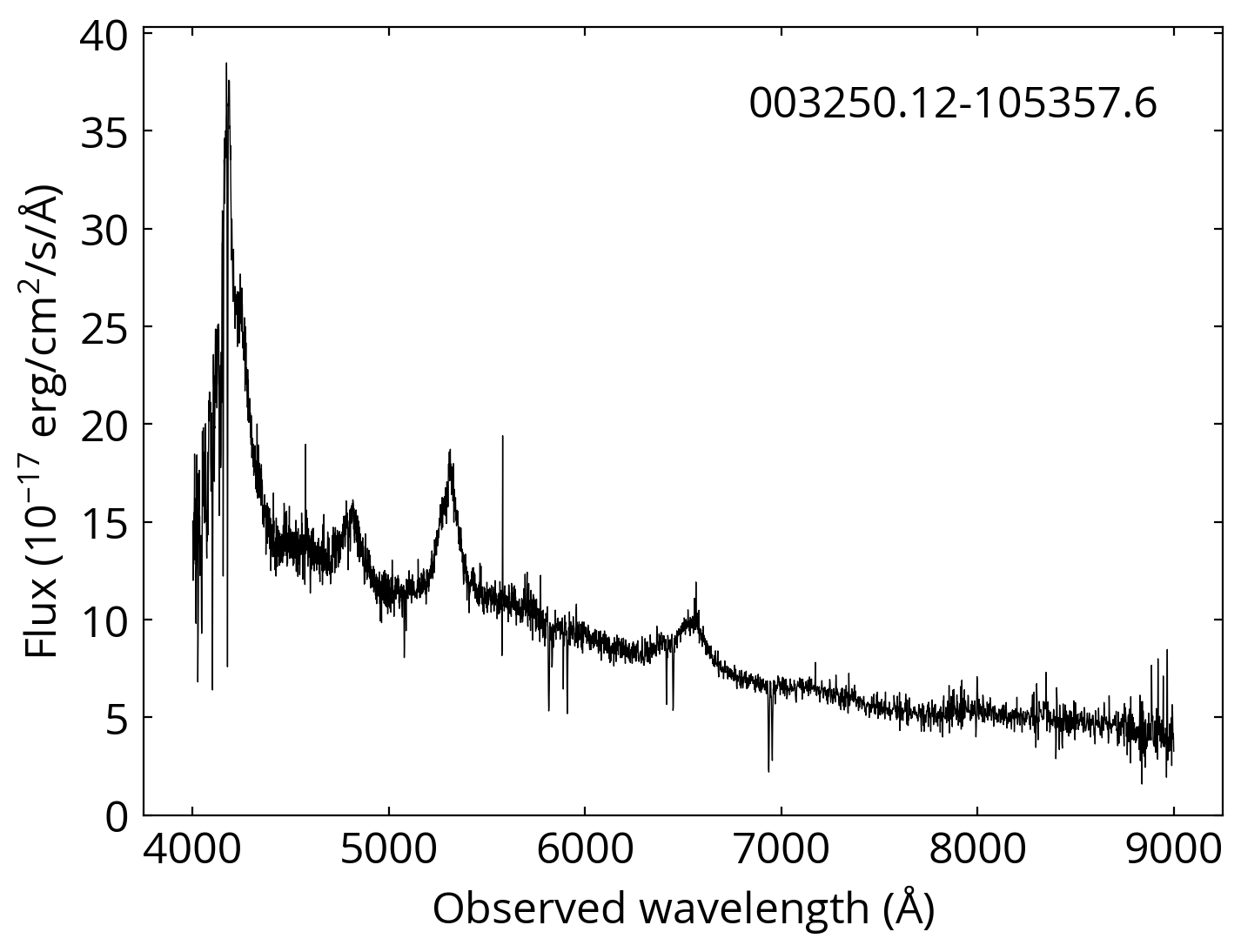}\hspace{0.01\textwidth}
\includegraphics[align=c,width=0.18\textwidth]{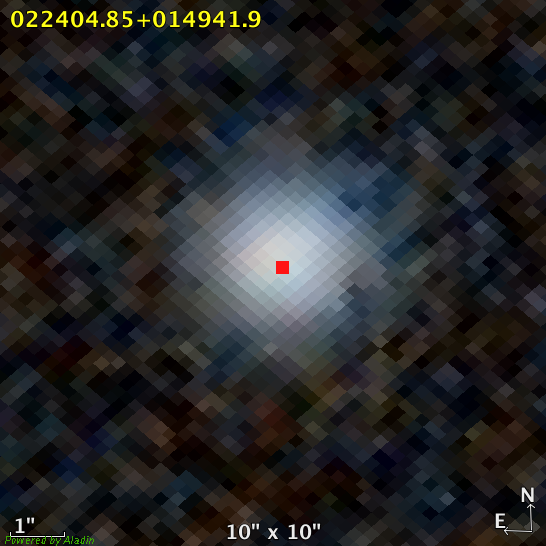}
\includegraphics[align=c,width=0.28\textwidth]{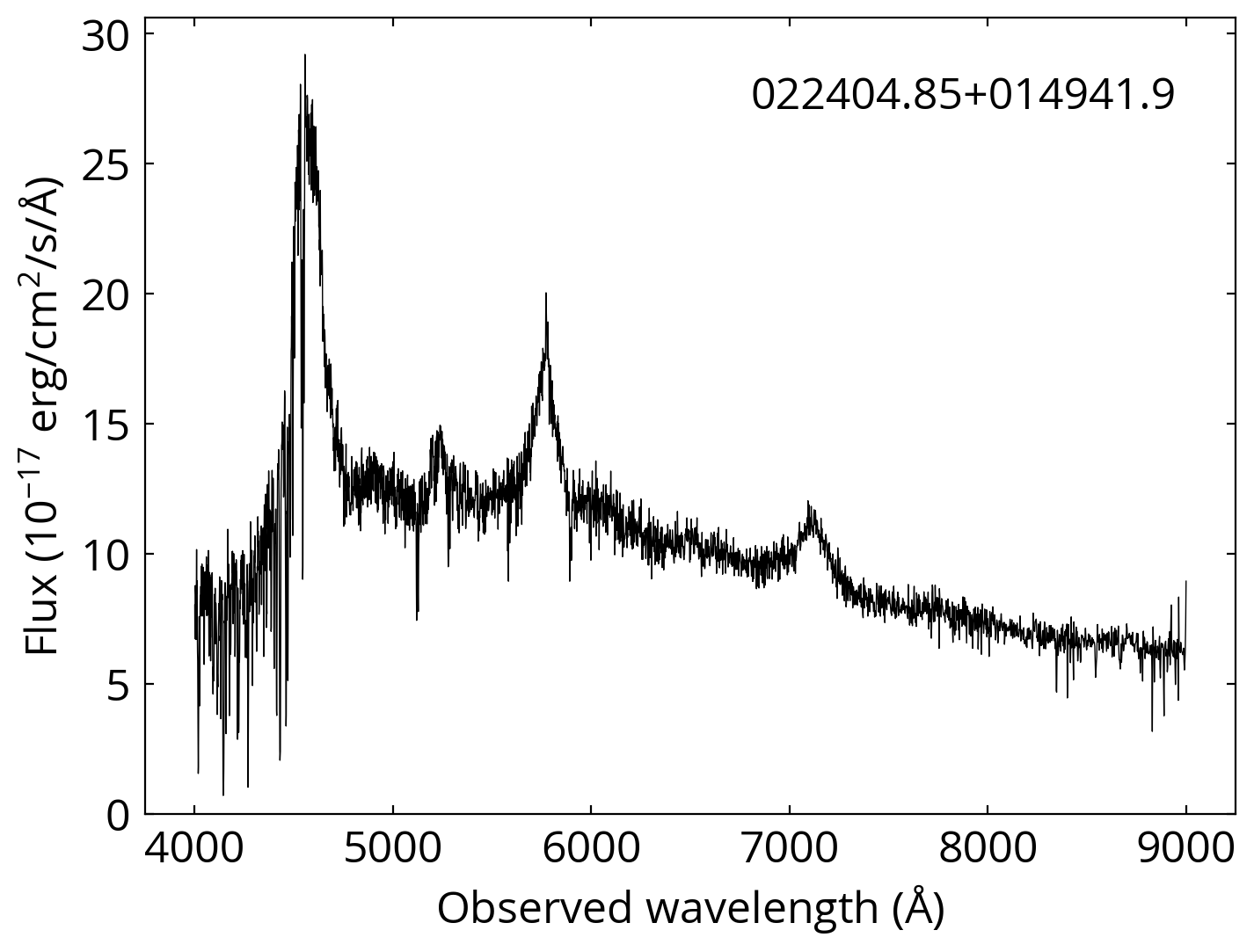}
\includegraphics[align=c,width=0.18\textwidth]{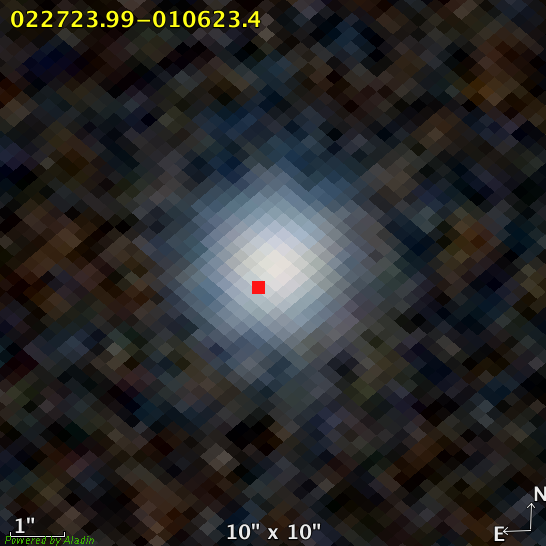}
\includegraphics[align=c,width=0.28\textwidth]{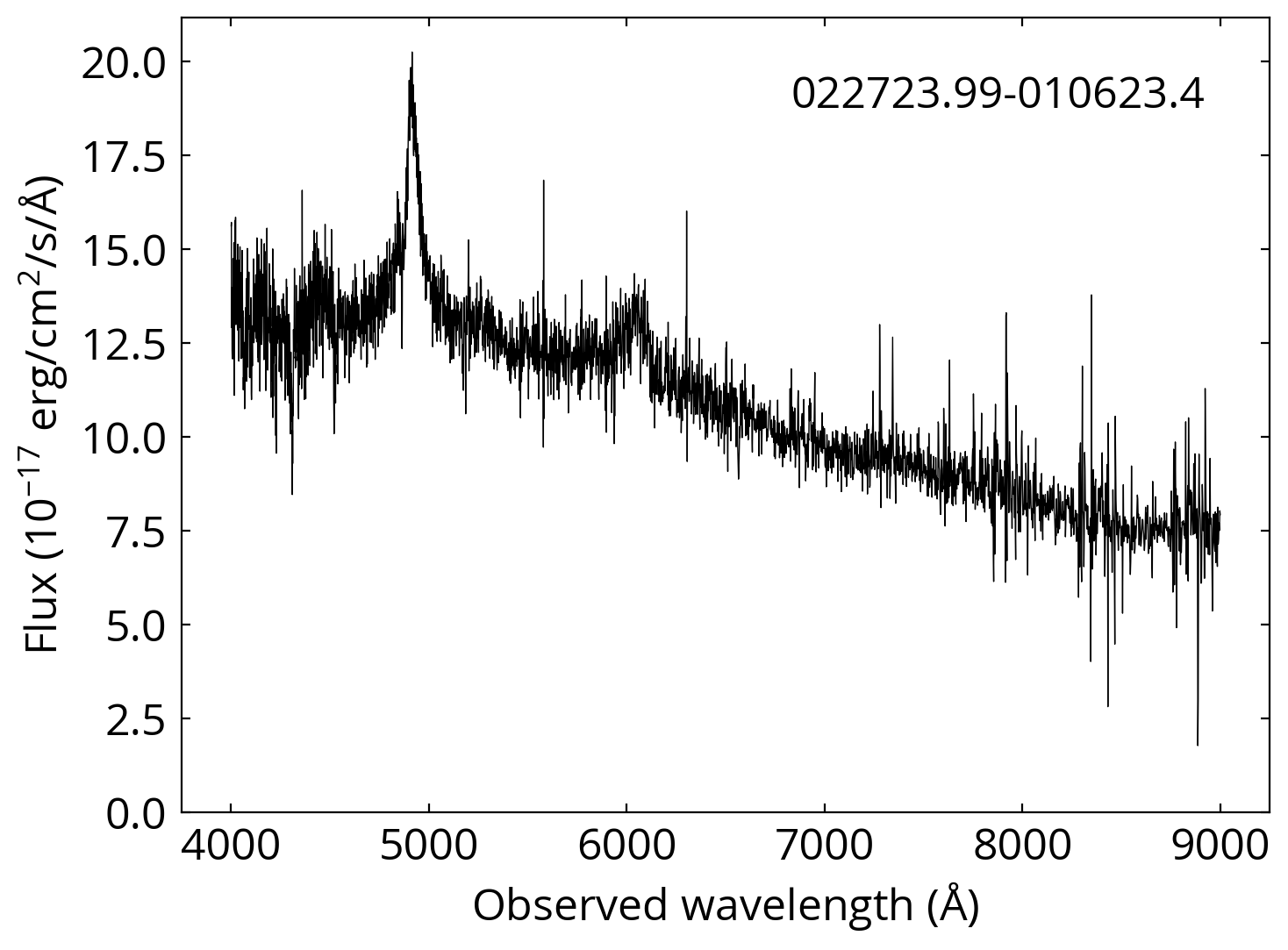}\hspace{0.01\textwidth}
\includegraphics[align=c,width=0.18\textwidth]{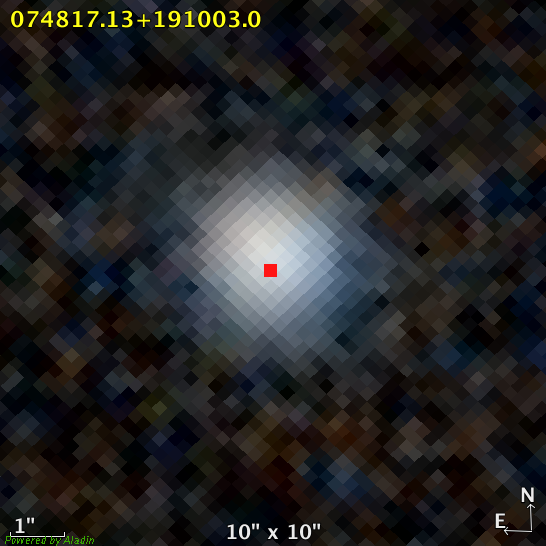}
\includegraphics[align=c,width=0.28\textwidth]{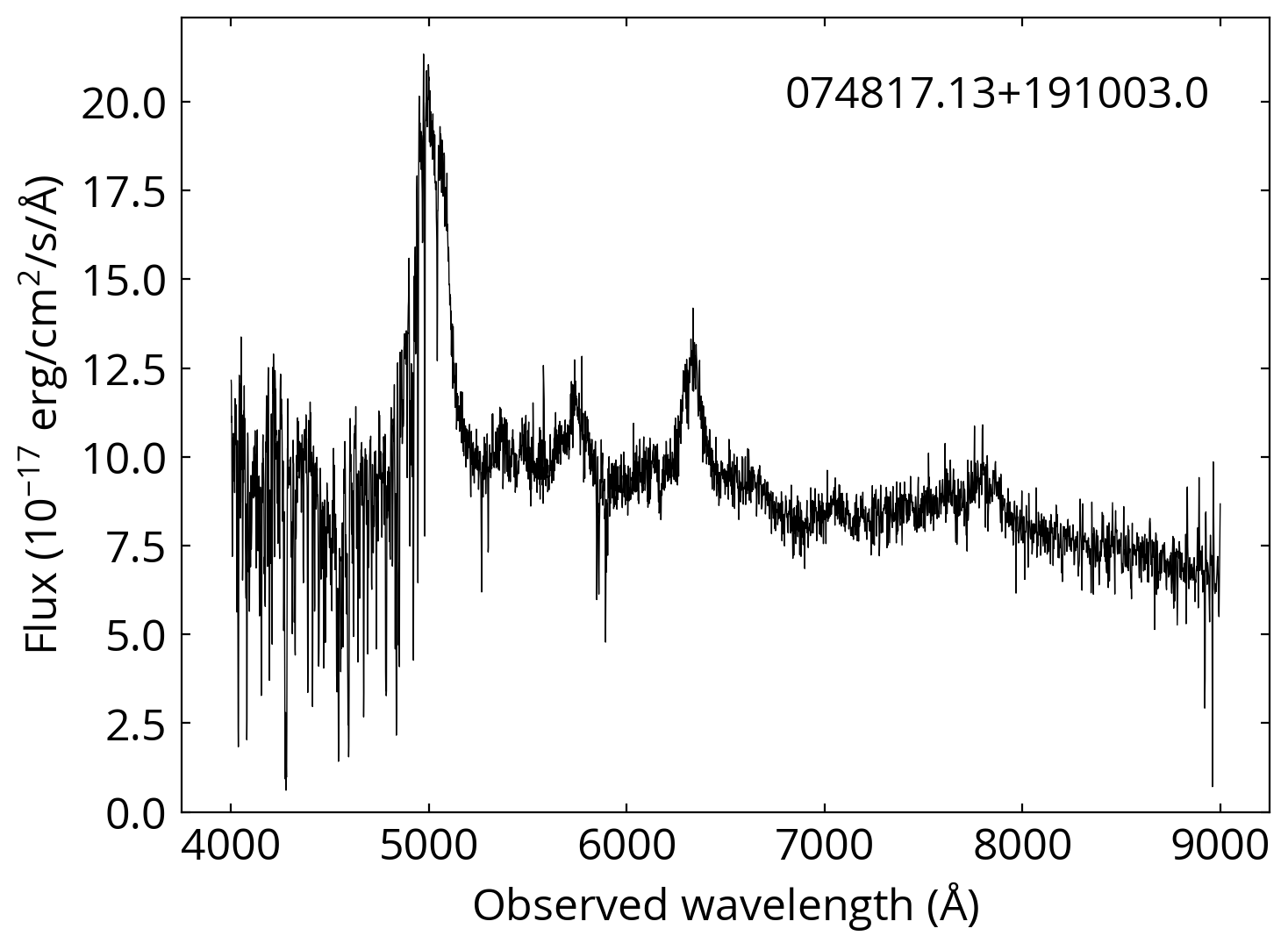}
\includegraphics[align=c,width=0.18\textwidth]{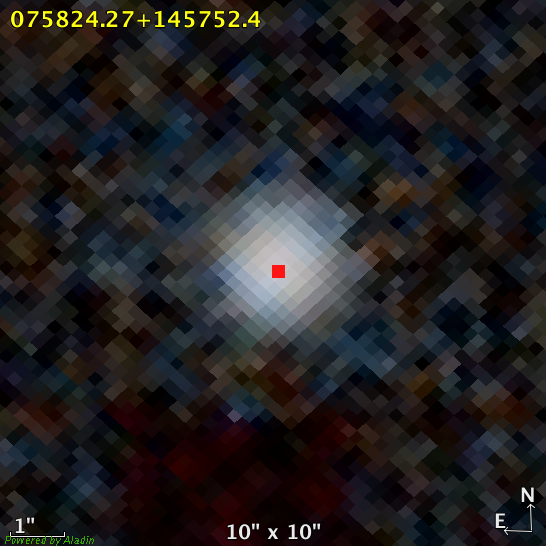}
\includegraphics[align=c,width=0.28\textwidth]{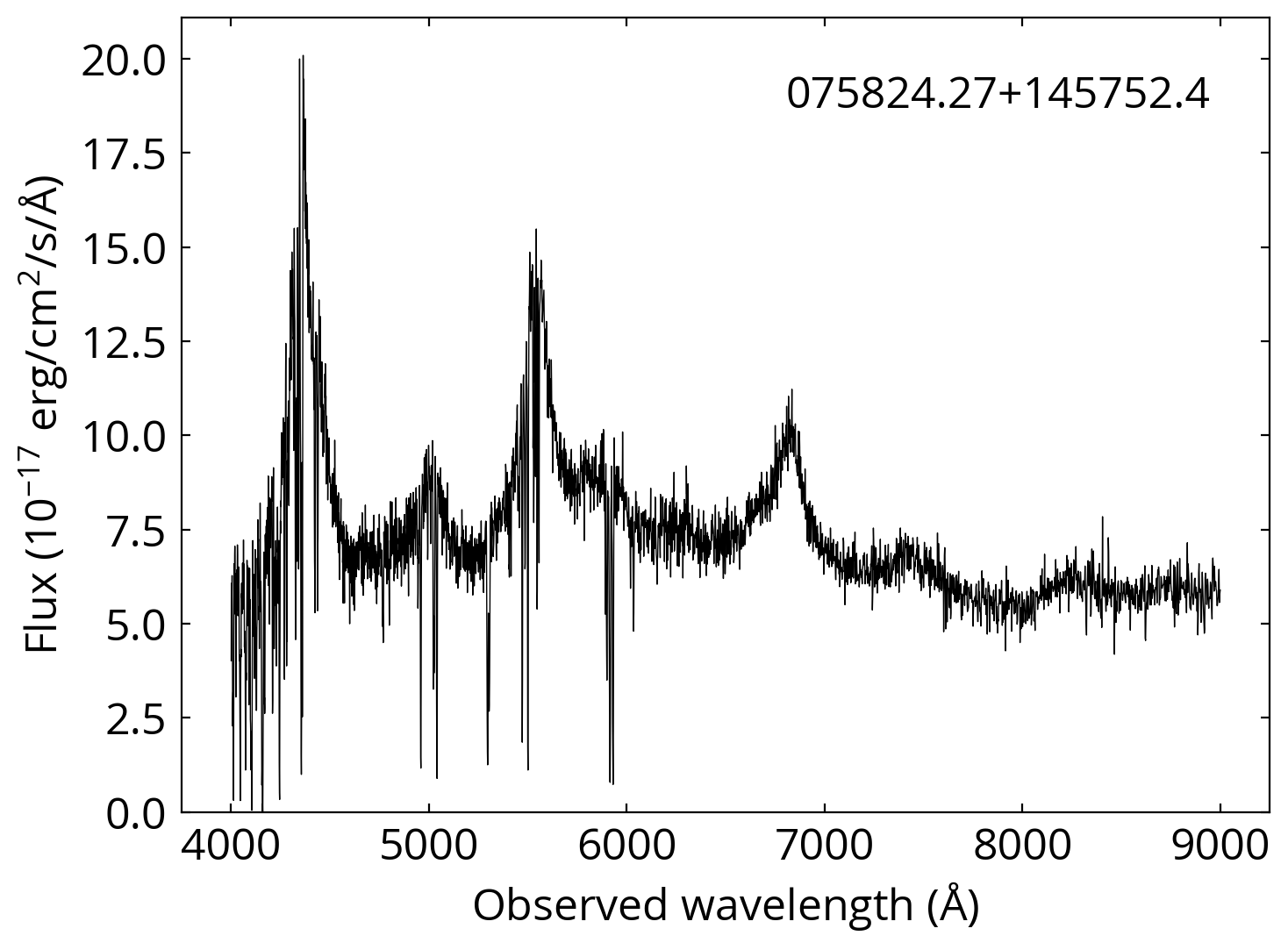}\hspace{0.01\textwidth}
\includegraphics[align=c,width=0.18\textwidth]{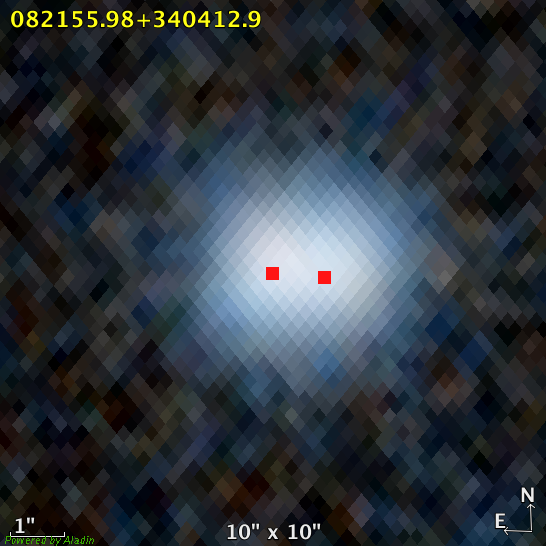}
\includegraphics[align=c,width=0.28\textwidth]{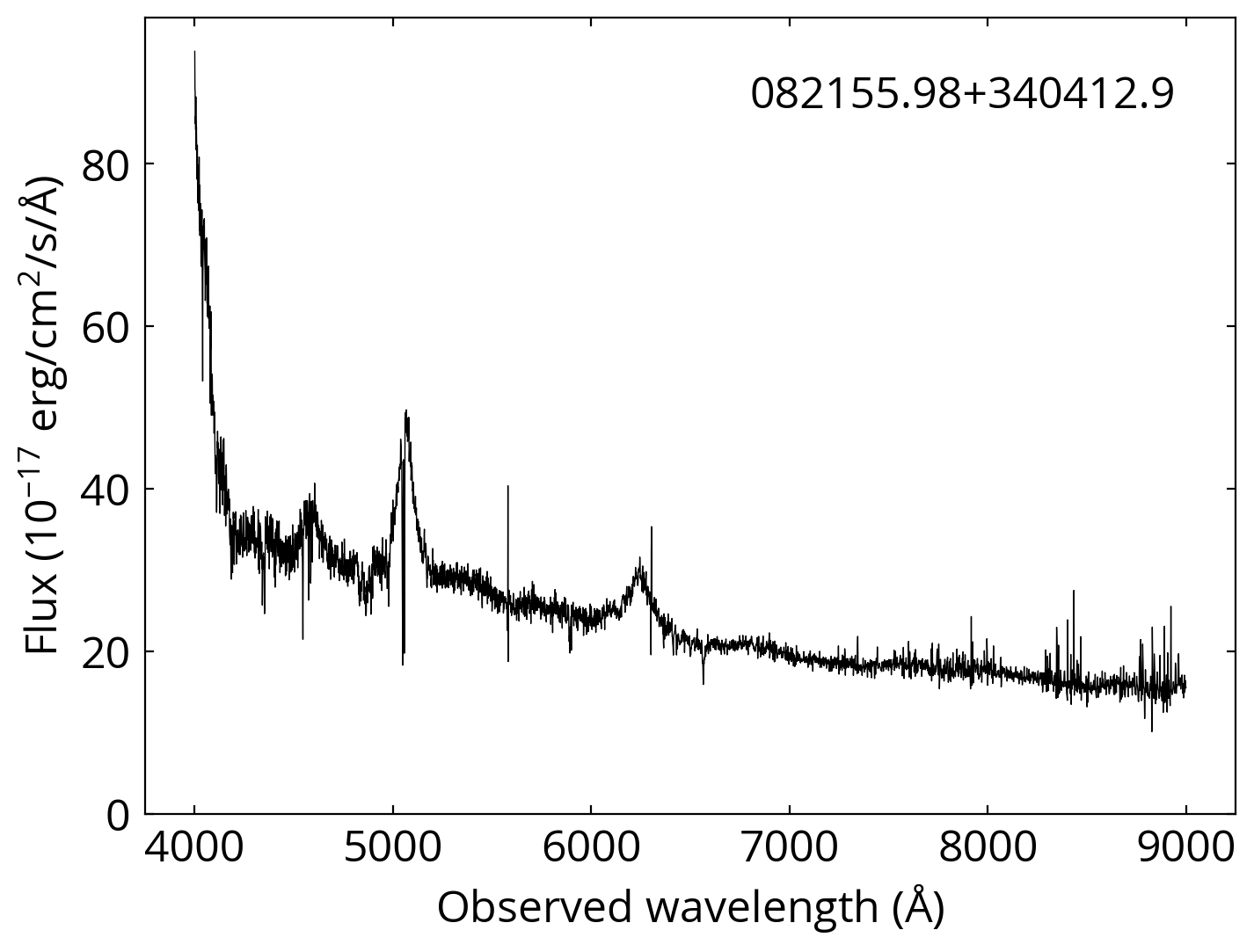}
\includegraphics[align=c,width=0.18\textwidth]{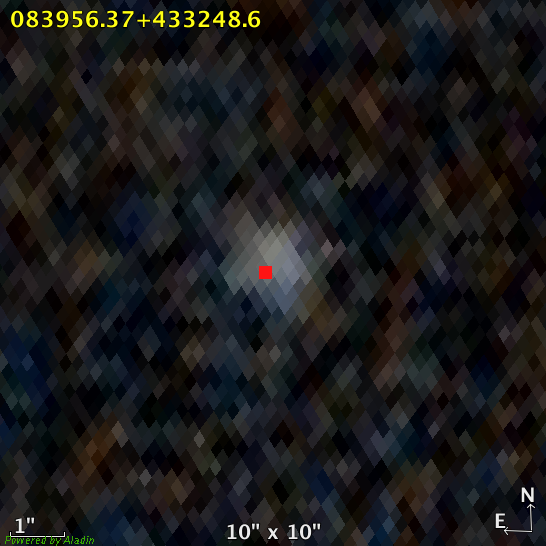}
\includegraphics[align=c,width=0.28\textwidth]{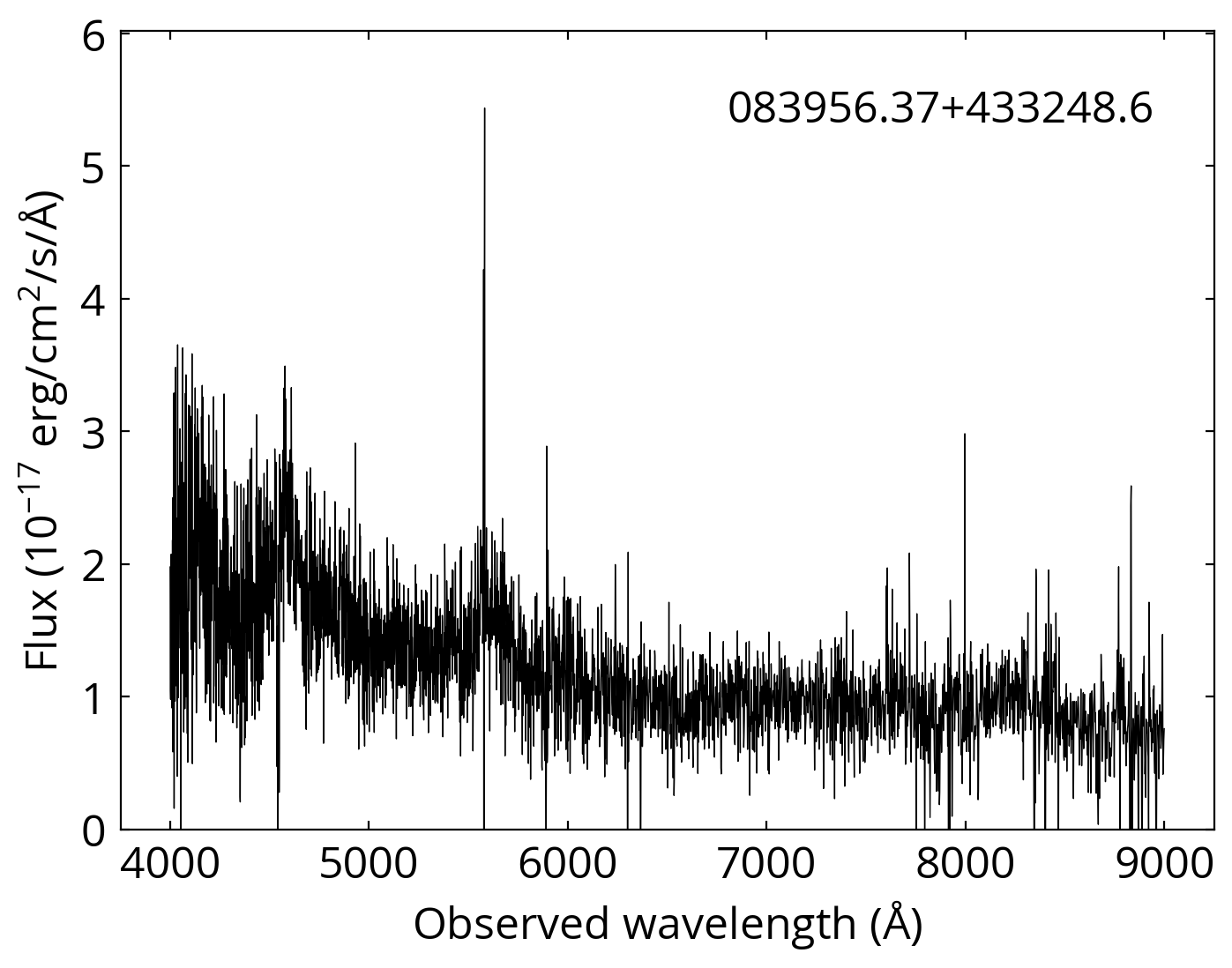}\hspace{0.01\textwidth}
\includegraphics[align=c,width=0.18\textwidth]{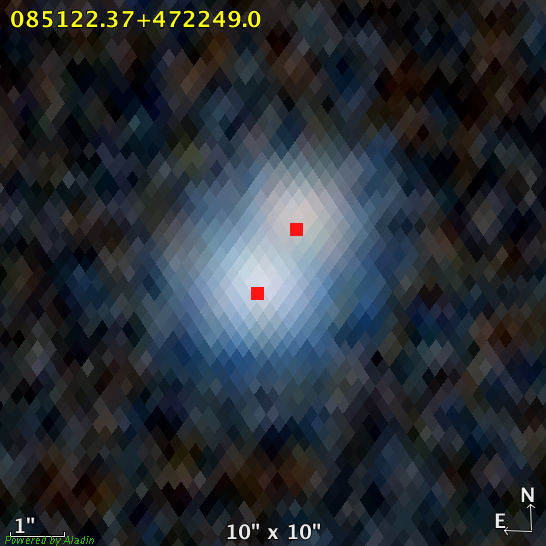}
\includegraphics[align=c,width=0.28\textwidth]{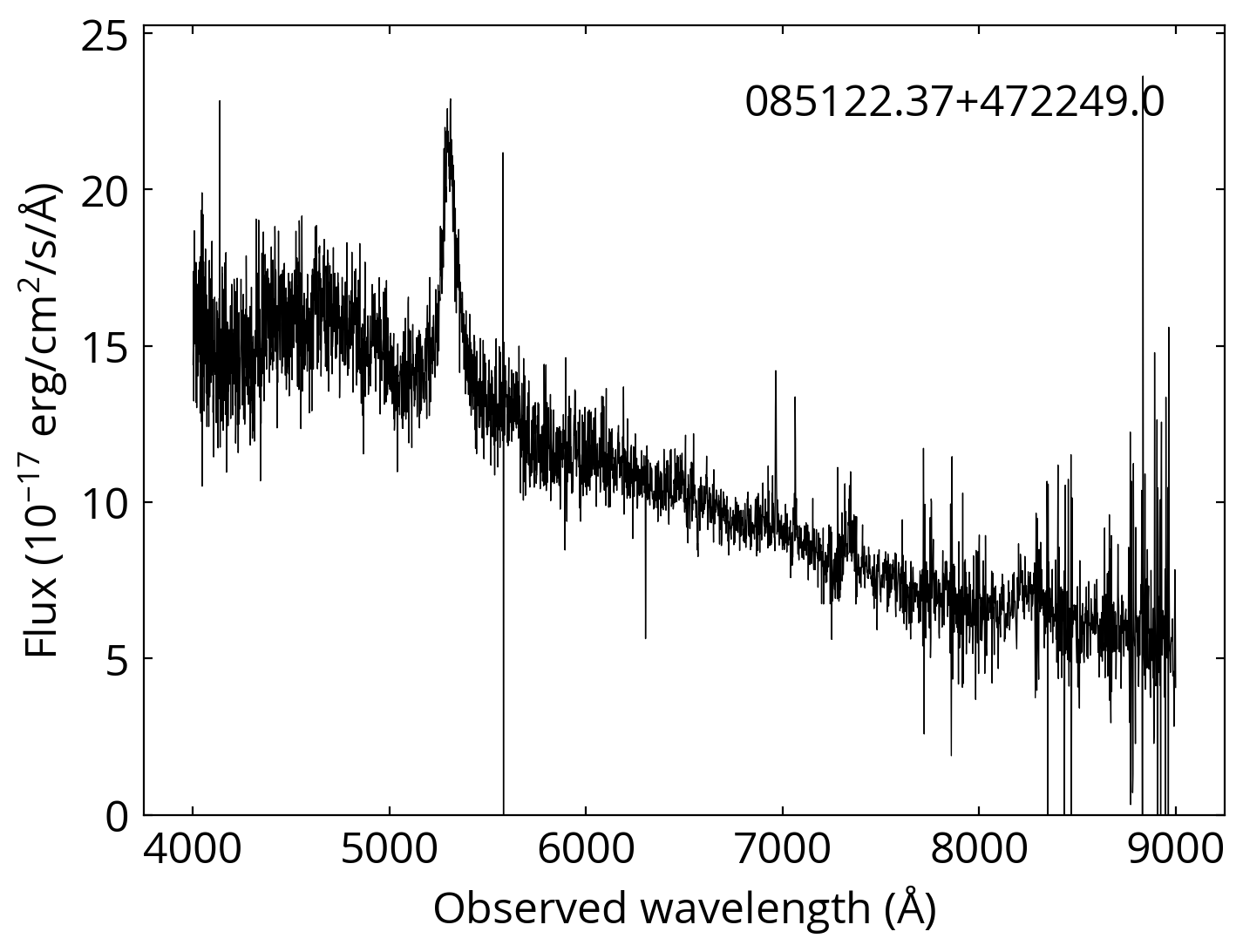}
\includegraphics[align=c,width=0.18\textwidth]{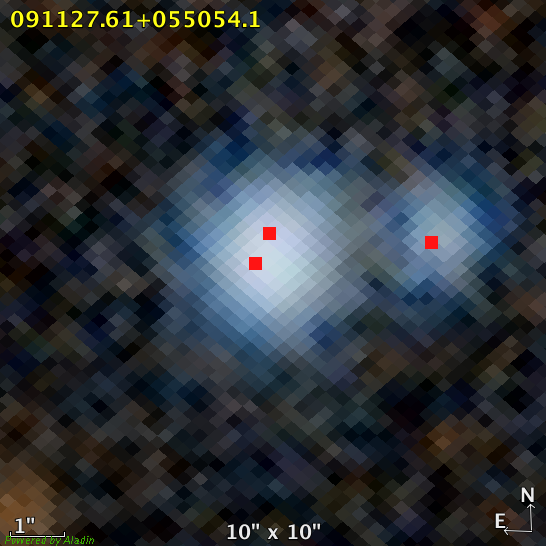} \hspace{0.012\textwidth}
\includegraphics[align=c,width=0.28\textwidth]{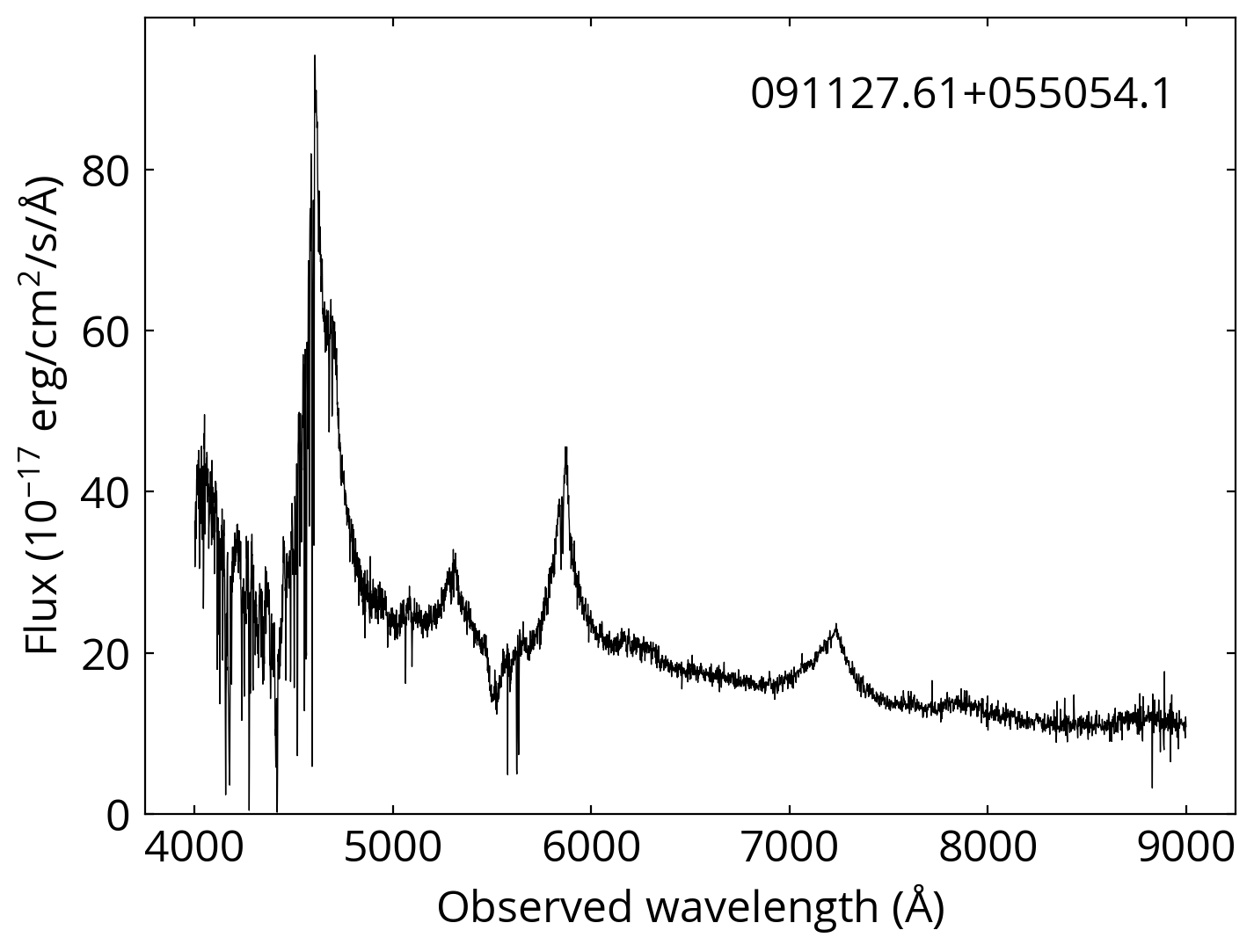}\hspace{0.025\textwidth}
\includegraphics[align=c,width=0.18\textwidth]{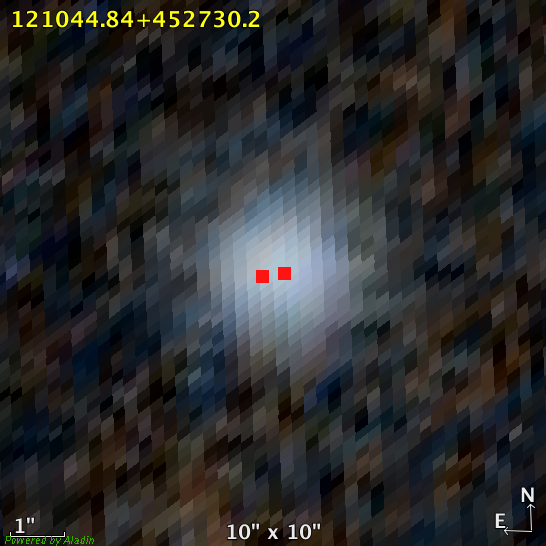} \hspace{0.012\textwidth}
\includegraphics[align=c,width=0.28\textwidth]{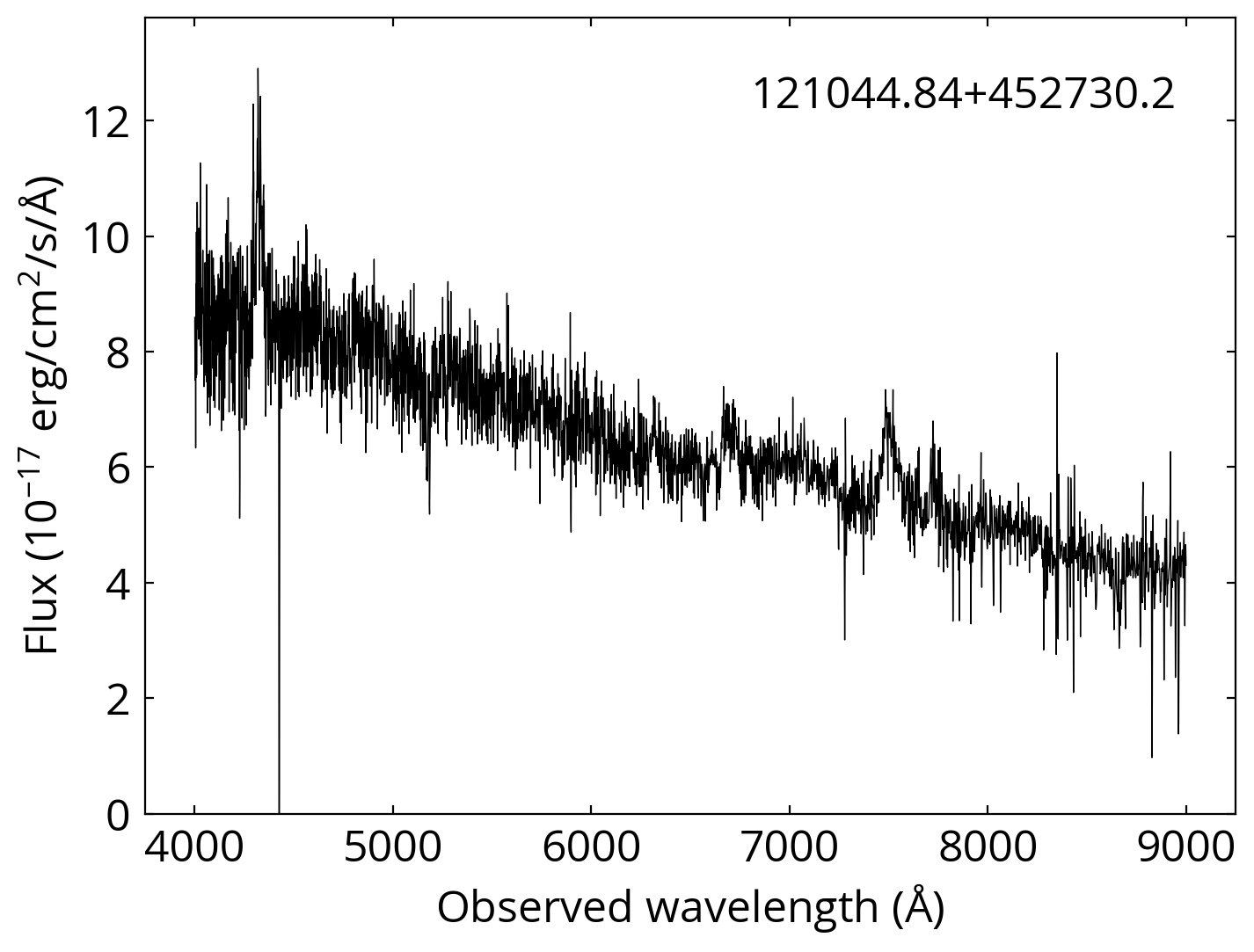}
\caption{Pan-STARRS 1 color-composite images (left) and SDSS spectra (right) for genuine quasars with non-zero proper motions from Gaia DR2. The red dots in the Pan-STARRS images are the Gaia detections at the J2015.5 epoch. Images have 10\arcsec\ on each side, and north (east) is up (left). }\label{fig:prop_1}
\end{figure*}

\begin{figure*}
\includegraphics[align=c,width=0.18\textwidth]{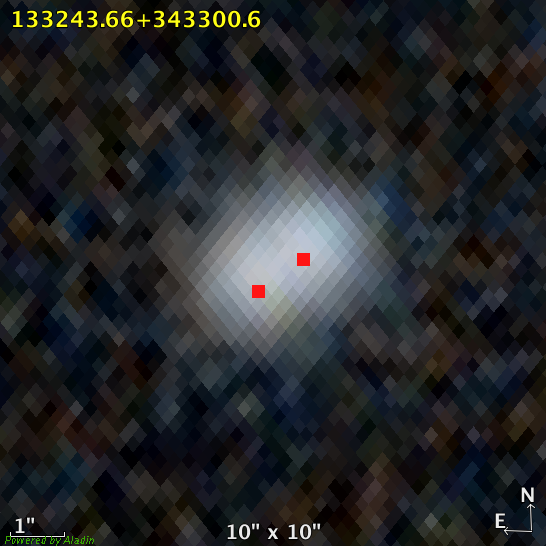}
\includegraphics[align=c,width=0.28\textwidth]{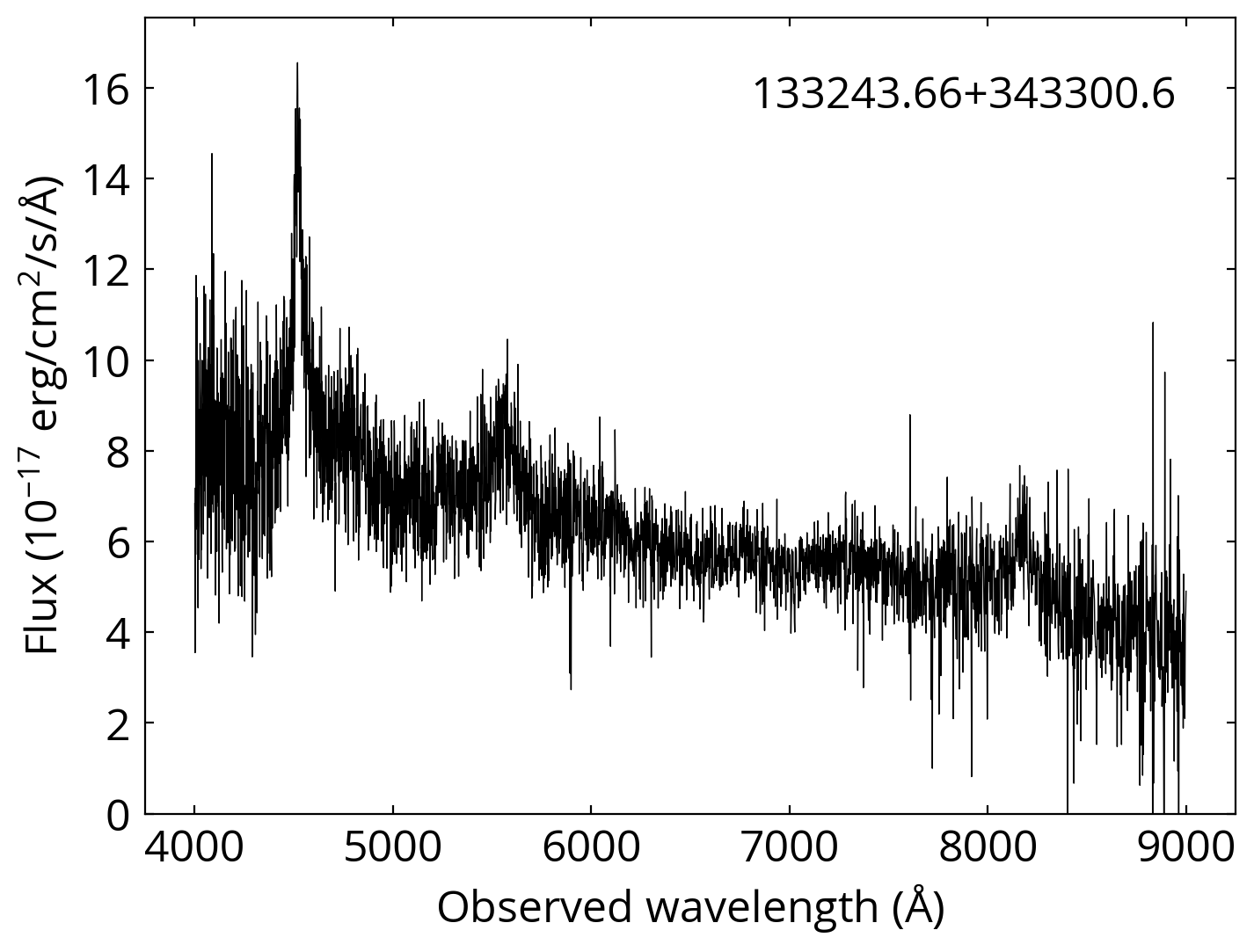}\hspace{0.01\textwidth}
\includegraphics[align=c,width=0.18\textwidth]{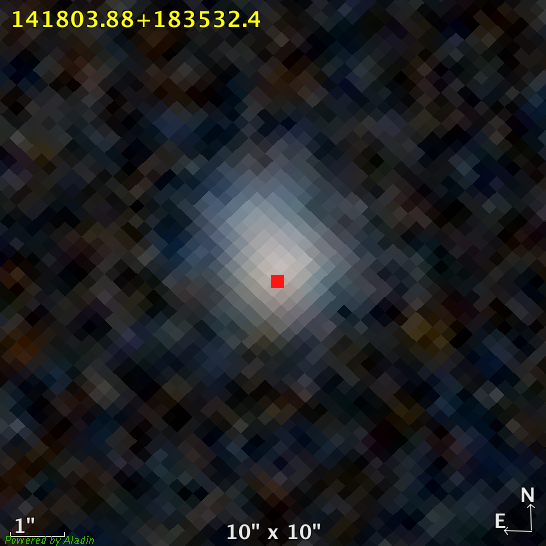}
\includegraphics[align=c,width=0.28\textwidth]{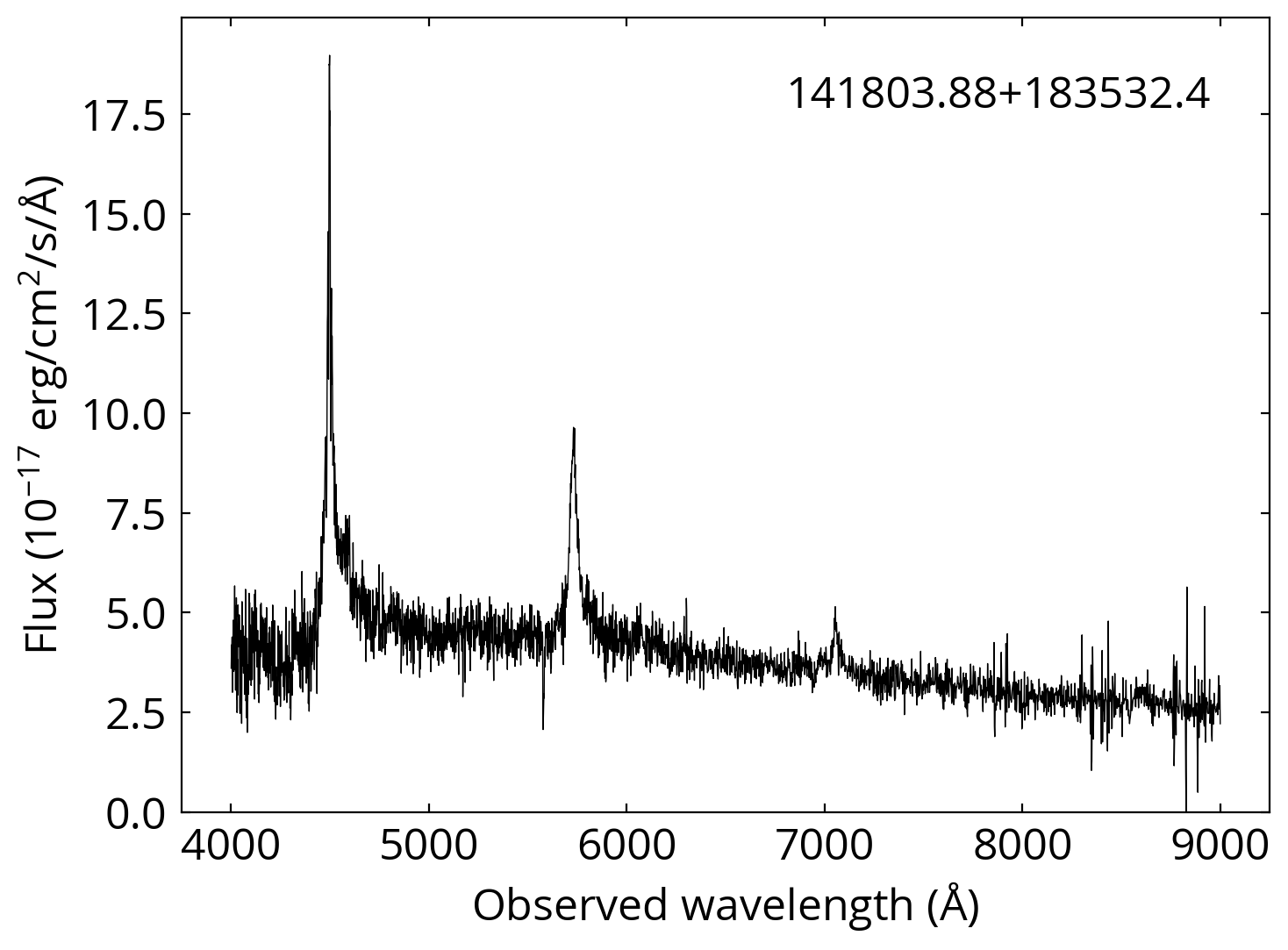}
\includegraphics[align=c,width=0.18\textwidth]{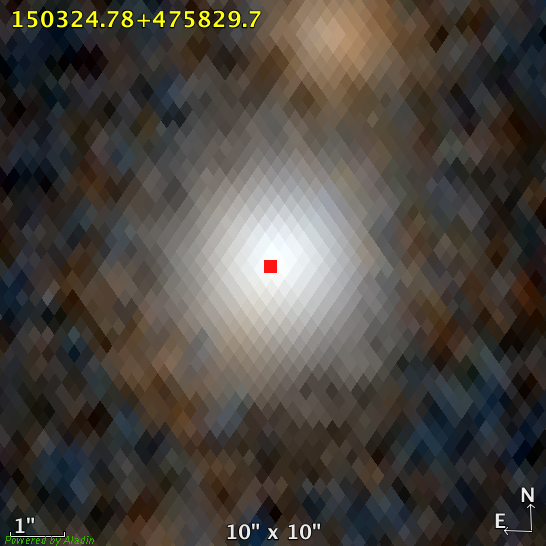}
\includegraphics[align=c,width=0.28\textwidth]{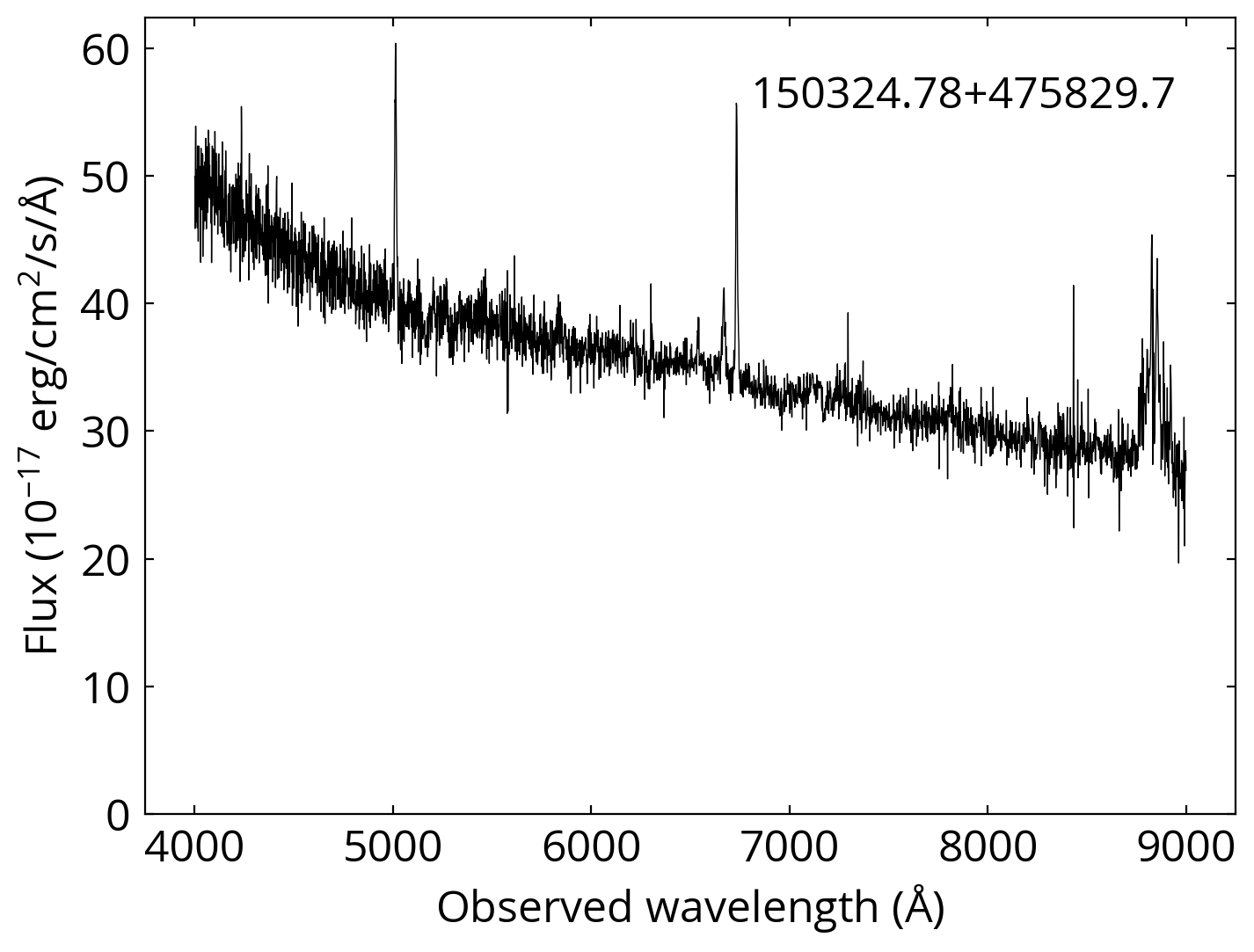}\hspace{0.01\textwidth}
\includegraphics[align=c,width=0.18\textwidth]{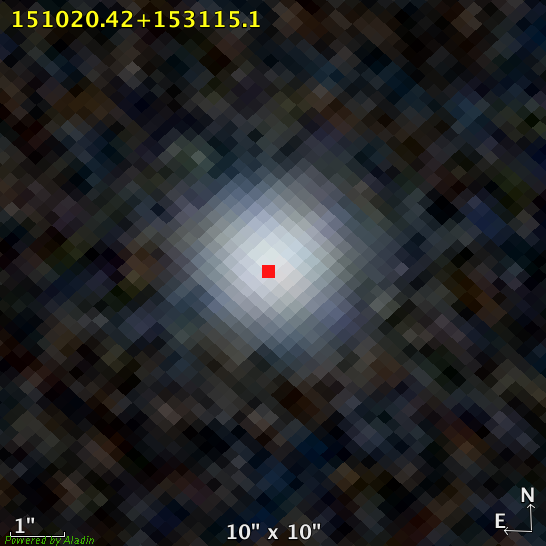}
\includegraphics[align=c,width=0.28\textwidth]{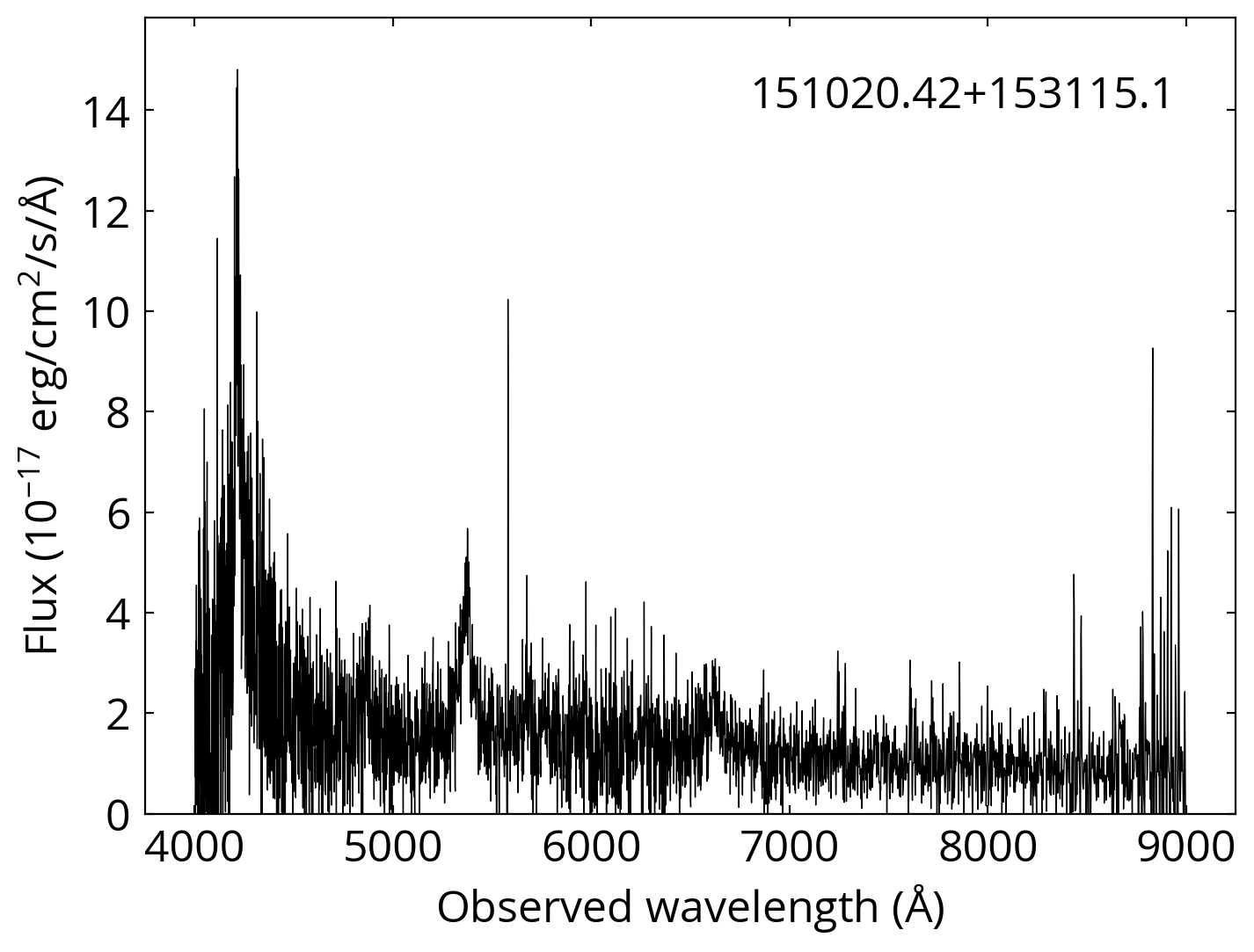}
\includegraphics[align=c,width=0.18\textwidth]{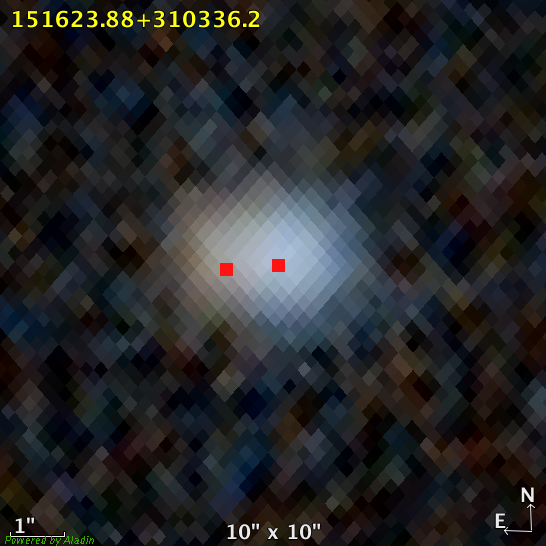}
\includegraphics[align=c,width=0.28\textwidth]{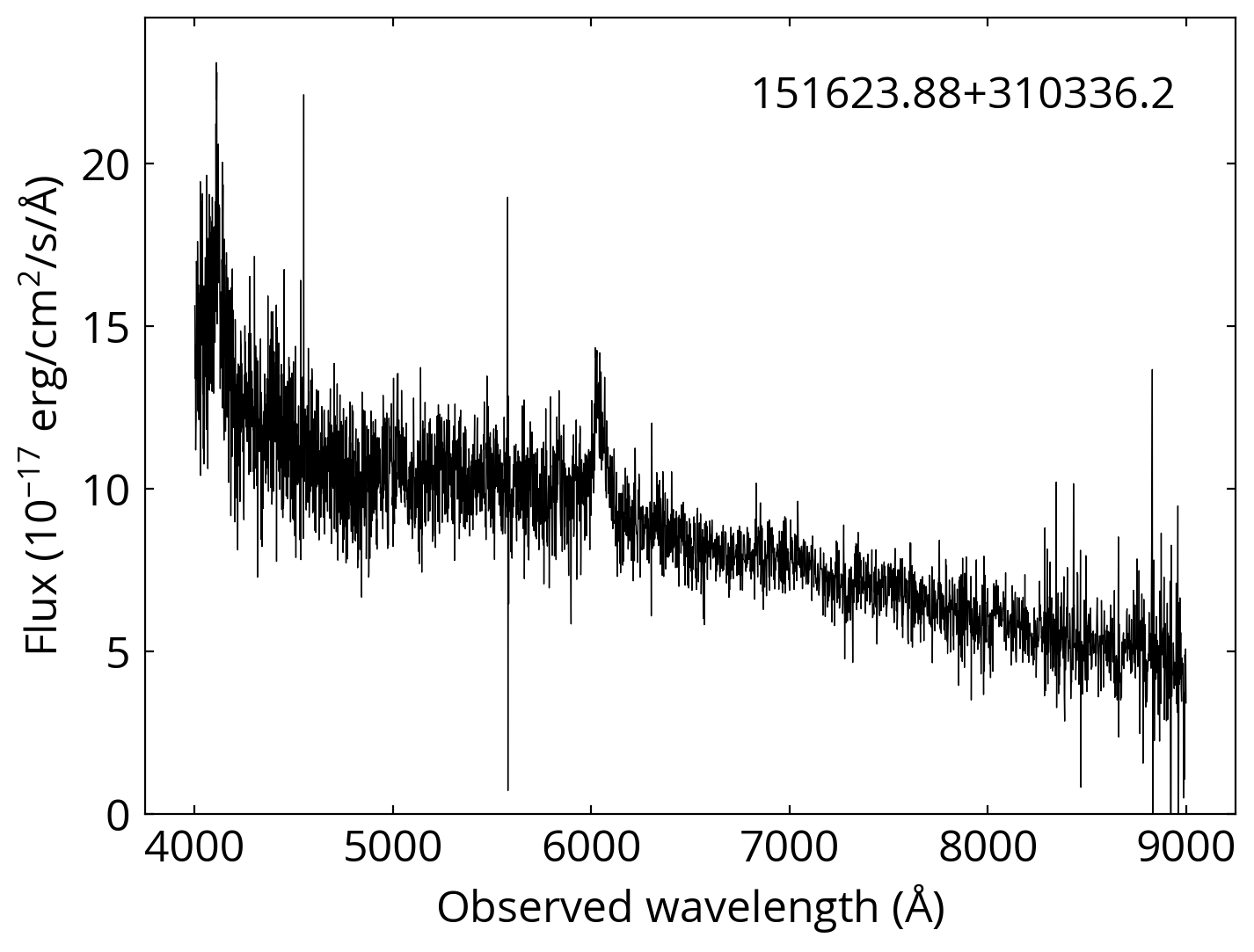}\hspace{0.01\textwidth}
\includegraphics[align=c,width=0.18\textwidth]{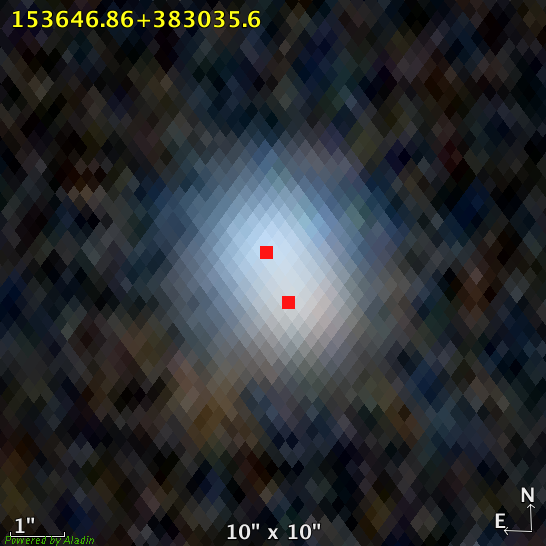}
\includegraphics[align=c,width=0.28\textwidth]{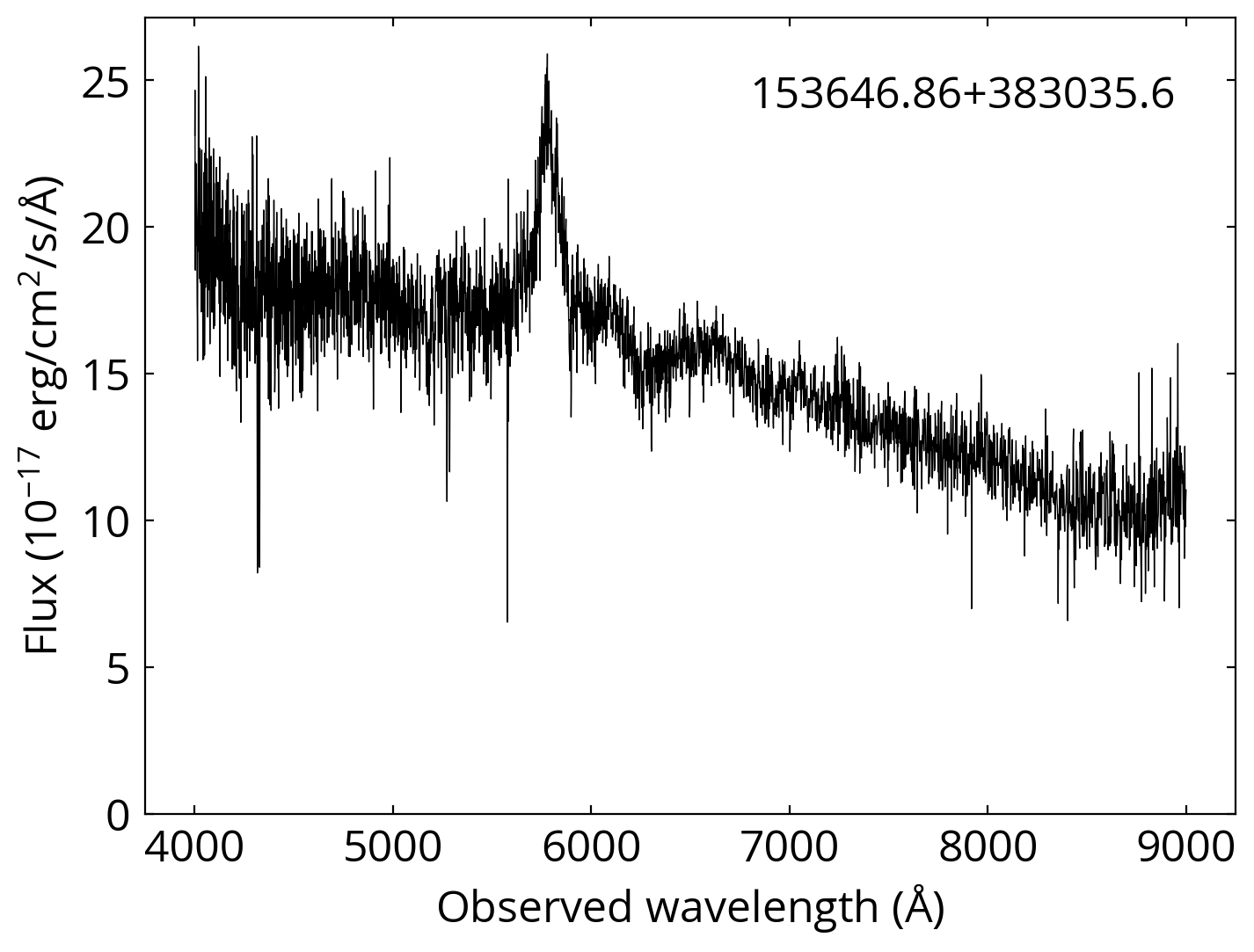}
\includegraphics[align=c,width=0.18\textwidth]{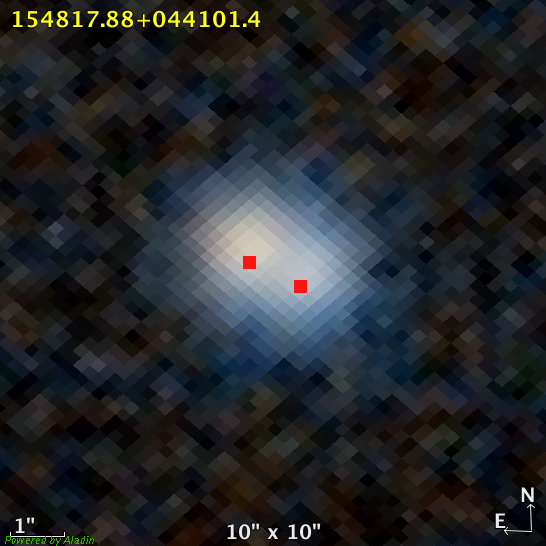}
\includegraphics[align=c,width=0.28\textwidth]{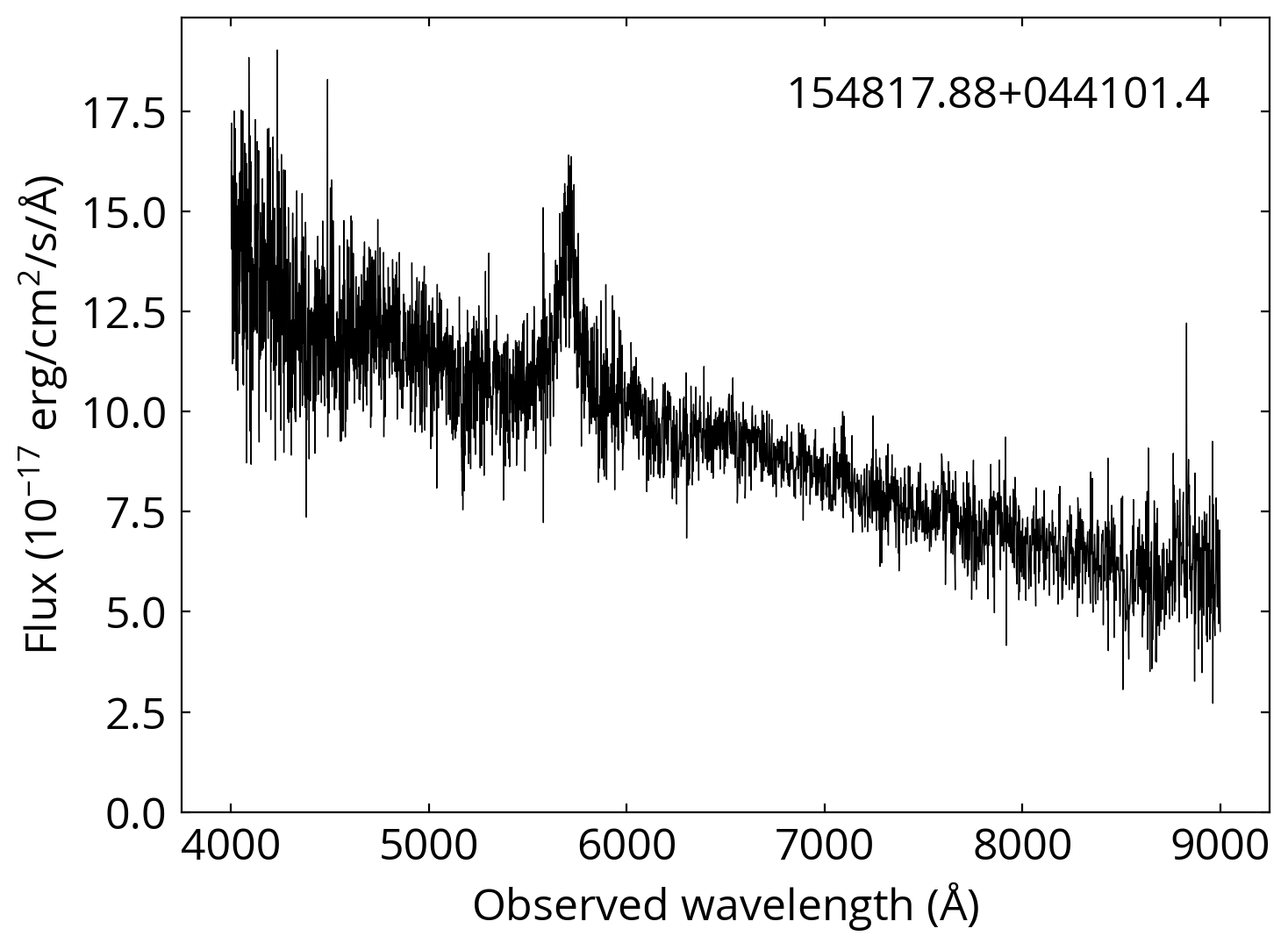}\hspace{0.01\textwidth}
\includegraphics[align=c,width=0.18\textwidth]{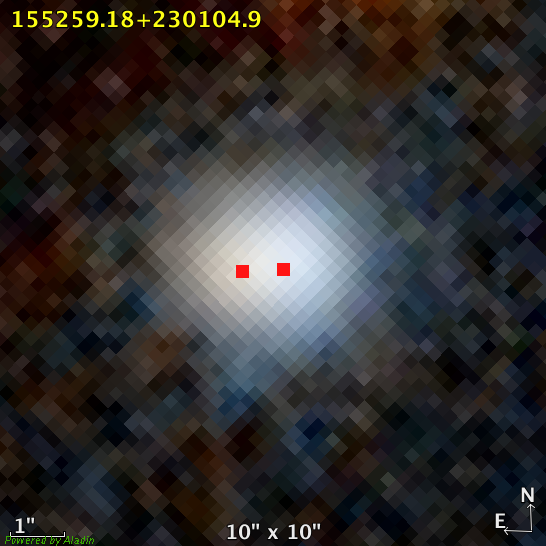}
\includegraphics[align=c,width=0.28\textwidth]{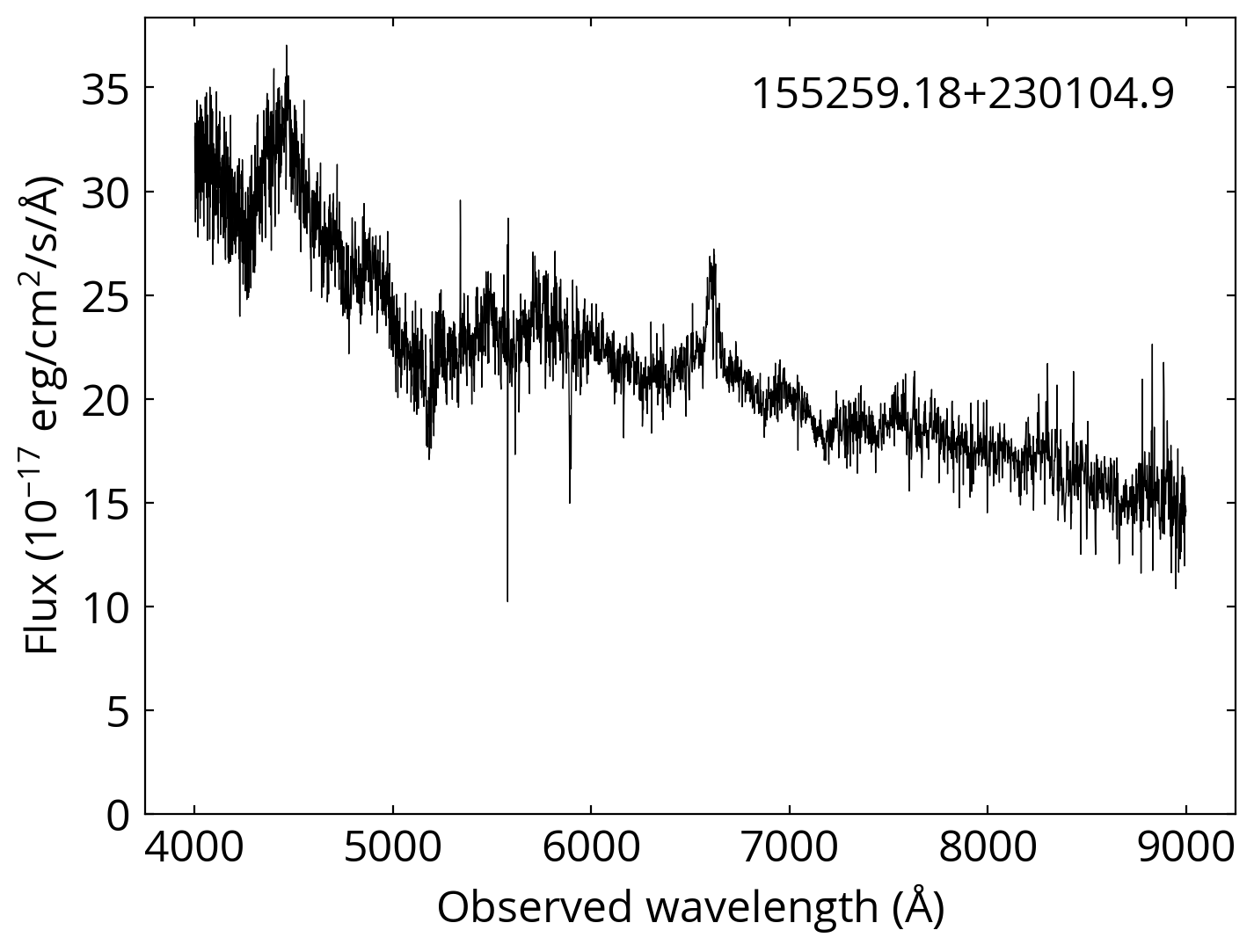}
\includegraphics[align=c,width=0.18\textwidth]{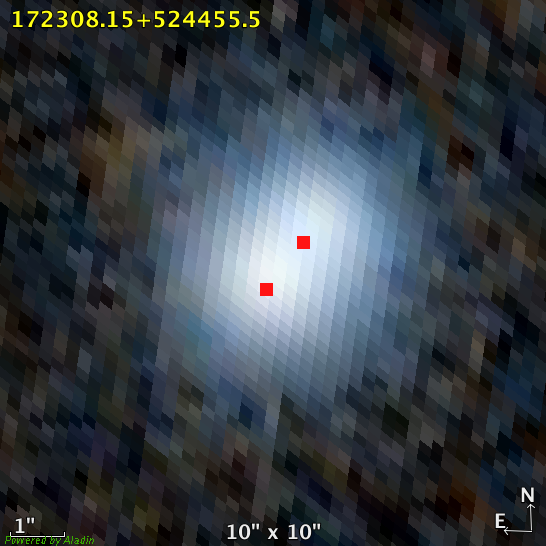} \hspace{0.012\textwidth}
\includegraphics[align=c,width=0.28\textwidth]{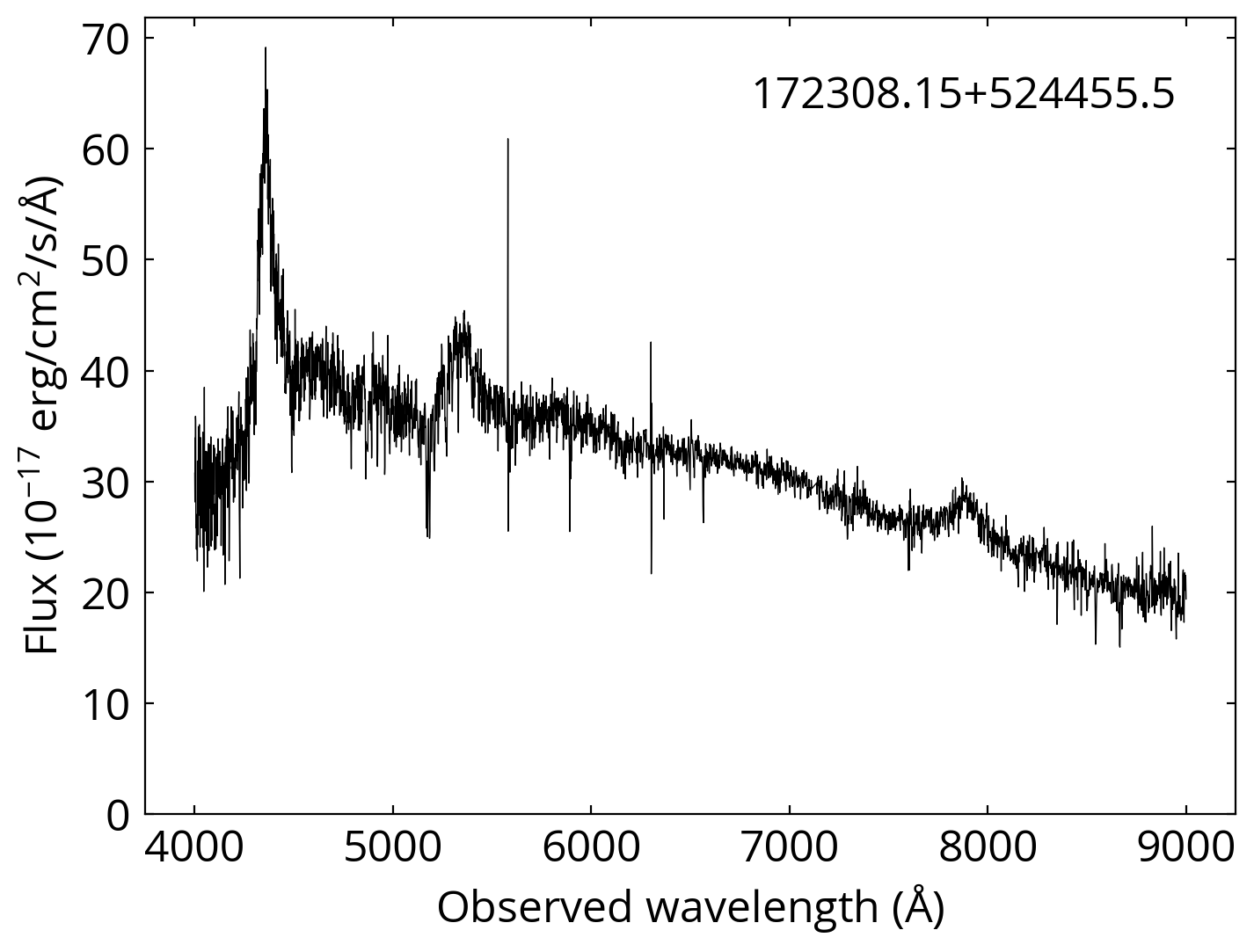}
\caption{Same as Fig.~\ref{fig:prop_1} but for a different set of objects. }
\end{figure*}

\end{document}